\documentclass[twocolumn]{aastex701}
\usepackage{multirow}
\usepackage{subcaption}  

\usepackage{enumitem}

\usepackage{amsmath}
\usepackage{amssymb}
\usepackage{bm}      
\usepackage{comment}

\usepackage{hyperref}
\usepackage[capitalise]{cleveref}

\usepackage{etoolbox}
\makeatletter
\patchcmd\H@refstepcounter{\protected@edef}{\protected@xdef}{}{}
\makeatother

\date{\today}

\begin{document}

\title{Rapid and robust simulation-based inference for kilonovae}

\author[0000-0003-2111-048X]{Stephanie M. Brown}
\affiliation{Oskar Klein Centre for Cosmoparticle Physics, Department of Physics, Stockholm University, Stockholm SE-106 91, Sweden}
\affiliation{Gravitation Astroparticle Physics Amsterdam (GRAPPA), University of Amsterdam, \\
Science Park 904, 1098 XH Amsterdam, The Netherlands}
\email{stephanie.brown@fysik.su.se}  

\author[0000-0002-8255-5127]{Mattia Bulla}
\affiliation{Department of Physics and Earth Science, University of Ferrara, via Saragat 1, I-44122 Ferrara, Italy}
\affiliation{INFN, Sezione di Ferrara, via Saragat 1, I-44122 Ferrara, Italy}
\affiliation{INAF, Osservatorio Astronomico d’Abruzzo, via Mentore Maggini snc, I-64100 Teramo, Italy}
\email{mattia.bulla@unife.it}

\author[0000-0002-2519-584X]{Hiranya V. Peiris}
\affiliation{Oskar Klein Centre for Cosmoparticle Physics, Department of Physics, Stockholm University, Stockholm SE-106 91, Sweden}
\affiliation{Institute of Astronomy and Kavli Institute for Cosmology, University of Cambridge, Madingley Road, Cambridge CB3 0HA, UK}
\affiliation{Cavendish Laboratory, Department of Physics, University of Cambridge, JJ Thomson Avenue, Cambridge, CB3 0HE, UK}
\email{hiranya.peiris@ast.cam.ac.uk}  

\author[0000-0003-2700-1030]{Nikhil Sarin}
\affiliation{Institute of Astronomy and Kavli Institute for Cosmology, University of Cambridge, Madingley Road, Cambridge CB3 0HA, UK}
\email{nsarin.astro@gmail.com} 

\author[0000-0002-0041-3783]{Daniel Mortlock}
\affiliation{Astrophysics Group, Imperial College London, Blackett Laboratory, Prince Consort Road, London, SW7 2AZ, UK}
\affiliation{Department of Mathematics, Imperial College London, London SW7 2AZ, UK}
\email{d.mortlock@imperial.ac.uk} 

\author[0009-0005-6323-0457]{Stephen Thorp}
\affiliation{Oskar Klein Centre for Cosmoparticle Physics, Department of Physics, Stockholm University, Stockholm SE-106 91, Sweden}
\affiliation{Institute of Astronomy and Kavli Institute for Cosmology, University of Cambridge, Madingley Road, Cambridge CB3 0HA, UK}
\email{sjt202@cam.ac.uk}

\author[0009-0004-7935-2785]{Gurjeet Jagwani}
\affiliation{Institute of Astronomy and Kavli Institute for Cosmology, University of Cambridge, Madingley Road, Cambridge CB3 0HA, UK}
\affiliation{Research Computing Services, University of Cambridge, Roger Needham Building, 7 JJ Thomson Ave, Cambridge CB3 0RB, UK}
\email{gj329@cam.ac.uk}

\author[0000-0002-3833-8520]{Stephan Rosswog}
\affiliation{Hamburg Observatory, Department of Physics, University of Hamburg, D-21029 Hamburg, Germany}
\affiliation{Oskar Klein Centre for Cosmoparticle Physics, Department of Astronomy, Stockholm University, Stockholm SE-106 91, Sweden}
\email{stephan.rosswog@astro.su.se}

\author[0000-0001-6573-7773]{Samaya Nissanke}
\affiliation{Deutsches Elektronen-Synchrotron DESY, Platanenallee 6, D-15738 Zeuthen, Germany}
\affiliation{Deutsches Zentrum für Astrophysik DZA, Postplatz 1, D-02826 Görlitz, Germany}
\affiliation{Institut für Physik und Astronomie, Universität Potsdam, D-14476 Potsdam, Germany}
\affiliation{Gravitation Astroparticle Physics Amsterdam (GRAPPA), University of Amsterdam, \\
Science Park 904, 1098 XH Amsterdam, The Netherlands}
\affiliation{Nikhef, Science Park 105, 1098 XG Amsterdam, The Netherlands}
\email{samaya.nissanke@desy.de}


\begin{abstract}

With the next generation of both electromagnetic and gravitational wave observatories beginning to come online, rapid analysis methods for kilonova data are becoming increasingly important in astronomy. Traditional Bayesian parameter estimation using Markov chain Monte Carlo (MCMC) is time-consuming and relies on explicit likelihood approximations that can break down when modeling uncertainties are significant. We develop a simulation-based inference (SBI) framework for kilonova parameter estimation using density-estimation likelihood-free inference. The framework uses a Gaussian process emulator trained on $\sim\!1300$ \textsc{possis} simulations. We demonstrate that SBI provides a rapid alternative to MCMC that is robust to likelihood misspecification. The standard Gaussian likelihood approximation fails to capture the non-Gaussian, correlated structure of emulator uncertainty; SBI learns this structure directly from forward simulations.
Simulation studies show that the SBI method accurately recovers injected parameters, while the MCMC suffers from systematic bias caused by likelihood misspecification. This problem persists when analyzing AT2017gfo, where a subset of the MCMC posteriors pile up at prior boundaries and the SBI posteriors do not. The SBI framework infers a total ejecta mass of $\sim 0.087 {\rm M}_{\odot}$ dominated by lanthanide-poor ejecta and excludes toroidal and peanut ejecta geometries at the 99th percentile for both components. The SBI framework generates $\sim\!2 \times 10^{4}$ posterior samples in seconds.

\end{abstract}

\section{Introduction}

The discovery of the binary neutron star merger GW170817 \citep{Abbott_2017a}, the coincident short gamma-ray burst GRB170817A \citep{Goldstein_2017,Savchenko_2017, LAT_2017}, and the associated kilonova AT2017gfo \citep{Coulter_2017,Villar_2017} marked the beginning of gravitational wave-based multimessenger astronomy \citep{Abbott_2017c}.

Subsequent analyses yielded constraints on the neutron star equation of state \citep{Bauswein_2017b,Ruiz_2018,Radice_2018a,Most_2018,Annala_2018, Hinderer_2019, Capano_2019, Raaijmakers_2020, Dietrich_2020}, independent measurements of the Hubble constant \citep{Abbott_2017d,Guidorzi_2017,Hotokezaka_2018,Dietrich_2020,Coughlin_2020a,Wang_2020}, confirmation that binary neutron star mergers produce at least a subset of short gamma-ray bursts \citep{Abbott_2017b}, and insight into the chemical evolution of our universe through $r$-process nucleosynthesis \citep{Cowperthwaite_2017,Evans_2017,Kasen_2017,Rosswog_2018,Kasliwal_2022}. 

Kilonovae are transients powered by rapid neutron capture nucleosynthesis with thermal emission in the ultraviolet, optical, and near-infrared bands \citep{Lattimer_1976, Li_1998, kulkarni_2005, Metzger_2010}. Despite extensive searches, only a small number of kilonova candidates have been identified \citep{Tanvir_2013, Troja_2018b, Troja_2019, Rastinejad_2022, Levan_2023, Stratta_2025}, and AT2017gfo remains the only event with a confirmed gravitational-wave counterpart. This landscape will change in the coming decades.
Next-generation electromagnetic facilities such as the Vera C.~Rubin Observatory Legacy Survey of Space and Time (LSST; \citealp{Ivezic_2019}), the James Webb Space Telescope \citep{Gardner_2023}, and the Nancy Grace Roman Space Telescope \citep{Schlieder_2024} will drastically increase the number of detected events. For instance, LSST is projected to detect $\mathcal{O}(10^2)$--$\mathcal{O}(10^3)$ kilonovae \citep{Andreoni_2021, Ragosta_2024, Scolnic_2019} during the duration of the survey. 
The next-generation of gravitational-wave detectors such as Cosmic Explorer \citep{Reitze_2019} 
and the Einstein Telescope \citep{Abac_2026} are expected to detect at least $\sim\!10^4$ binary neutron star mergers annually. Of these events $\sim \! 10^2$ events should have observable kilonovae \citep{Colombo_2025}. 

Analyzing such large samples will require inference frameworks that are both accurate and computationally efficient. Both traditional Markov chain Monte Carlo (MCMC; see e.g.\ \citealp{Geyer_2011}) and nested sampling algorithms \citep{Skilling_2004, Skilling_2006, Feroz_2009, Handley_2015, Speagle_2020} are commonly used when studying kilonova data. These samplers can require minutes to hours per event depending on the speed of the likelihood evaluation. A simple four parameter kilonova model can be fit in minutes \citep{Peng_2024, Desai_2025}, but computation time increases with model complexity \citep{Cole_2022, Ashton_2022}. New scalable methods that are capable of producing reliable posterior samples in seconds will be valuable for future data sets \citep{Darc_2024, Desai_2025}.

Kilonova emission is difficult to model due to the complex physics of the ejecta. Key complications include the ejecta geometry and composition (particularly the effect of neutrino transport on the electron fraction), wavelength-dependent opacities from $r$-process elements, viewing-angle dependencies, and uncertainties in radioactive heating and thermalization \citep{Metzger_2019}. State-of-the-art radiative transfer codes---including \textsc{Possis} \citep{Bulla_2019,Bulla_2023}, \textsc{SuperNu} \citep{Wollaeger_2014, Korobkin_2021}, and others \citep{Kasen_2013, Kasen_2017, Tanaka_2013, Collins_2024}---provide high-fidelity predictions but are too computationally expensive for direct likelihood-based inference. 

Surrogate models have therefore been developed to enable Bayesian inference on manageable timescales. Both neural network \citep{Lukoiute_2022,Anand_2023} and Gaussian process (GP; \citealp{Coughlin_2018, Coughlin_2019, Dietrich_2020, Pang_2023}) emulators have been used to predict kilonova light curves. However, due to the cost of generating training simulations, parameters such as ejecta geometry or electron fraction are often fixed \citep{Coughlin_2018, Dietrich_2020, Pang_2021, Pang_2023, Lukoiute_2022, Anand_2023, Peng_2024, King_2025}. In addition, many analyses are restricted to a subset of photometric bands (e.g.\ \textit{grizy} and \textit{JHK}) to reduce computational overhead. For example, AT2017gfo was observed in up to 37 distinct bands~\citep{Villar_2017}, but the commonly-used 8 bands include only $72\%$ of data points. An ideal framework should avoid this information loss.

The existing likelihood-based analyses rely on explicit $\chi^2$ or Gaussian likelihoods \citep{Coughlin_2018, Anand_2023, Pang_2023, Ristic_2023, Peng_2024, Jhawar_2025}. Often these neglect correlations in emulator and data uncertainties across time and wavelength. Most commonly, these analyses use a fixed systematic uncertainty (typically 1 magnitude) to account for emulator error \citep{Coughlin_2018, Coughlin_2019, Ristic_2022, 
Anand_2023, Pang_2023, Ristic_2023, Peng_2024, Jhawar_2025}. Recent work by \citet{Jhawar_2025} demonstrates the benefit of incorporating band- and time-dependent uncertainties, though their treatment simplifies the time dependence through piecewise interpolation.
These analytical likelihoods approximate the true likelihood, which may be non-Gaussian or highly correlated. As we show in this paper, the misspecification of the likelihood can lead to biased or overconfident posterior predictions.

Likelihood-free inference, also referred to as simulation-based inference (SBI; for a review see \citealp{Cranmer_2020}), offers an alternative. Density-estimation likelihood-free inference is an SBI method that directly trains a neural density estimator (e.g.\ a normalizing flow or diffusion model; \citealp{Papamakarios_2021, Kobyzev_2021, Yang_2023, Arruda_2025}) using forward simulations of data--parameter pairs \citep[see e.g.][]{Alsing_2019, Papamakarios_2019a}. An amortized neural posterior estimator (ANPE; \citealp{Papamakarios_2016, Lueckmann_2017}) learns the conditional density, $\mathcal{P}(\bm{\Phi}|\bm{d})$, of parameters $\bm{\Phi}$ given data $\bm{d}$, and once trained can be applied to new data to rapidly generate posterior samples without explicit likelihood evaluation.

In the kilonova context, SBI remains largely unexplored. \citet{Desai_2025} presents an SBI framework for inferring intrinsic parameters of kilonovae observed by the Zwicky Transient Facility (ZTF; \citealp{Bellm_2018}), but reduces the dimensionality by assuming a spherically symmetric single component model \citep{Kasen_2017} and marginalizing over distance and time, inferring only three parameters. Similarly, \citet{Darc_2024} applies SBI to kilonova spectra, but uses a two-component model from \textsc{KilonovaNet} \citep{Lukoiute_2022}, that only has free four parameters. These works demonstrate SBI's potential, but they rely on simplifying assumptions (e.g.\ fixing ejecta shape or electron fraction) that limit their ability to capture the full complexity of multi-component kilonova models.

In this paper we develop a GP emulator to enable SBI to be applied to a broader range of kilonova models. Our starting point is to focus on AT2017gfo as a canonical real world example in \cref{sec:at2017gfo}. We then present the \textsc{possis} kilonova simulations used here in \cref{sec:POSSIS}. \cref{sec:GP} describes the construction of the GP-based emulator, which is optimized for parameter inference through Bayesian emulator optimization (BEO). \cref{sec:likelihood} discusses the limitations of the standard likelihood approximations used in the MCMC analysis, and \cref{sec:SBI} presents the flow-matching based SBI framework. We compare the performance of SBI to likelihood-based MCMC on both simulated data and AT2017gfo in \cref{sec:results} and conclude with a discussion in \cref{sec:discussion}. For reference, Appendix~\ref{app:notation} provides a comprehensive list of symbols and 
notation used throughout the paper, organized by category.

The results here demonstrate that the analytical likelihood in standard MCMC  approaches leads to biased and overconfident posteriors due to likelihood misspecification; SBI avoids this bias and provides both faster and more robust parameter estimation. 

\begin{figure*}[t]
    \centering
    \includegraphics[width=\textwidth]{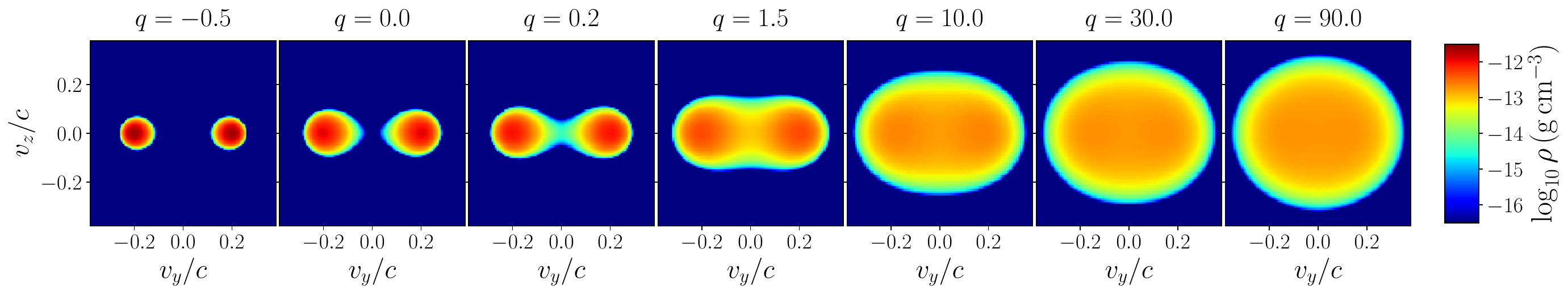}
    \caption{Velocity densities in the $v_y$-$v_z$ plane for single-component ejecta models 1~day after the merger, with densities described by Cassini ovals as in Eq.~\ref{eq:cassini}. Models are shown for increasing values of the shape parameter $q$ (from left to right) and share the same total ejecta mass, $m_{\rm ej}=0.1\,{\rm M}_\odot$, and mass-weighted averaged ejecta velocity, $v_{\rm ej}=0.2\,c$. The models are symmetric about the $z$ axis and the merger plane is defined by the $x$-$y$ plane.}
    \label{fig:cassini}
\end{figure*}

\section{AT2017gfo Data}
\label{sec:at2017gfo}

AT2017gfo is the most thoroughly observed kilonova to date. Therefore, we use AT2017gfo to compare the performance of SBI and likelihood-based methods on real data. Additionally, we take advantage of the multi-messenger nature of this event and incorporate constraints on the extrinsic parameters (distance and viewing angle) from other works.

The optical and near-infrared photometric data \citep{Andreoni_2017, Arcavi_2017, Coulter_2017, Cowperthwaite_2017, Diaz_2017, Drout_2017, Evans_2017, Hu_2017, Kasliwal_2017, Lipunov_2017, Pian_2017,  Shappee_2017, Smartt_2017, Tanvir_2017, Troja_2017, Utsumi_2017, Valenti_2017, Pozanenko_2018} were initially compiled by 
\citet{Villar_2017, Villar_2018}. The data is processed using the {\sc Redback}\footnote{\url{https://github.com/nikhil-sarin/redback}} package \citep{Sarin2024} and taken from the Open Astronomy Catalog API~\citep{oac_api}. 
This dataset contains 620 observations across 25 ultraviolet, optical, and near-infrared photometric bands (\textit{w, u, U, B, g, V, r, R, i, I, z, y, J, H, K} and F336W, F475W, F606W, F625W, F775W, F814W, F850W, F105W, F110W, and F160W) and spans 0.45 to 29.4 days post-merger. The Open Astronomy Catalog API catalog \citep{oac_api} provides a publicly available and reproducible data set with associated bandpass definitions and measurement uncertainties while containing all but 27 of the 647 measurements in the first compilation \citep{Villar_2017}. These points are excluded due to uncertain filter information.

The multi-messenger nature of this event provides independent constraints on the extrinsic parameters. The most precise distance measurement ($d_L = 40.7 \pm 2.36$ Mpc) comes from surface brightness fluctuation measurements of the host galaxy NGC 4993 \citep{Cantiello_2018}. Very long baseline interferometry (VLBI) observations of the radio signal constrain the viewing angle to $\theta_{\rm v} = 21.3 \pm 2.5$ degrees \citep{Mooley_2022}. These 
measurements are incorporated as informative priors during inference (see the prior specifications in \cref{sec:SBI}).

\section{Kilonova Models}
\label{sec:POSSIS}

Neutron star mergers eject matter via two main channels: dynamical ejecta and post-merger wind ejecta \citep{Nakar_2020}. These components differ in origin, time scale, and composition and can both contribute significantly to the observable kilonova emission.

Dynamical ejecta are caused by tidal torques and shock interactions that occur when the two neutron stars come to contact and are launched within milliseconds of coalescence. The ejecta are fast ($\sim\! 0.1$-$0.3 \, c$), anisotropic, and the tidal tails are particularly neutron-rich, making them a robust site for heavy $r$-process nucleosynthesis, including lanthanides \citep{Bauswein_2013, Sekiguchi_2015, Radice_2018b}. 

Post-merger wind ejecta are launched on longer timescales ($\sim\! 0.01$--$1$~s) from the remnant accretion disk and are driven by neutrino irradiation, viscous/angular-momentum transport, and magnetically-powered winds. Compared to the dynamical component, these ejecta are typically slower ($\sim\! 0.03$--$0.1~c$) and have a broader range of electron fractions; see \citet{Metzger_2019} for a review of kilonova ejecta components.

The relative contributions of the two components to the kilonova depend sensitively on the properties of the binary and the lifetime of the merger remnant.

We simulate neutron star mergers by modeling each component separately (\cref{subsec:1comp}) using the radiative transfer code \textsc{possis} (\cref{subsec:possis}); we then combine the two contributions to obtain the predicted kilonova emission (\cref{subsec:2comp}).

\subsection{Single Component Model}
\label{subsec:1comp}

Both the dynamical and wind ejecta processes can, taken in isolation, be described by a single model with four intrinsic physical parameters: 
the total ejecta mass, $m_{\rm ej}$;
the electron fraction, $Y_{\rm e}$; 
the mass-weighted averaged ejecta velocity, $v_{\rm ej}$; 
and the shape parameter, $q$, which controls whether the emission is toroidal ($q \simeq 0$) or spherical ($q \gg 1$). 
Each physical configuration is then observed from a viewing angle, $\theta_{\rm v}$, which is the polar angle defined from the axis of symmetry. The complete list of single-component parameters is hence $\bm{\phi} = (m_{\rm ej}, Y_{\rm e}, v_{\rm ej}, q, \theta_{\rm v})$.

The shape of the ejecta is defined by the Cassini oval prescription introduced in \citet{Korobkin_2021} and enforcing an oblate morphology. Specifically, the ejecta are assumed to be symmetric about an axis orthogonal to the merger plane, and their density distribution is given by\footnote{Equation~1 in \cite{Korobkin_2021} has a sign error (O.~ Korobkin, private communication).}
\begin{equation}
\rho(v_r,\theta)=\rho_0\bigg(\frac{t}{t_0}\bigg)^3\,\bigg[q-\bigg(\frac{v_r}{v_0}\bigg)^4-2 \,\bigg(\frac{v_r}{v_0}\bigg)^2\,\cos 2\theta\bigg]^3 
\label{eq:cassini}
\end{equation}
where $v_r$ and $\theta$ are axisymmetric coordinates, $\rho_0$ is the density at a reference time $t_0$ (determined by $m_{\rm ej}$), and $v_0$ sets the velocity scale (determined by $v_{\rm ej}$). The resultant geometry is shown in Fig.~\ref{fig:cassini} for several different values of $q$. 

\subsection{POSSIS}
\label{subsec:possis}

We use the 3D Monte Carlo radiative transfer code \textsc{possis}
\citep{Bulla_2019,Bulla_2023} to solve the single component model described above. In \textsc{possis}, the ejecta are represented on a three-dimensional Cartesian grid with $n_{\rm grid} = 100$ cells on each side, with spatial coordinates, densities, and composition in each cell defined at a reference time, and then evolved under the assumption of homologous expansion. Monte Carlo photon packets are then generated within the ejecta model according to the energy distribution from the decay of $r$-process nuclei and its subsequent thermalization. The distribution is evaluated by using pre-computed heating rates \citep{Rosswog_2024} and calculating thermalization efficiencies \citep{Barnes_2016,Wollaeger_2018}, which both depend on the local properties of the ejecta. The photon packets are then propagated as they diffuse through the expanding ejecta and interact with matter, using pre-computed wavelength-dependent opacities  \citep{Tanaka_2020}, which depend on local ejecta properties such as density, temperature, and electron fraction $Y_{\rm e}$. The computation time required per \textsc{possis} simulation was a minimum of 4.7 CPUh and a maximum of 3939 CPUh with a median of 76.1 CPUh. Multithreading with 36 threads was used to reduce computation time. 

Although \textsc{possis} adopts local prescriptions for heating rates, thermalization efficiencies, and opacities---quantities known to play a key role in kilonova modeling \citep{Bulla_2023,Sarin_2024,Brethauer_2024}---their implementation is subject to systematic uncertainties. For instance, opacities from \cite{Tanaka_2020}, computed up to ionization stage IV, are likely underestimated at early times ($t\lesssim0.5$--$1$\,d) when the ejecta can be hotter than $\sim \! 20\,000$\,K and more highly-ionized \citep{Banerjee_2020,Banerjee_2022}, potentially biasing the kilonova to higher luminosities, especially in near-UV and blue optical bands.
Moreover, the opacities are computed under the assumption of local thermodynamic equilibrium (LTE), which is likely to fail starting from a few days after the merger \citep{Pognan_2022}. There is also evidence \citep{Kato_2024} that the opacities from \cite{Tanaka_2020} are underestimated for singly-ionized lanthanides. These limitations are present in the forward model used to train the ANPE, and are therefore shared between the likelihood-based and the SBI analyses.

Spectral energy distributions (SEDs) are extracted for different viewing angles, whose locations are provided as inputs at the start of the simulation. SEDs are computed ``on the fly'' by spawning ``virtual'' packets at each interaction, sending them to the observers, and weighting their contribution by the probability that such an interaction would contribute to the escaping radiation \citep{Kerzendorf_2014}. This approach reduces numerical noise and significantly speeds up simulations compared to the standard approach of binning escaping photon packets \citep{Bulla_2015}. 

Given parameters ${\bm \phi}$ and a set of $J$ post-merger times $t_{1:J}$, \textsc{possis} outputs full SEDs for a pre-specified set observers at an (arbitrary) luminosity distance of $1$~Mpc. For each band $\beta \in \{w, u, U, \ldots, {\rm F110W}, {\rm F160W} \}$ we use \texttt{sncosmo} \citep{sncosmo_2025} to integrate the SEDs over the filter response to obtain a time series of photon fluxes\footnote{The photon flux is the rate at which photons pass through unit area at the observer, so has cgs units of photons/cm$^2$/s.}, $\Gamma_\beta(t_1, \bm{\phi}), \Gamma_\beta(t_2, \bm{\phi}), \ldots, \Gamma_\beta(t_J, \bm{\phi})$, at a (different) reference distance of $10$~pc.

\subsection{Two Component Models}
\label{subsec:2comp}

We obtain the final prediction for the kilonova emission by combining two single-component light curves corresponding to the dynamical and wind contributions. The full set of parameters for this two-component model is $\bm{\Phi} = ( \bm{\phi}_{\rm dyn}, \bm{\phi}_{\rm wind}, d_L) = (m^{\rm wind}_{\rm ej},m^{\rm dyn}_{\rm ej}, Y^{\rm wind}_{\rm e}, Y^{\rm dyn}_{\rm e}, v^{\rm wind}_{\rm ej}, v^{\rm dyn}_{\rm ej}, \theta_{\rm v}, q^{\rm wind}, q^{\rm dyn}, d_L)$, where $d_L$ is the luminosity distance to the system. The viewing angle $\theta_{\rm v}$ is the same for the two components, which reduces the effective number of parameters from 11 to 10. 

The components are differentiated by imposing two conditions: $Y_{\rm e}^{\rm wind} > Y_{\rm e}^{\rm dyn}$, so that the wind ejecta are relatively lanthanide-poor and the dynamical ejecta are lanthanide-rich; and $m_{\rm ej}^{{\rm wind}} \geq 0.01\,{\rm M}_\odot$ while $m_{\rm ej}^{\rm dyn} \leq 0.02\,{\rm M}_\odot$. 

The total predicted flux\footnote{The energy flux is the rate at which energy passes through unit area at the observer, so has cgs units of erg/cm$^2$/s.} in band $\beta$ at time $t$ is
\begin{equation}
\label{eq:distance_scaling}
f_\beta(t, \bm{\Phi})
  =  \frac{f_{\beta, 0}}{\Gamma_{\beta,0}} \left[
  \Gamma_\beta(t, \bm{\phi}_{\rm dyn})
  +
 \Gamma_\beta(t, \bm{\phi}_{\rm wind})
  \right] \left(\frac{d_L}{10 \; \textrm{pc}}\right)^{-2},
\end{equation} 
where $\Gamma_\beta(t, \bm{\phi}_{\rm dyn})$ and $\Gamma_\beta(t, \bm{\phi}_{\rm wind})$ are the photon fluxes for the dynamical and wind components, respectively, 
pre-computed by \textsc{possis} (\cref{sec:POSSIS}), and $\Gamma_{\beta,0}$ and $f_{\beta,0}$ are the AB zero-point photon and energy fluxes integrated over the bandpass for band $\beta$, obtained from {\sc sncosmo} and {\sc Redback} \citep{Sarin2024} respectively.

Equation~\ref{eq:distance_scaling} implicitly assumes that the interplay between the two components can be neglected, i.e., that photons generated in one component will eventually escape to the observer without propagating to and interacting with the other ejecta component. This simplifying assumption is common in the literature \citep[e.g.][]{Villar_2017}, but it can lead to biases in the predicted SEDs and light curves \citep[e.g.][]{Kawaguchi_2018,Bulla_2019}. However, the size of the training grid increases exponentially with the number of parameters, making it infeasible to produce a grid of two-component models where all physical parameters of both components are free. This modeling misspecification is relevant when analyzing observational data  (see \cref{sec:results}).

\section{Gaussian Process Emulator}
\label{sec:GP}

Both training an SBI framework and running likelihood-based inference can require a million or more model evaluations. The computational cost of the \textsc{possis} models makes this infeasible. Therefore, we develop a GP emulator (motivated by the work of \citealp{Rogers_2019, Bird_2019, Setzer_2023}) for the single-component kilonova simulations. These single-component emulators are then combined into a two-component model (as in \cref{subsec:2comp}).

\begin{table*}
    \centering
    \caption{Parameter values defining the original \textsc{possis} training grid. Each model is evaluated at $J = 82$ times logarithmically spaced between 0.31 and 29.18 days.
    }
    \label{tab:pars}
    \begin{tabular}{llr} \hline \hline
        parameter & symbol & values \\ \hline
        ejecta mass & $m_{\rm ej} $ [M$_{\odot}$] & 0.001, 0.003, 0.01, 0.03, 0.1  \\ 
        electron fraction & $Y_{\rm e}$ & 0.1, 0.2, 0.3, 0.4 \\ 
        ejecta velocity & $v_{\rm ej}$ $[c]$ & 0.05, 0.1, 0.15, 0.2, 0.25 \\ 
        shape &$q$ & $-$0.5, $-$0.2, 0, 0.2, 1, 1.5, 2, 2.5, 3, 10, 30, 90 \\ 
        viewing angle & $\theta_{\rm v}$ [$^{\circ}$] & 0.0, 25.9, 36.9, 45.6, 53.1, 60.0, 66.4, 72.5, 78.5, 84.3, 90.0 \\ \hline \hline
    \end{tabular}
\end{table*}

\subsection{Training Data}
\label{subsec:train_dat}

Our initial training set consists of $N = 1200$ single-component \textsc{possis} simulations evaluated at $11$ viewing angles equally spaced in cosine from the merger plane at $\cos\theta_{\rm v}=0$ to the pole where $\cos\theta_{\rm v}=1$. The parameter values defining this grid are listed in Table~\ref{tab:pars}. Each simulated light curve is evaluated at $J=82$ logarithmically spaced time points between 0.31 and 29.18 days. Earlier times were not included as \textsc{possis} models are less accurate at early times and the first observation for AT2017gfo was at 0.45 days.

This grid is the initial training set. We augment this training set using BEO \citep{Auer_2002a, Auer_2002b,Dani_2008, Rogers_2019} to improve model accuracy and parameter recovery during inference (see \cref{subsec:BEO}). 

The photon fluxes predicted by the \textsc{possis} simulations span many orders of magnitudes. Peak photon fluxes across the grid range from $4\times10^{10}$ to $3\times10^{13}\,{\rm photons\,cm^{-2}\,s^{-1}}$. For a single simulation, the typical flux values span $\sim \mathcal{O}(10^{-1}) -\mathcal{O}(10^{13})$, but time--band bins with a single photon packet can have fluxes as low as $10^{-40}\,{\rm photons\, cm^{-2}\, s^{-1}}$. Training directly on the fluxes would not only cause the Gaussian process regression to be numerically unstable but also cause the Gaussian-process fit to be entirely dominated by the brightest region of the light curves. Logarithmic transforms are commonly used to compress such dynamic ranges. However, the \textsc{possis} fluxes can be both zero and negative due to the Monte Carlo noise, making logarithmic transformation unsuitable. We hence transform the photon fluxes using an asinh compression, similar to the asinh ``magnitudes'' proposed by \cite{Lupton_1999}. This compresses large fluxes to a logarithmic scale while preserving linearity at low and negative fluxes. Under this prescription, an output photon flux $\Gamma_\beta(t, \bm{\phi})$ in band $\beta$ at time $t$ for single-component parameters $\bm{\phi}$ is transformed to be
\begin{equation}
\label{eq:mag_to_flux}
    \mu_{\beta}(t, \bm{\phi}) = - 2.5 \log_{10} (e) \left[ {\rm asinh}\! \left( \frac{\Gamma_{\beta}(t, \bm{\phi})/\Gamma_{\beta,0}}{2} \right) \right].
\end{equation}
For training, the band magnitudes are normalized by subtracting the mean light curve (computed for each time step using all light curves in the training set for that band) and then scaling by the global range (maximum minus minimum magnitude) across the entire training set for that band. 

The above process is reversed when predictions are made.  Given the mean magnitude prediction from the GP, $\bar{\mu}_\beta(t,\bm{\phi})$, the corresponding photon flux is 
\begin{equation}
\label{eq:asinh_to_flux}
\Gamma_\beta(t,\bm{\phi}) = 2 \Gamma_{\beta,0} \sinh \left[
- \frac{ \bar{\mu}_\beta(t,\bm{\phi}) }{2.5 \log_{10} e} \right].
\end{equation}
The non-linearity of this relationship must also be accounted for when propagating model uncertainty into flux space, as discussed in \cref{sec:likelihood}.

\subsection{Emulator Construction}
\label{subsec:gp}
We model each single-component kilonova using Gaussian process (GP) regression to map a five-dimensional vector of parameters, $\bm{\phi}$, 
to a vector of band-dependent asinh magnitudes evaluated at $J=82$ fixed time points, $t_{1:J}=(t_1,\dots,t_J)$. In each filter $\beta$, the light curve is $$\bm{\mu}_{\beta}(\bm{\phi})=[\mu_{\beta}(t_1;\bm{\phi}), \mu_{\beta}(t_1;\bm{\phi}), \dots, \mu_{\beta}(t_J;\bm{\phi})]^{\top} \; .$$
For training stability, we take the log of $m_{ej}$ and $q+1$ and use the cosine of $\theta_{\rm v}$.
Unlike standard GP formulations, which map inputs to scalar outputs, here each parameter vector maps to a light curve (vector output). Each photometric band is emulated independently.

A GP is fully specified by its mean function and covariance matrix $\bm{K}(\bm{\phi},\bm{\phi}';\bm{\psi})$. In this work, we model the normalized magnitudes in band $\beta$ as
\begin{equation}
\Tilde{\bm{\mu}}_{\beta}(\bm{\phi},t_j) \sim 
\mathcal{GP}\!\left[\bm{0},\, \bm{K}_{\beta}(\bm{\phi},\bm{\phi}';\bm{\psi}_{\beta})\right],
\end{equation}
where $\bm{\psi}_{\beta} = (\sigma_{0,\beta}^2, \ell_{\beta,1:5}, \sigma^2_{w,\beta})$ are the kernel's hyperparameters in a given band (shared among all time points), and the zero-mean condition is enforced by subtracting the mean light curve across the training set in band $\beta$. We adopt a radial basis function kernel,
\begin{equation}
\label{eq:RBF_apjl}
k_{\beta}(\bm{\phi}, \bm{\phi}') =
\sigma_{0,\beta}^2
\exp\!\left[- \frac{1}{2} \sum_{p=1}^5
\frac{({\phi}_p-{\phi}_p')^2}{\ell_{\beta,p}^2}
\right],
\end{equation}
where $\ell_{\beta,p}$ is the characteristic length scale for parameter $p$ and $\sigma_{0,\beta}$ is the amplitude. A white-noise term $\sigma_{w,\beta}^2$ accounts for Monte Carlo noise in the \textsc{possis} simulations such that the covariance matrix in each band is defined by
\begin{equation}
(\bm{K}_{\beta})_{mn} = k_{\beta}(\bm{\phi}_m,\bm{\phi}_n) + \delta_{mn}\sigma_{w,\beta}^2 \quad m,n = 1, \dots, N \; , 
\end{equation}
where $N$ is the number of training simulations and $\bm{K}_{\beta} \in \mathbb{R}^{N \times N}$.
The kernel is independent of time. Temporal smoothness in predictions arises because the shared kernel leads to similar magnitudes at adjacent times.

The hyperparameters are optimized using \texttt{GPyTorch} \citep{Gardner_2018}. We minimize the average negative marginal log-likelihood across all time steps. Each time point is weighted equally to ensure the emulator performs consistently for all times. The loss is
\begin{multline}
\label{eq:gp_loss_all}
\mathcal{L}(\bm{\psi}_\beta) =  -\frac{1}{J}\sum_{j=1}^{J}
\bigg[ -\frac{1}{2} \Tilde{\bm{\mu}}_\beta(t_j)^{\top} \bm{K}_\beta^{-1}
\Tilde{\bm{\mu}}_\beta(t_j) \\- \frac{1}{2}\log|\bm{K}_\beta| - \frac{N}{2}\log(2\pi) \bigg] \;.
\end{multline}
Optimization is performed using the \texttt{AdamW} algorithm \citep{Kingma_2017,Loshchilov_2019}.

Given a test point $\bm{\phi}_*$, the predicted mean light curve $\bar{\bm{\mu}}_{\beta,*}$ and the associated covariance $\mathrm{Cov}(\bar{\bm{\mu}}_{\beta,*})$ are

\begin{align} 
\label{eq:gp_mean} 
\bar{\bm{\mu}}_{\beta,*} &= \bm{K}^{\top}_{\beta,*} \bm{K}^{-1}_{\beta} \bm{Y}_{\beta}, \\
\label{eq:gp_cov} 
\mathrm{Cov}(\bar{\bm{\mu}}_{\beta,*}) &= \bm{K}_{\beta,**} - \bm{K}_{\beta,*}^\top \bm{K}_{\beta}^{-1} \bm{K}_{\beta,*} ,
\end{align}
where $\bm{K}_{\beta,*} = \bm{K}_{\beta}(\bm{\phi}_*,\bm{\phi}; \bm{\psi}_{\beta}) $,
$\bm{K}_{\beta,**} = \bm{K}_{\beta}(\bm{\phi}_*,\bm{\phi}_*; \bm{\psi}_{\beta})$, and $\bm{Y}_{\beta} \in \mathbb{R}^{N \times J}$ is the training set in a given band and is defined $\bm{Y}_{\beta,ij} = \Tilde{\bm{\mu}}_{\beta}(t_j;\bm{\phi}_i)$. The predictive variance provides a direct estimate of interpolation uncertainty, which is propagated through the two-component model (see Eq.\ \ref{eq:var_prop}). Because the emulator is trained on a fixed time grid, predictions at arbitrary observation times are obtained via linear interpolation between adjacent grid points. The time sampling is dense enough that time interpolation errors are much less than model uncertainty and have a negligible effect on the light curves or inferred parameters.

To enable rapid generation of multi-band light curves, we implement a custom batched GP in \texttt{PyTorch}. Covariance matrices $\bm{K}_{\beta} \in \mathbb{R}^{N \times N}$ for all $B=25$ bands are stacked into a tensor $\bm{\mathcal{K}} = (\bm{K}_{1}, \dots ,\bm{K}_{25}) \in \mathbb{R}^{B \times N \times N}$.
This allows us to precompute $\bm{\mathcal{K}}^{-1}$ and $\bm{\mathcal{K}}^{-1}\bm{\mathcal{Y}}$, where $\bm{\mathcal{Y}} = (\bm{Y}_1, \dots, \bm{Y}_{25}) \in \mathbb{R}^{B \times N \times J}$ are the stacked training sets for all bands. The predictions at an arbitrary test point $\bm{\phi}_*$ in all bands are then evaluated using batched tensor operations. This enables simultaneous evaluation across all 25 bands and 82 times with significantly reduced memory requirements. An implementation with a separate covariance matrix for every band--time combination would require $25 \times 82 = 2050$ covariance factorizations, which could not be loaded simultaneously on a single NVIDIA HGX A100 GPU for the training-set size used here. By sharing one covariance structure across all 82 time outputs within each band, our implementation reduces the dominant covariance storage by a factor of $82$. By precomputing $\bm{\mathcal{K}}^{-1}\bm{\mathcal{Y}}$, we decrease the time at each mean prediction by $0.22\,{\rm s}$. Simultaneous multi-band mean predictions are approximately $2\times$ faster than sequential per-band evaluation.

For the likelihood-based analysis, the GP mean prediction is used (see \cref{sec:likelihood}). In contrast, the forward simulations (\cref{subsec:forward_model}) used to train the SBI framework depend on random draws from the GP posterior\footnote{A vector of random draws $\bm{x}$ can be drawn from a multivariate Gaussian with mean $\bm{m}$ and covariance $\bm{C}$ by computing $\bm{x} = \bm{m} + \bm{L} \bm{z}$, where $\bm{L}$ is the Cholesky decomposition of $\bm{C}=\bm{L}\bm{L}^\top$, and $\bm{z}$ is a vector of random draws from $\mathcal{N}(0,1)$.}:
\begin{equation}
\label{eq:gp_pred_dist}
    \bm{\mu}_{\beta}(\bm{\phi}_*) \sim 
\mathcal{N}\!\left[ \bar{\bm{\mu}}_{\beta}(\bm{\phi}_*),\, \mathrm{Cov}(\Tilde{\bm{\mu}}_{\beta,*}) \right].
\end{equation}
The resulting asinh magnitudes are interpolated to the observation times.

\subsection{Bayesian Emulator Optimization}
\label{subsec:BEO}

The emulator was initially trained on the model grid described in \cref{subsec:train_dat}. We then iteratively augment the training set using BEO. At each iteration, we start with a GP trained on the existing training data, estimate the best parameter values using an acquisition function, and add \textsc{possis} simulations with those parameters to the training set.

Following \citet{Rogers_2019}, we adopt a modified GP upper confidence bound acquisition strategy \citep{Auer_2002a,Auer_2002b,Dani_2008} that is designed to improve posterior recovery rather than solely minimizing GP interpolation error. As AT2017gfo is used in this work to compare the performance of our SBI framework to a likelihood-based MCMC analysis, we optimize our training set to recover the likelihood-based posterior of AT2017gfo (see \cref{sec:likelihood}).

\begin{figure}[ht]
    \centering
    \includegraphics[width=\linewidth]{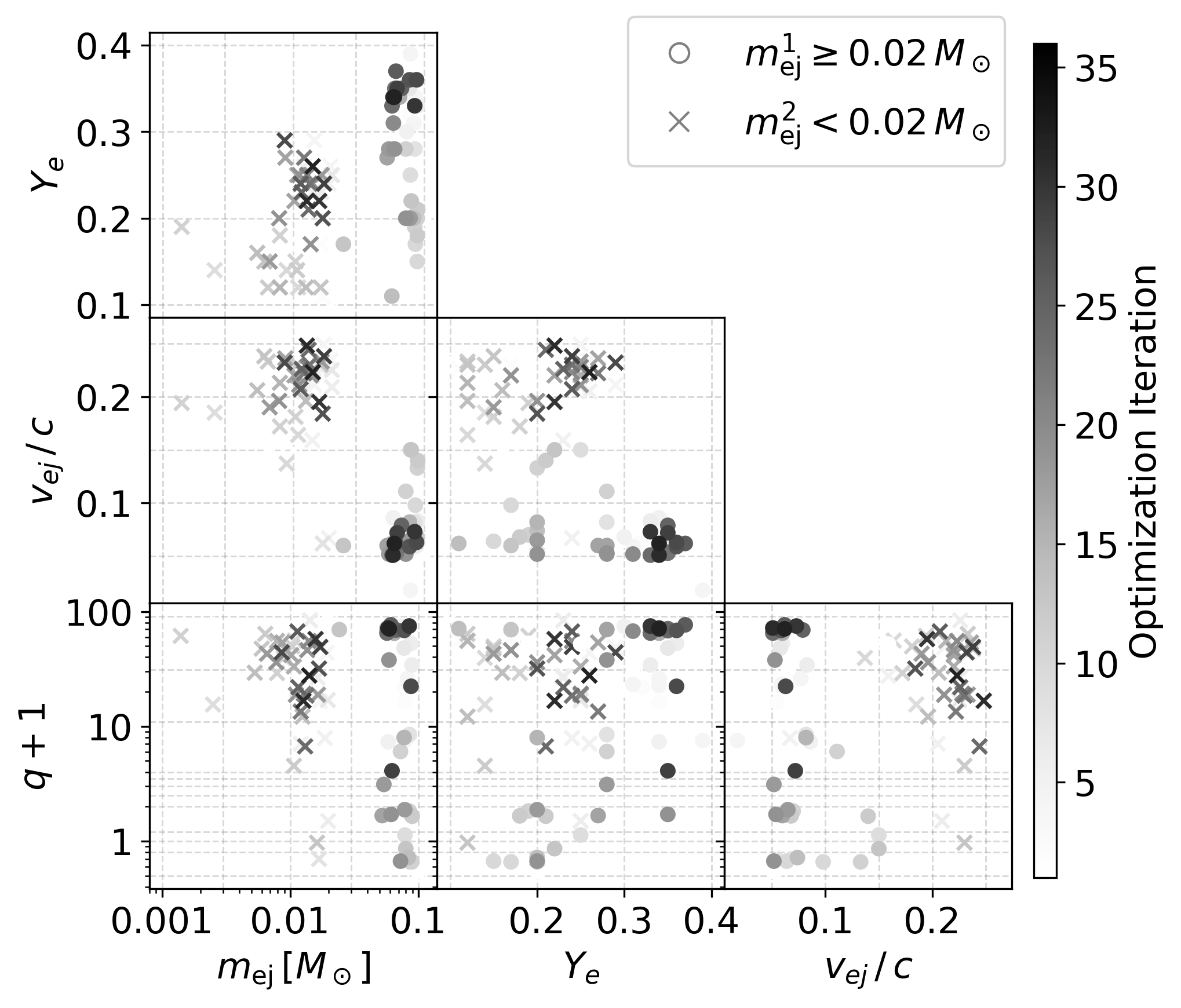}
    \caption{Physical parameters (mass, $m_{\rm ej}$; electron fraction, $Y_{\rm e}$; velocity, $v_{\rm ej}$; and shape, $q$) of points added to the training set as a function of optimization iteration. Wind ejecta points ($m_{\rm ej} \geq 0.02 \, {\rm M}_{\odot}$) are shown as dots, and dynamical ejecta points ($m_{\rm ej} < 0.02 \, {\rm M}_{\odot}$) are shown as crosses. The original training grid is shown as lines.}
    \label{fig:bo_params}
\end{figure}

\begin{figure*}[ht]
    \centering
    \includegraphics[width=\linewidth]{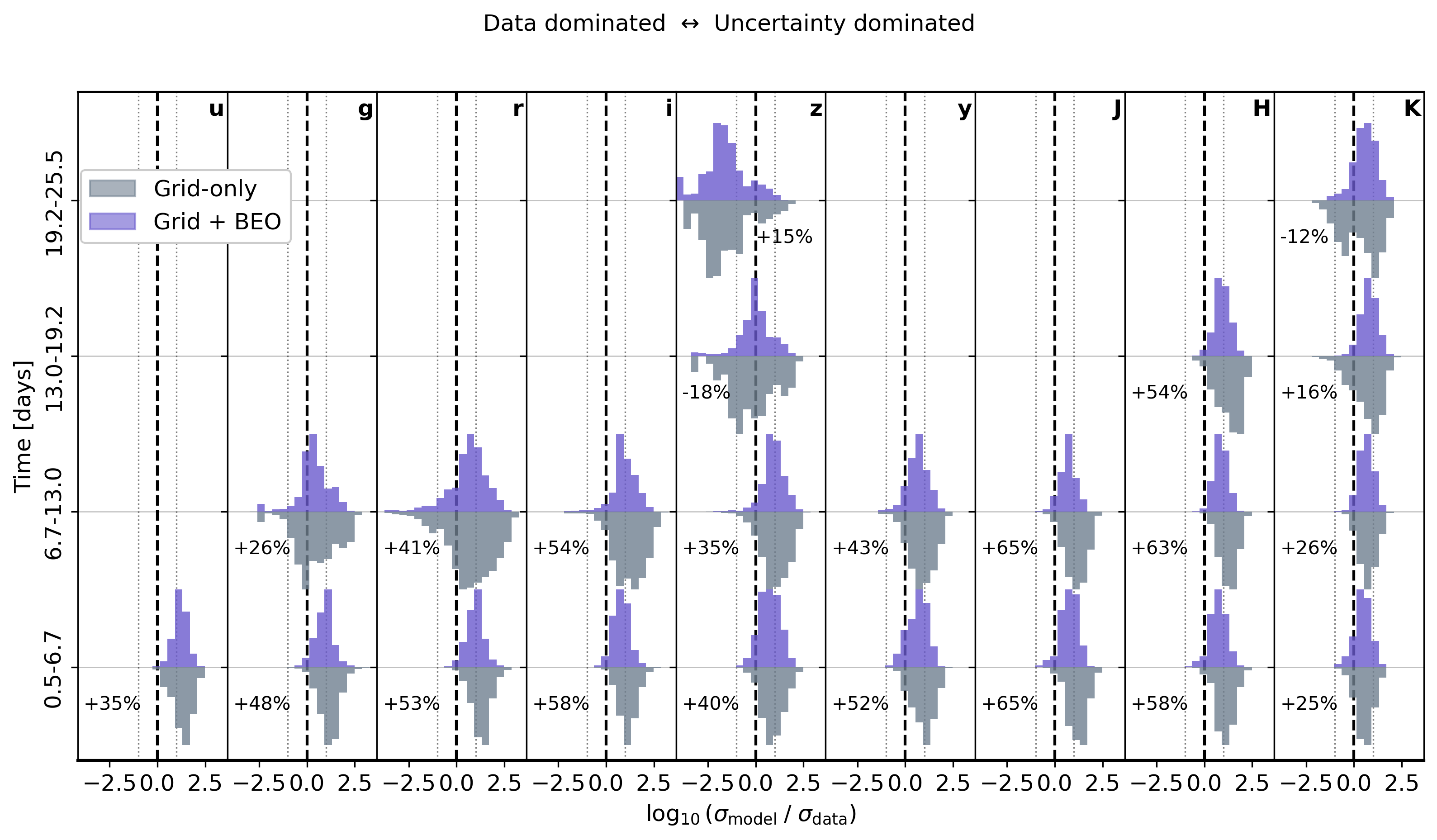}
    \caption{Logarithm of the ratio of emulator-to-data uncertainty in flux, sorted by wavelength (left to right) and binned time (bottom to top). Values $>0$ indicate that emulator error dominates the likelihood, while values $<0$ indicate that observational uncertainty dominates. Predicted emulator error for BEO models with $m_{\rm ej}>0.02\, {\rm M}_{\odot}$ for grid-only model (grey) and grid+BEO (purple).  Light vertical dashed lines indicate $\log_{10} (\sigma_{\rm{model}}/\sigma_{\rm{data}})= -1$ and $1$ and the heavy dashed line shows $\log_{10} (\sigma_{\rm{model}}/\sigma_{\rm{data}})= 0$ or equivalently $\sigma_{\rm{model}} = \sigma_{\rm{data}}$. The distributions contain 62 BEO parameter sets, each of which is evaluated at 13 viewing angles. The number of data points differs between panels because the AT2017gfo observations are not uniform in time or wavelength. We therefore show density-normalized histograms using 20 bins in each panel. The median improvement is indicated.}
    \label{fig:lvr_8x5}
\end{figure*}

The acquisition function for the BEO process is
\begin{equation}
\label{eq:acq_func}
    \mathcal{A}(\bm{\Phi}) = \ln \mathcal{P}(\bm{\Phi}|\bm{d}_\text{obs})
    + \alpha\, \bm{\sigma}_m^\top \Sigma^{-1} \bm{\sigma}_m ,
\end{equation}
where $\mathcal{P}(\bm{\Phi}|\bm{d}_{\rm obs})$ is the posterior probability, $\bm{\sigma}_m = \sqrt{\mathrm{Var}[f(\bm{\Phi})]}$ is the vector of GP-predicted standard deviations in flux units, $\bm{d}_\text{obs}$ is the observed AT2017gfo data, $\Sigma = \mathrm{diag}(\bm{\sigma}_d^2)$ is the observational uncertainty matrix for AT2017gfo, and the hyperparameter $\alpha$ balances exploration (large $\alpha$) with refinement near high-posterior-density regions (small $\alpha$). We begin with $\alpha=0.007$ to balance exploitation and exploration, then decrease $\alpha$ according to $\alpha = 0.007\sqrt{\nu}$ where $\nu$ linearly decreases until $\alpha$ reaches $0.002$. The posterior $\mathcal{P}(\bm{\Phi}|\bm{d}_{\rm obs})$ is approximated using a Gaussian likelihood with uncorrelated model uncertainty (Eq.~\ref{eq:likelihood}), the limitations of which are discussed in  \cref{sec:likelihood}.

Models can be selected using the acquisition function either serially (retraining between each acquisition) or in batches (proposing multiple models simultaneously without retraining). Batch acquisition is computationally faster but less optimal, as later proposals in a batch are less informed about the true posterior. We employ a hybrid approach: 14 batches followed by 12 serial acquisitions. This allows us to balance computational efficiency with optimization quality as emulator uncertainty decreases (see \citet{Rogers_2019} for details on batch optimization).

We evaluate the acquisition function using the full two-component kilonova model discussed in \cref{subsec:2comp}, but the training set consists of single-component \textsc{possis} simulations defined by the physical parameters $m_{\rm ej}, Y_{\rm e}, v_{\rm ej},$ and $ q$. At each BEO iteration, we therefore add two physically distinct single-component models: one with $m_{\rm ej} ^{\rm wind}\geq0.02\,{\rm M}_{\odot}$ (wind-like ejecta); and one with $m_{\rm ej}^{\rm dyn} \leq0.02\,M_{\odot}$ (dynamical ejecta). Each added model is evaluated at 13 viewing angles to improve constraints on $\theta_{\rm v}$. We added a total of 124 simulations to the training set during the BEO process. The parameters for the models added using BEO are shown in \cref{fig:bo_params}. 

To assess the performance of the BEO process, we check for both reduced uncertainty and improved calibration using similar methods to those presented in \citet{Pedersen_2021} and \citet{Rogers_2021}.

\begin{figure*}[ht]
    \centering
    \includegraphics[width=\linewidth]{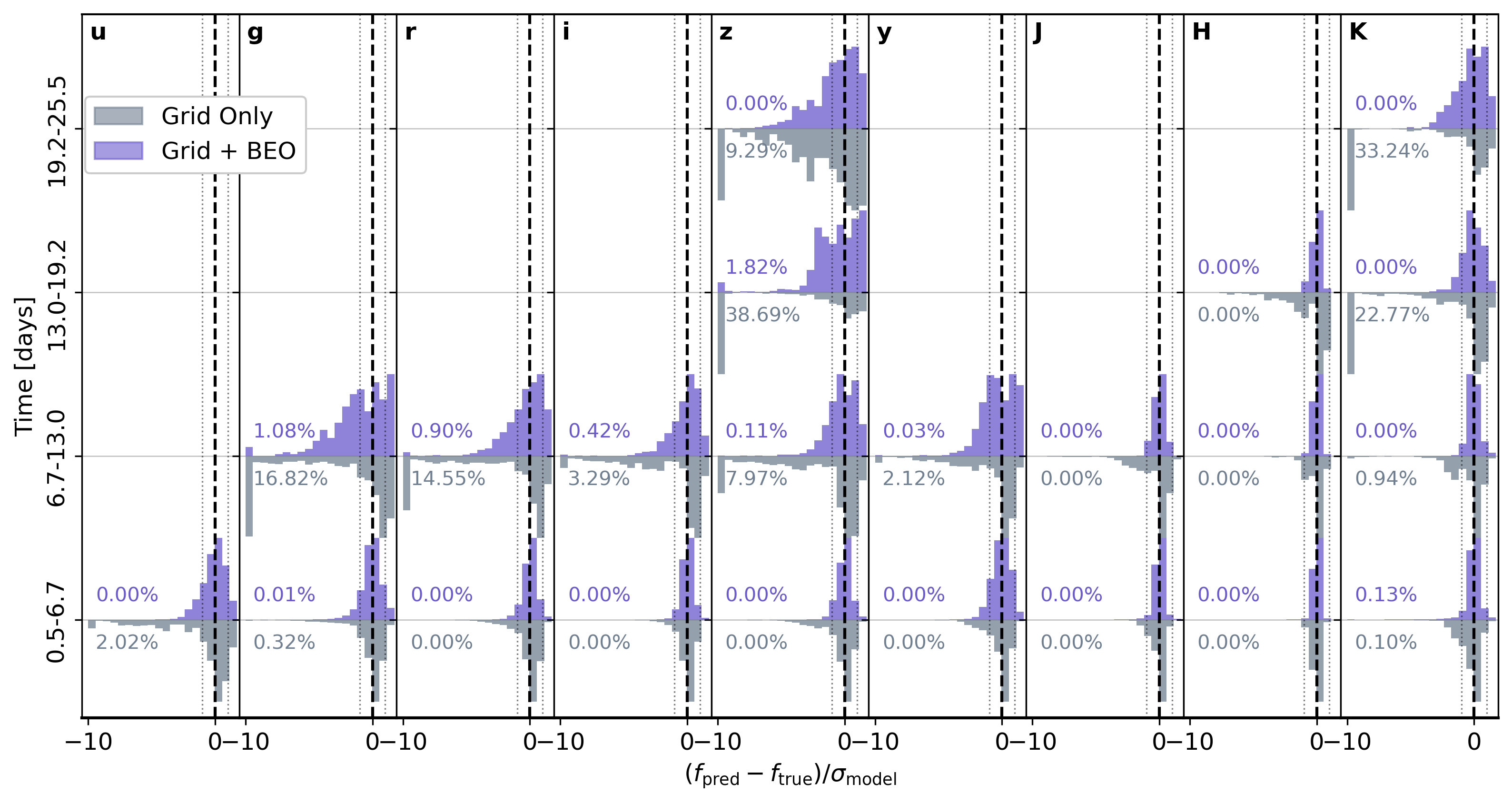}
    \caption{Distribution of the ratio of empirical-to-predicted emulator error, $(f_\mathrm{pred}-f_\mathrm{true})/\sigma_\mathrm{model}$ for the points in Fig.~\ref{fig:lvr_8x5} where $f_\mathrm{true}$ is the true value, $f_\mathrm{pred}$ is the GP prediction, and $\sigma_\mathrm{model} = \sqrt{\mathrm{Var}(f_\mathrm{pred})}$. Values are binned by wavelength (left to right) and observation time (bottom to top). Grey: grid-only model; purple: grid+BEO model. Numbers beside each histogram indicate the percentage of values with $(f_\mathrm{pred}-f_\mathrm{true})/\sigma_\mathrm{model} < -10$. Light vertical dashed lines indicate $(f_{\rm{pred}} - f_{\rm{true}})/\sigma_{\rm{model}}= -1, 1$ and the heavy dashed line shows $(f_{\rm{pred}} - f_{\rm{true}})/\sigma_{\rm{model}}= 0$. The distributions contain 62 BEO parameter sets, each of which is evaluated at 13 viewing angles. The number of data points differs between panels because the AT2017gfo observations are not uniform in time or wavelength. We therefore show density-normalized histograms using 20 bins in each panel. }
    \label{fig:emp_err_ratio}
\end{figure*}

We examine the logarithm of the ratio between predicted emulator uncertainty and observational uncertainty in flux for newly added models with $m_{\rm ej} \geq 0.02\,{\rm M}_{\odot}$. Only models with $m_{\rm ej} \geq 0.02\,{\rm M}_{\odot}$ are shown because the two-component model uncertainty is dominated by the uncertainty of the wind component. \cref{fig:lvr_8x5} compares this ratio for the GP trained on the original training grid and the optimized training set as a function of band and time (shown for the \textit{ugrizy}+\textit{JHK} filters). The optimized training set reduces the predicted emulator uncertainty in the regions of parameter space most relevant to parameter inference. 
In each time--band bin, we calculate the median change in the ratio of emulator uncertainty to observational uncertainty across the $m_{\rm ej} \geq 0.02\,{\rm M}_{\odot}$ BEO simulations. This number is indicated in the figure. The mean (median) improvement over all times and bands is $24\%$ ($35\%$). 

The late-time $z$-band has an unusually small model-to-data uncertainty ratio. The AT2017gfo fluxes in this time–band bin are very faint and have large observational uncertainties. At the same time, the propagated model variance, given by \cref{eq:var_prop}, generally decreases toward lower fluxes. These effects result in a model uncertainty that is smaller than the observational uncertainty in flux units.

We evaluate the effect of BEO on emulator calibration by examining the ratio of empirical error to predicted uncertainty (\cref{fig:emp_err_ratio}) for the same models shown in \cref{fig:lvr_8x5}. When trained only on the original grid, the emulator often significantly under-predicts its own uncertainty and systematically underestimates brightness in most bands. The BEO procedure not only reduces model error but also improves the accuracy of the predicted variances. Figures~\ref{fig:lvr_8x5} and \ref{fig:emp_err_ratio} provide complementary diagnostics. The first shows that the change in predicted error relative to data, while the second assesses whether the predicted error is consistent with the true error.

To quantify the improvement in empirical model accuracy across the data set, we calculate median improvement in $|f_{\rm pred}-f_{\rm true}|/\sigma_{\rm data}$ in each time--band bin. Across all times and bands, the mean (median) improvement is $65\%$ ($67\%$).

The improvement in predicted model uncertainty is not uniform across all bands and times. Bands and epochs (such as late-time $z$ band) that were poorly calibrated before BEO show little to no decrease in predictive uncertainty but significant improvement in the ratio of empirical error to predicted error. Models that were already well calibrated, such as the early time $y,J,$ and $H$ bands, show greater reduction in predicted uncertainty.

Two time–band bins (the $z$ band between 13 and 19 days and the $K$ band between 19 and 25 days) show a negative median improvement in the predicted uncertainty after applying BEO. This does not necessarily indicate a failure of the BEO. First, the acquisition function is evaluated on all band-time data simultaneously, and each band is trained using all time points. The median uncertainty in the $z$ and $K$ bands decreases when all times are considered, and the median model uncertainty across all times and bands also improves. Second, both of these bins were poorly calibrated before the BEO simulations were added. \cref{fig:emp_err_ratio} shows that in the $z$ band between 13 and 19 days, $39\%$ of the data points had predicted uncertainties at least 10 times smaller than the empirical errors before BEO; this fraction decreased to $2\%$ after BEO. Similarly, in the late-time $K$ band, the number of data points with  empirical uncertainties at least a factor of 10 larger than the predicted uncertainties decreases from $33\%$ before to $0\%$ after. Even though the median predicted variance increases slightly in these two bins, the BEO process improved the overall calibration and reduced the model uncertainty.

\subsection{Emulator Validation}

\begin{figure}[h]
    \centering
    \includegraphics[width=\linewidth]{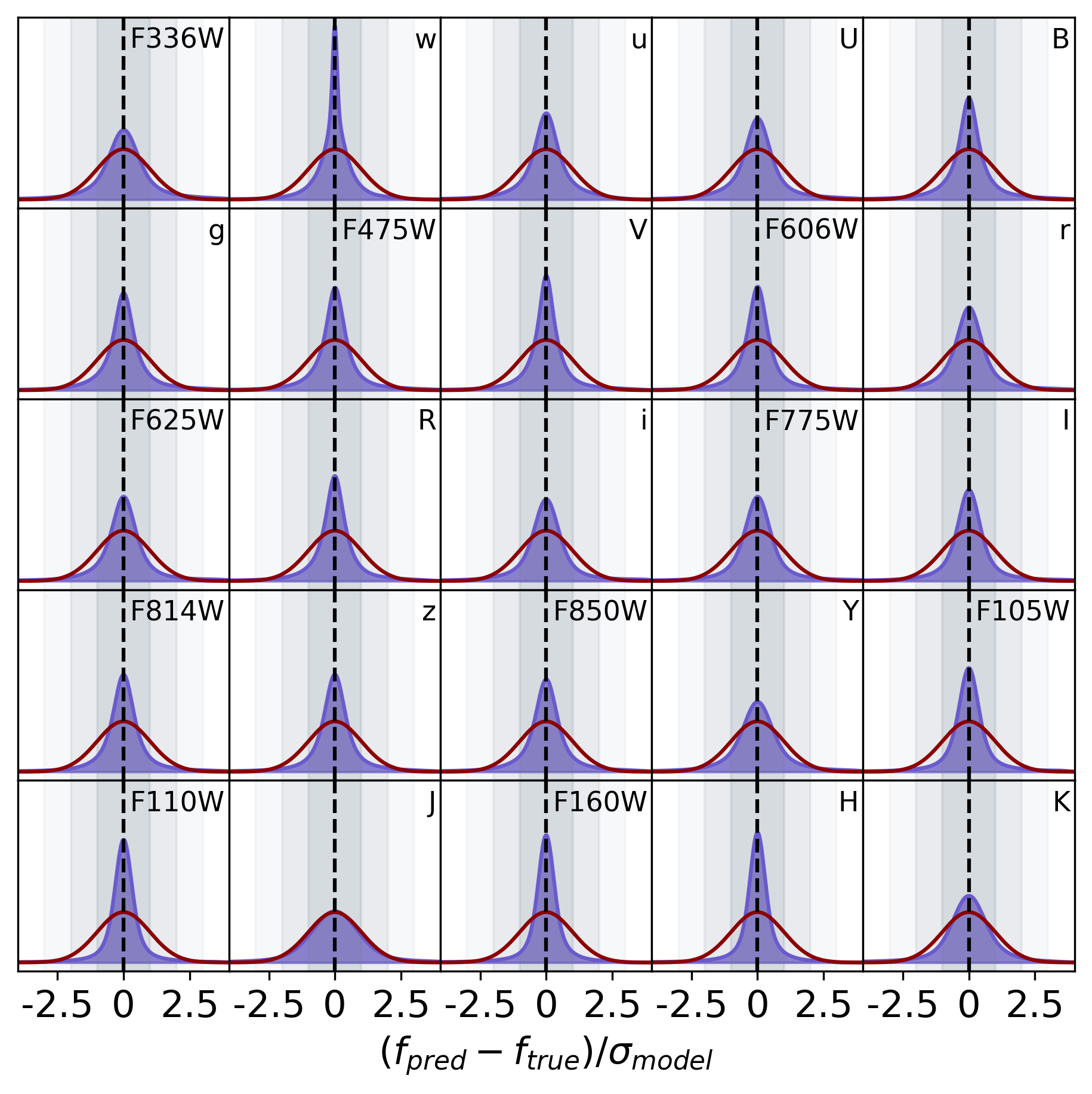}
    \caption{Ratio of empirical error, defined as the difference between predicted flux $f_{\mathrm{pred}}$ and true flux $f_{\mathrm{true}}$, to predicted uncertainty, $\sigma_\mathrm{model} = \sqrt{\mathrm{Var}(f_\mathrm{pred})}$ for all bands in the emulator. The red curve shows a unit normal distribution for comparison, and the shaded bands show the $1,2,$ and $3 \, \sigma$ intervals.}
    \label{fig:loo_ppt_25band}
\end{figure}

We assess emulator performance using leave-one-out posterior predictive tests on the single-component light curves. For each iteration, we remove a single physical configuration defined by $(m_{\rm ej}, Y_{\rm e}, v_{\rm ej}, q)$ and simultaneously remove all associated viewing angles\footnote{This could be considered a $K$-fold ``leave-eleven-out'' test.}. The kernel hyperparameters are held fixed.

For each withheld model, we compute the empirical error (difference between GP mean prediction and true \textsc{possis} light curve in flux units) and compare it to the predicted emulator uncertainty, which is given by the square root of the predicted variance. The ratio of empirical error to predicted uncertainty should follow a unit normal distribution if the emulator is unbiased and well calibrated.

\cref{fig:loo_ppt_25band} shows this distribution across all bands and times for the final GP, which is trained on the combined grid+BEO training set (see \cref{subsec:BEO}). We find no systematic bias in the mean predictions. In most bands, the distribution is narrower than a unit Gaussian, indicating that the model is conservative and overestimates its uncertainty. This is preferable to underestimation and suggests the white-noise term adequately captures Monte Carlo noise in the \textsc{possis} simulations, which is larger at early times than late times.

\section{Likelihood-Based inference}
\label{sec:likelihood}

We perform Bayesian parameter estimation using the standard likelihood-based approach with an MCMC sampler\footnote{\url{https://github.com/justinalsing/affine}} on AT2017gfo and simulated data. The inference is conducted in flux space and uses a data-set $\bm{d}_\text{obs}$ comprising $n_{\text{obs}}=620$ measurements which irregularly span all 25 photometric bands and $J = 82$ observation times; the band of the $i$'th measurement is $\beta_i$ and corresponds to time is $t_i$, for which the two-component GP prediction has mean $\bar{f}_{\beta_i}(t_i, \bm{\Phi})$ and model uncertainty $\sigma_{{\rm m},i}(\bm{\Phi})$. We adopt a Gaussian (log-)likelihood defined by 
\begin{align}
\label{eq:likelihood}
& \ln \mathcal{L}(\bm{d}_\text{obs}|\bm{\Phi}) = \\ \nonumber
& -\frac{1}{2}
\sum_{i=1}^{n_{\rm obs}}
\left\{
\frac{[\bar{f}_{\beta_i}(t_i, \bm{\Phi}) - d_i]^2}
{\sigma_{d,i}^2 + \sigma_{m,i}^2(\bm{\Phi})}
+ \ln 2\pi[\sigma_{d,i}^2 + \sigma_{m,i}^2(\bm{\Phi})]
\right\},
\end{align}
where $\sigma_{d,i}$ is the observational uncertainty and it is implicit that the uncertainties are uncorrelated (i.e.\ the total covariance is diagonal). This likelihood is also used in the BEO process (\cref{subsec:BEO}) to construct the approximate posterior $\mathcal{P}(\bm{\Phi}|\bm{d}_{\mathrm{obs}})$ in the acquisition function (Eq.\ \ref{eq:acq_func}).

The focus of this discussion is likelihood misspecification, the failure of the likelihood to capture the true sampling distribution. This is distinct from model misspecification, which arises from missing physics in \textsc{possis} models.

The flux model variance $\bm{\sigma}_m^2$ for the two-component model must be computed from the predicted asinh magnitude $\bar{\bm{\mu}}_c$ and uncertainty $\mathrm{Var}(\bar{\bm{\mu}}_c)$ of the individual components. Uncertainties on calculated quantities are typically propagated using a first-order Taylor expansion of the propagation function. This linear approximation works well for weakly nonlinear functions. However, our flux calculation involves an asinh function (Eq.\ \ref{eq:asinh_to_flux}), which is highly nonlinear. We found that the first-order Taylor expansion underestimated the propagated variance by up to $40\%$ relative to the second-order expansion in some regions of parameter space. Including third-order terms did not significantly improve the variance estimate, so we adopt a second-order Taylor expansion \citep{Mekid_2008},
\begin{equation}
\label{eq:var_prop}
\bm{\sigma}_{m}^2 \approx 
 \sum_{c=1}^{2}
\Bigg[ \left( \frac{\partial \bar{\bm{f}}}{\partial \bar{\bm{\mu}}_c}
\right)^2 \mathrm{Var}(\bar{\bm{\mu}}_c)
+
\frac{1}{2}
\left(
\frac{\partial^2 \bar{\bm{f}}}{\partial \bar{\bm{\mu}}_c^2}
\right)^2
\mathrm{Var}(\bar{\bm{\mu}}_c)^2
\Bigg] ,
\end{equation}
where $\rm{Var}(\bm{\mu}_c) =  \rm{diag} \; \mathrm{Cov}(\bm{\mu}_c)$ is the diagonal of the GP covariance matrix for component $c$. 

The GP enters the likelihood only through the mean $\bar{\bm{f}}(\bm{\Phi})$ and variance $\bm{\sigma}_{m}^2(\bm{\Phi})$, meaning that the likelihood-based analysis does not depend on the full GP predictive distribution, only on an approximation of it. In contrast, the SBI framework is trained on random samples from the GP, allowing it to implicitly marginalise over the emulator uncertainty.

While this likelihood includes time- and band-dependent uncertainties, the inclusion of which was recently shown to improve parameter recovery \citep{Jhawar_2025}, it neglects correlations in emulator uncertainties across time and wavelength and between parameters. In principle, the full GP predictive covariance should enter the likelihood. However constructing a fully correlated likelihood (see e.g. \citealp{Pedersen_2021}) requires constructing and inverting the full $(n_{\rm obs} \times n_{\rm obs})$ covariance matrix for $\mathcal{O}(10^4)$ walkers in each likelihood evaluation, which is prohibitively expensive computationally. For 620 observations and 1000 walkers, storing the full covariance would require $\mathcal{O}(10\,{\rm GB})$ of memory per likelihood evaluation, compared to $\mathcal{O}(10\,{\rm MB})$ under the diagonal approximation. 

To assess the validity of the Gaussian approximation in the likelihood, we look at the sampling distribution. For a parameter vector from the AT2017gfo ANPE posterior, we generate 1000 light curves by drawing from the GP predictive distribution. Then Gaussian observational noise based on AT2017gfo is added by drawing randomly from $\mathcal{N}(0,\sigma_{d,i}^2)$ for the $i$th observation.

\begin{figure}[t]
    \centering
    \includegraphics[width=\linewidth]{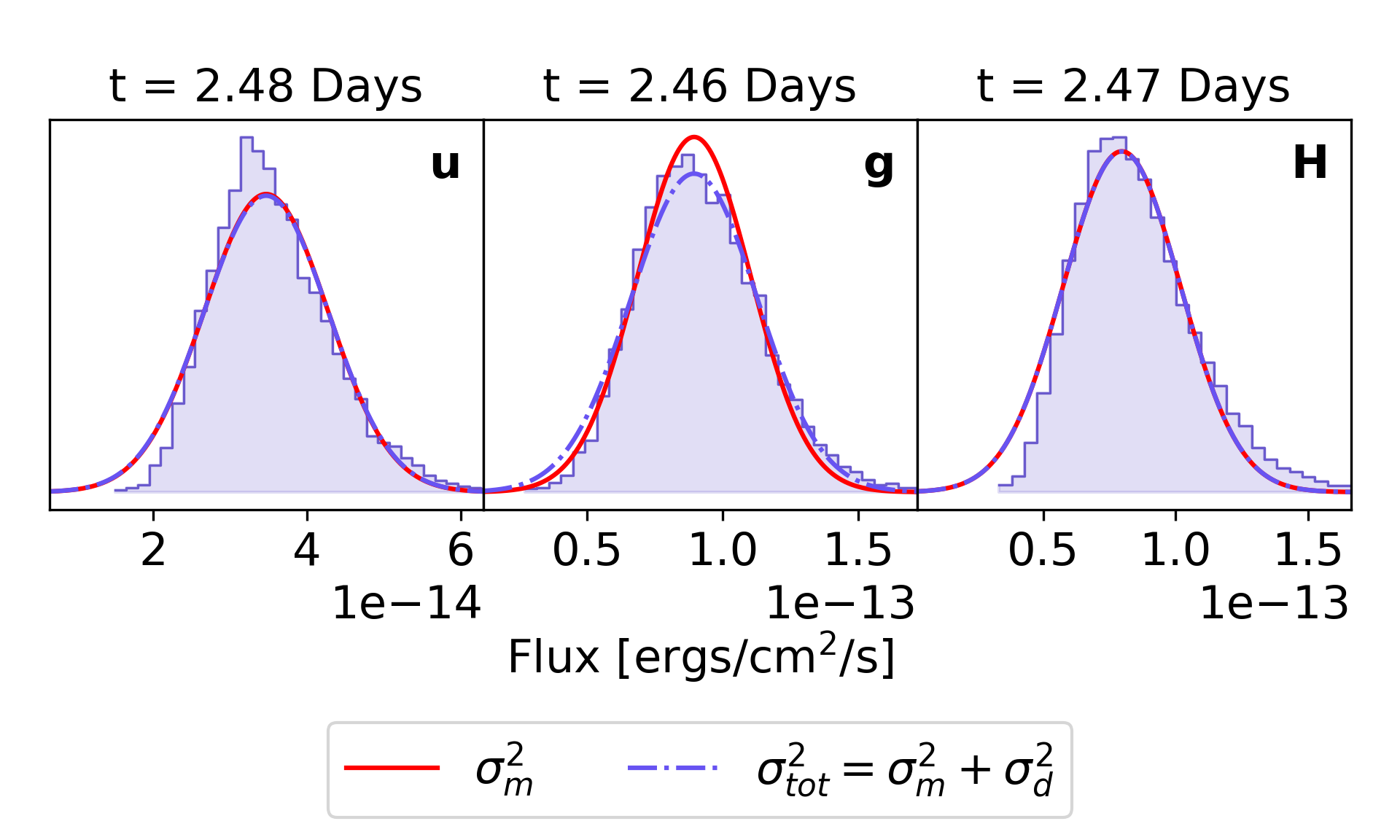}
    \caption{Sampling distribution (blue histogram) around $t_{obs} \simeq 2.5$~d in the \textit{u}, \textit{g}, and \textit{H} bands. The model uncertainty (red) and the total uncertainty (blue) are shown as Gaussian distributions with mean equal to the empirical mean of the sampling distribution and variance equal to $\sigma^2_{m}$ and $\sigma^2_\text{tot} = \sigma^2_{m} + \sigma^2_{d}$, respectively.} 
    \label{fig:sampling_dist}
\end{figure}

Figure~\ref{fig:sampling_dist} shows the resulting sampling distributions in representative bands at observation time $t\simeq2.5$~d. Some of the empirical distributions have noticeable asymmetry relative to a Gaussian distribution. In particular, the Gaussian model over-predicts the probability of faint flux values and under-predicts bright fluxes when compared to the empirical sampling distribution. This sampling distribution is what the ANPE is trained on. Therefore, \cref{fig:sampling_dist} shows the difference between the Gaussian approximation in the analytical likelihood and the true distribution learned by the SBI framework.

Figure~\ref{fig:sampling_dist} captures the sampling distribution at one spot in parameter space. The approximation may break down even further in other regions (e.g.\ near the prior boundaries). GPs do not extrapolate well, and the uncertainty increases at the boundaries of parameter space. This can lead to increased likelihood misspecification and drive the MCMC walkers to prior boundaries. 

The differences between the true sampling distribution and the approximation arise from both the nonlinear magnitude-to-flux transformation and correlated emulator uncertainties, which are not captured by the diagonal of the GP covariance matrix. 

The effect of the likelihood misspecification is explored in the parameter inference results presented in \cref{sec:results}. 

\section{Simulation-Based Inference}
\label{sec:SBI}

In order to explore the posterior constraints on $\bm{\Phi}$ given $\bm{d}$ without relying on the likelihood assumptions discussed in \cref{sec:likelihood}, which were shown to break down in \cref{fig:sampling_dist}, we develop an SBI framework that sidesteps likelihood misspecification by learning the posterior distribution directly from forward simulations.

Our approach uses the GP emulator from \cref{sec:GP} as a forward model, with a separate neural network (the ANPE) trained on its outputs to learn the mapping from data to parameters. First, we generate a training set using a forward model based on random draws from the GP emulator (\cref{subsec:forward_model}). We then train the neural network density estimator using flow matching to learn the conditional distribution $\mathcal{P}(\bm{\Phi}|\bm{d})$ (\cref{subsec:delfi}). Finally, we validate the performance of the ANPE by checking for bias in parameter recovery (\cref{subsec:SBI_valid}).

\subsection{Forward Modeling}
\label{subsec:forward_model}

We construct a forward model that maps a parameter vector $\bm{\Phi}$ to synthetic observations evaluated at the same 620 combinations of bands and times as AT2017gfo. The forward simulation consists of three elements: a prior sampler over $\bm{\Phi}$; a GP-based two-component kilonova light curve; and the AT2017gfo observational noise model.

To enable a direct comparison between the SBI posterior and the likelihood-based posterior, we adopt the same priors in both analyses. The prior ranges for component masses, electron fractions, velocities, and shapes are set by the \textsc{possis} training grid (\cref{subsec:train_dat}), and the priors for the component masses are uniform in $\log_{10} m_{\rm ej}$, consistent with the grid spacing. As discussed in \cref{sec:at2017gfo}, the luminosity distance prior, which is a truncated Gaussian centered at $d_L = 40.7$~Mpc with a standard deviation of $\sigma = 2.36$~Mpc, is motivated by surface-brightness measurements of the host galaxy \citep{Cantiello_2018}. The viewing angle prior (Gaussian with mean $21.3^\circ$ and $\sigma=2.5^\circ$, with hard bounds at $5 \sigma$) is informed by VLBI observations \citep{Mooley_2022}. We impose $Y_{\rm e}^{\rm wind} > Y_{\rm e}^{\rm dyn}$ to preserve the definition of component one as wind ejecta and component two as the dynamical ejecta. The priors are summarized in \cref{tab:priors}.

For each parameter draw, a two-component light curve is generated using the GP emulator. The forward model uses random draws from the GP predictive distribution. This differs critically from the likelihood-based analysis in \cref{sec:likelihood}, which uses the GP mean. Using samples from the GP predictive distribution ensures that the training set accurately captures the emulator uncertainty.

Observational noise is incorporated by adding Gaussian noise consistent with the measured uncertainties of AT2017gfo,
\begin{equation}
\hat{f}_{\beta_i}(t_i,\bm{\Phi}) = f_{\beta_i}(t_i,\bm{\Phi}) + \epsilon_i, \quad \epsilon_i \sim \mathcal{N}(0,\sigma_{d,i}^2),
\end{equation}
where $\sigma_{d,i}$ is the reported uncertainty of the $i$-th observation. We generate $10^6$ forward simulations to train the conditional density estimator.

Prior to training the flow-matching network, both parameters and fluxes are normalized. Both parameter distributions and data (each time--band pair) are scaled to zero mean and unit variance to improve network training.

\begin{table}[h]
    \centering
    \caption{Prior distributions for the 10 model parameters.}
    \label{tab:priors}
    \begin{tabular}{cc} \hline \hline
        parameter & prior \\ \hline
        $d_L$ [Mpc] & $\mathcal{N}_T(40.7,\,2.35;\,[38.3,\,43.0])$ \\ 
        $m_{\rm ej}^{\rm wind}$ [M$_{\odot}$ ]& log uniform [0.01,\,0.1] \\ 
        $m_{\rm ej}^{\rm dyn}$ [M$_{\odot}$] & log uniform [0.001,\,0.02] \\ 
        $Y_{\rm e}^{\rm wind}$,$Y_{\rm e}^{\rm dyn}$  &  uniform [0.1,\,0.4] \\ 
        $v_{\rm ej}^{\rm wind}$, $v_{\rm ej}^{\rm dyn}$  [$c$] & uniform [0.05,\,0.25] \\ 
        $\theta_{v}$ [$^{\circ}$] & $\mathcal{N}_T(21.3,\,2.5;\,[8.8,\,33.8])$ \\ 
        $q^{\rm wind}$, $q^{\rm dyn}$ &  uniform [$-$0.5,\,90] \\ \hline \hline
\end{tabular}
\end{table}

\subsection{Conditional Density Estimation with Flow Matching}
\label{subsec:delfi}

We approximate the conditional density $\mathcal{P}(\bm{\Phi}|\bm{d})$ using density-estimation likelihood-free inference with flow matching \citep{Lipman_2023}, implemented through the \texttt{flowfusion}\footnote{\url{https://github.com/Cosmo-Pop/flowfusion}} package \citep{Alsing_2024, Thorp_2024, Thorp_2025, Leistedt_2026}. The neural density estimator is trained on parameter--light curve pairs $\{\bm{\Phi}, \bm{d} \}$ generated by the forward model to learn a global conditional distribution over parameters given light curves. For a specific set of observations $\bm{d}_{\rm obs}$, the posterior is obtained by evaluating the learned conditional density at $\bm{d} = \bm{d}_{\rm obs}$.

The conditional distribution is represented as a continuous normalizing flow \citep{Chen_2018, Grathwohl_2019}. The flow defines a pseudo-time-dependent transformation between a simple base density $\bm{z}_1 \sim \mathcal{N}(0,\bm{I})$ and samples $\bm{z}_0$ drawn from the target conditional density. The transport is governed by
\begin{equation}
\frac{d\bm{z}_\tau}{d\tau} = \bm{v}_{\vartheta}(\bm{z}_\tau, \tau, \tilde{\bm{f}}),
\end{equation}
where $\tau$ is the pseudo-time variable, $\bm{v}_{\vartheta}$ is a neural network that parameterizes the velocity field and $\tilde{\bm{f}}$ denotes the normalized light curve that serves as a conditional input.

Following \citet{Lipman_2023}, the network is trained using a flow-matching objective. For each simulated pair $(\bm{z}_0,\tilde{\bm{f}})$ from the forward model, we sample $\bm{z}_1 \sim \mathcal{N}(0,\bm{I})$ and define the interpolation
\begin{equation}
\bm{z}_\tau = (1-\tau)\bm{z}_1 + \tau \bm{z}_0 .
\end{equation}
The target velocity along this path is $\bm{u} = \bm{z}_0 - \bm{z}_1$. The network parameters $\vartheta$ are optimized by minimizing
\begin{equation}
\mathcal{L}(\vartheta) =
\mathbb{E}_{\tau,\bm{z}_0,\bm{z}_1}
 \left[
\left\|
\bm{v}_{\vartheta}(\bm{z}_\tau,\tau,\tilde{\bm{f}})
- (\bm{z}_0 - \bm{z}_1)
\right\|^2
\right],
\end{equation}
which regresses the neural velocity field toward the target transport field.

Training is performed using the \texttt{AdamW} optimizer \citep{ Kingma_2017, Loshchilov_2019}. After training, posterior samples for $\bm{d}_{\rm obs}$ are generated by drawing $\bm{z}_1 \sim \mathcal{N}(\bm{0},\bm{I})$ and integrating the learned differential equation backward to $\tau=0$ (we use an adaptive Runge--Kutta solver; \citealp{Dormand_1980, Chen_2018}), followed by transformation back to the physical parameter space. Once trained, this produces posterior samples in seconds, compared to the hours needed for likelihood-based inference.

\begin{figure*}[ht]
\centering
\includegraphics[width=\linewidth]{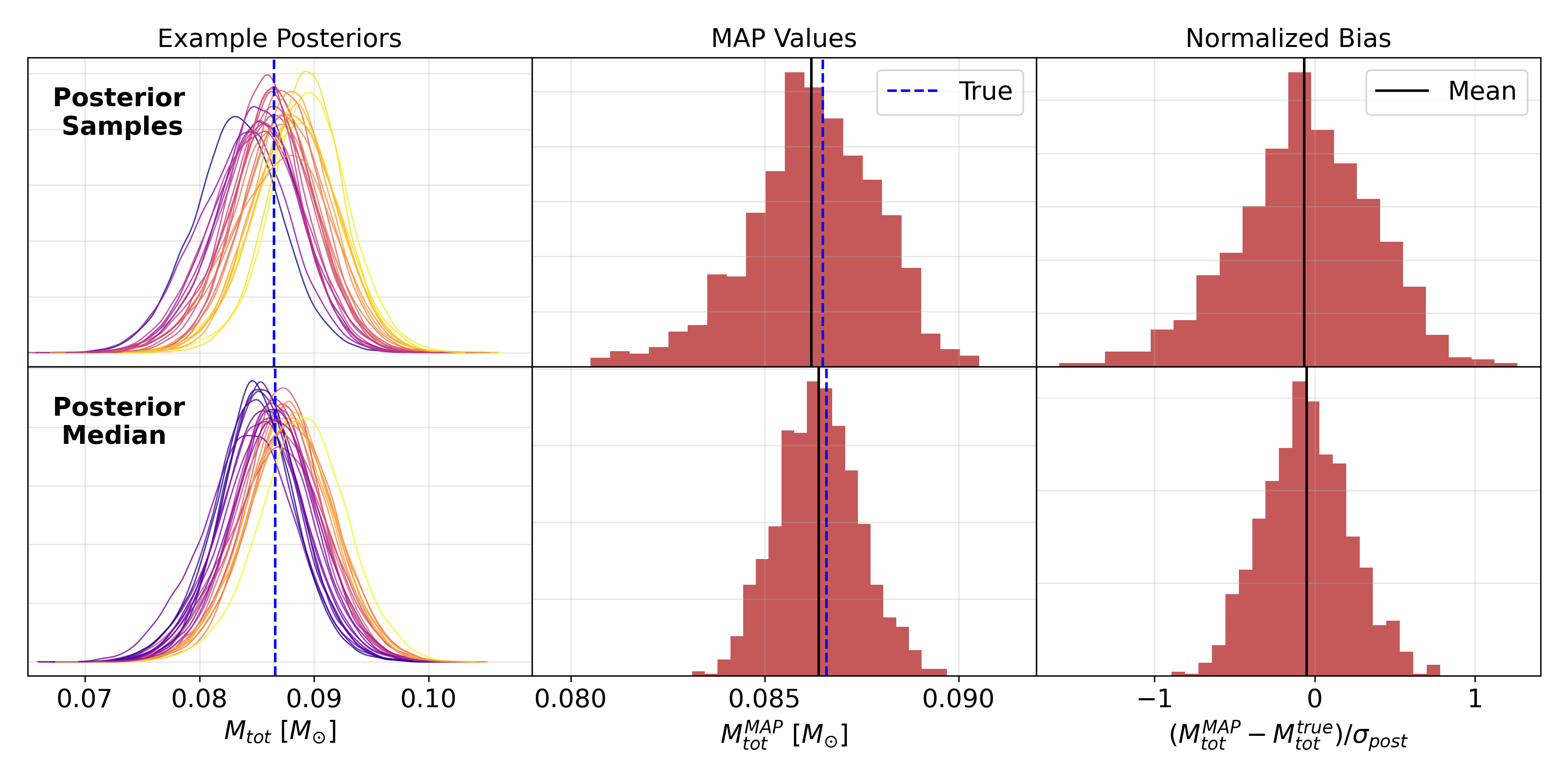}
\caption{Posterior bias assessment for the kilonova total mass. Each row shows results from 1,000 independent synthetic datasets with known injected parameters. 
Top row: Parameters for synthetic datasets are randomly drawn from the SBI posterior for AT2017gfo. Bottom row: Parameters for synthetic datasets are fixed to the median posterior values from AT2017gfo.
Left: kernel density estimations of posterior predictions for 25 random synthetic data sets. Center: distribution of 1,000 MAP values ($M_{\rm tot}^{\rm MAP}$) recovered from synthetic datasets. Right: distribution of normalized bias $(M_{\rm tot}^{\rm MAP} - M_{\rm tot}^{\rm true})/\sigma_{\rm post}$. Vertical dashed blue lines show the true (injected) parameter value; black lines show the distribution mean.}
\label{fig:map_bias_post}
\end{figure*}

\subsection{Validation and Posterior Bias Tests}
\label{subsec:SBI_valid}

We evaluate the performance of the conditional density estimator using two simulation studies, each of which is designed to test for bias in parameter recovery. We focus on total ejecta mass, $M_{\rm tot}=m_{\rm ej}^{\rm wind}+m_{\rm ej}^{\rm dyn}$ because it is less sensitive to limitations in the prescriptions for heating rates and opacities underlying \textsc{possis} than electron fraction, ejecta velocity, and shape.

In each study, synthetic light curves are generated using the two-component GP mean model and Gaussian noise from the observational data model described in \cref{subsec:forward_model}. Then $2\times10^4$ posterior samples are generated for each simulated dataset $\bm{d}_{\rm sim}$ using the trained ANPE. We compute the maximum-a-posteriori (MAP) estimate of $M_{\rm tot}$ and the posterior standard deviation using the posterior samples.

We evaluate performance across the ANPE posterior for AT2017gfo by fixing the total mass to its posterior median and drawing the remaining parameters from the posterior. The component masses are determined using the sampled mass ratio. For 1,000 posterior draws, we generate synthetic datasets and analyze each with the trained ANPE. Examples of the resultant posterior densities for the total mass are shown in the top row of \cref{fig:map_bias_post}, along with the distribution of MAP estimates and posterior biases (normalized by the posterior standard deviation), which confirm that the procedure is unbiased.

We also test the recovery at a fixed parameter vector corresponding to the median posterior values for AT2017gfo. We generate 1,000 independent noise realizations and analyze each dataset with the ANPE. The results, shown in the bottom row of \cref{fig:map_bias_post}, confirm that the ANPE mass posterior is minimally biased in this region of parameter space.

In both studies, there is a very small mean offset in the total mass. The right-hand panel shows that the average biases are $-0.065\sigma$ and $-0.05\sigma$, respectively, indicating that the MAP predictions are well within the $1\sigma$ posterior bounds. The narrower than Gaussian distributions in the right-hand panels suggest that the ANPE is somewhat conservative (overestimating the uncertainty).

\begin{table}[ht]
    \centering
    \caption{Parameter bias recovery summary for model parameters excluding masses, which are shown in \cref{fig:map_bias_post}. Columns list the true parameter value, the mean MAP estimate over 1,000 realizations, and the mean normalized bias $(\mathrm{MAP} - \mathrm{truth})/\sigma_{\mathrm{post}}$. For all parameters except $d_L$, the absolute mean normalized bias is less than 1, indicating no significant systematic bias. }
    \label{tab:bias_recovery}
    \begin{tabular}{cccc} \hline \hline
        parameter & truth & mean MAP & mean norm.\ bias  \\ \hline
        $d_L$ [Mpc] & 41.202 & 42.277 & 1.105 \\
        $Y_{\rm e}^{\rm wind}$ & 0.367 & 0.367 & 0.121 \\
        $Y_{\rm e}^{\rm dyn}$ & 0.223 & 0.230 & 0.177 \\
        $v_{\rm ej}^{\rm wind}$ [$c$] & 0.074 & 0.073 & $-$0.430 \\
        $v_{\rm ej}^{\rm dyn}$ [$c$] & 0.184 & 0.187 & 0.342 \\
        $\theta_{v}$ [$^{\circ}$] & 21.480 & 21.389 & $-$0.038 \\
        $q^{\rm wind}+1$ & 83.994 & 85.890 & 0.233 \\
        $q^{\rm dyn}+1$ & 23.718 & 27.134 & 0.142 \\
     \hline \hline 
    \end{tabular}
\end{table}

The results for recovery at a fixed parameter vector for the remaining parameters are summarized in \cref{tab:bias_recovery}. The only parameter where the absolute bias is greater than $0.5 \sigma$ is distance. This is, in part, due to how the MAP estimate is computed. The MAP estimate is obtained using Gaussian kernel density estimation to the posterior samples. The distance posterior spans the full prior range and is partially truncated at the prior bounds (see \cref{fig:corner_sim}), which is not well represented by the Gaussian, leading to an artificially small sigma and therefore an inflated normalized bias.

\section{Results}
\label{sec:results}

We compare parameter estimates obtained using our ANPE framework and a likelihood-based analysis. Both methods utilize the two-component kilonova emulator described in \cref{sec:GP}. However, the MCMC likelihood is evaluated with respect to the GP mean prediction, whereas the ANPE is trained on random samples from the GP, which incorporates the full predictive covariance. We first test the ability of the ANPE and MCMC to recover accurate posteriors using simulated data with known parameters (\cref{subsec:sim_study}). Then both inference methods are applied to the observed light curves of AT2017gfo (\cref{subsec:at2017gfo}) in order to compare the performance of the two methods and highlight biases that arise due to likelihood misspecification. 

\subsection{Simulation Study}
\label{subsec:sim_study}

\begin{figure*}
    \centering
    \includegraphics[width=\textwidth]{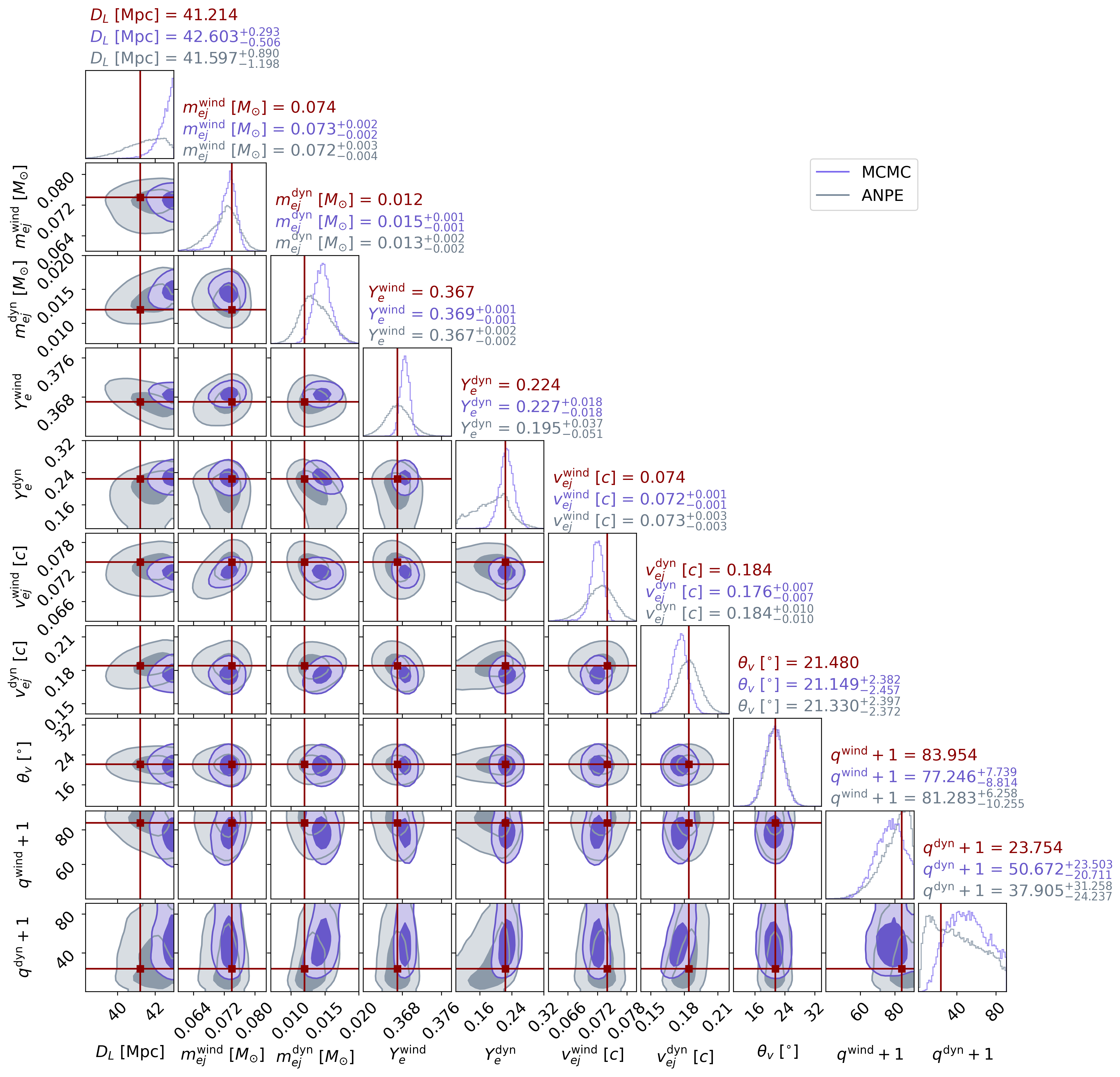}
    \caption{Posterior distributions from a simulation recovery using an MCMC sampler (purple) and the ANPE (grey). The true parameter values are shown in red. Titles report the median and the 16th and 84th percentiles of each posterior. Contours indicate the 1$\sigma$ and 2$\sigma$ credible regions. The true injected parameter values are written in red.}
    \label{fig:corner_sim}
\end{figure*}

To evaluate the accuracy of the inference methods, we perform two simulation studies. If the Gaussian likelihood does not accurately capture the sampling distribution, we expect to see pathologies in the likelihood-based parameter recovery. These can include biased posteriors, over- or under-confident posteriors, or in extreme cases, posterior predictive light curves that do not match the data.

Synthetic light curves are generated using the GP predictive mean with observational uncertainties drawn from the data model described in \cref{subsec:forward_model}. The same synthetic data set is analyzed using both the ANPE and MCMC sampler.

\subsubsection{Single Parameter Injection}

For the first simulation study, we generate synthetic data using the two-component model with parameters set to the posterior median values recovered for AT2017gfo using the ANPE. The resulting posterior distributions are shown in \cref{fig:corner_sim}.

The ANPE posterior distributions recover all injected parameters within the $1\sigma$ credible regions. However, the posteriors for the luminosity distance $d_L$ and viewing angle $\theta_{\rm v}$ are similar to their Gaussian priors, which reflects the fact that the simulated kilonova data only weakly constrain these parameters under the adopted priors.

The likelihood-based analysis recovers most of the injected parameters. However, we see evidence of the likelihood misspecification in the biased posteriors of $d_L$ and $m_{\rm ej}^{\rm dyn}$. This behavior persists even when this simulation recovery test is repeated with different noise realizations, confirming a systematic bias at this point in parameter space. The inferred luminosity distance is railing against the upper boundary of the prior range and away from the true value. This differs from the ANPE distance bias discussed in \cref{subsec:SBI_valid}, which arises because the MAP and posterior width are computed using Gaussian kernel density estimation applied to posterior samples. Since the distance posterior is truncated at the prior boundaries, the Gaussian approximation underestimates the width, which inflates the normalized bias. Applying the MAP estimate procedure from \cref{subsec:SBI_valid} to the likelihood-based distance posterior gives a recovery bias estimate of $\sim\! 4 \sigma$. In the likelihood-based analysis, the mass of the second component is also consistently overestimated. This is likely due to the degeneracy between the luminosity distance and the mass. The ANPE posteriors are broader than those obtained with the MCMC sampler. This reflects both the incorporation of the full GP predictive covariance and the over-estimate of uncertainty shown in \cref{subsec:SBI_valid}.  

\begin{figure}[ht]
    \centering
    \includegraphics[width=0.9\linewidth]{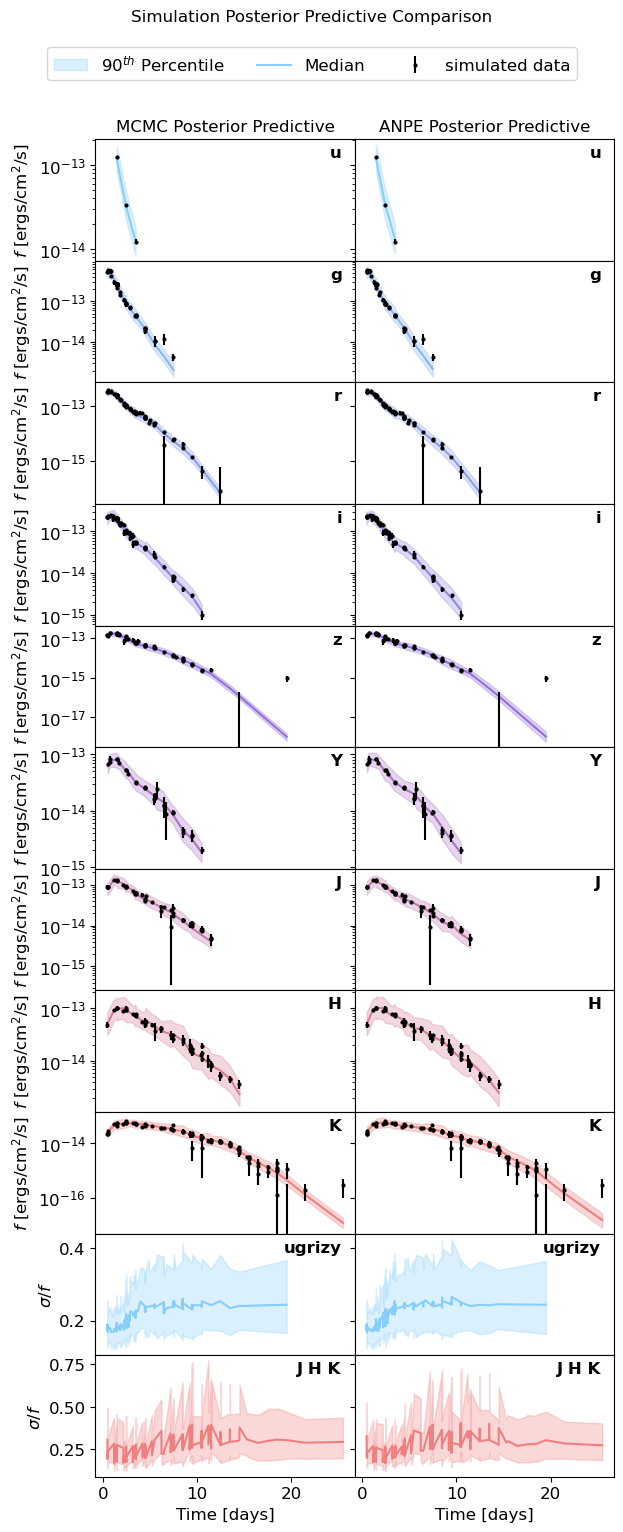}
    \caption{Posterior predictive distributions for a simulated light curve, comparing inference using MCMC (left) and ANPE (right). Predicted fluxes are generated from posterior samples, with the median light curve shown as a solid line and the 90th percentile range shown as a shaded band. The simulated data used for inference are shown in black. The bottom two panels show the distribution of the flux-to-uncertainty ratio for the reddest bands (\textit{JHK}) and for the remaining optical bands (\textit{ugrizy}).}
    \label{fig:postpred_sim}
\end{figure}

\Cref{fig:postpred_sim} shows posterior predictive light curves for a commonly analyzed subset of bands (\textit{ugrizy}+\textit{JHK}). Random samples are first drawn from each posterior, and corresponding light curves are then generated using random draws from the GP emulator. For both inference methods, the simulated data lie near the median predicted light curves and largely within the $90^{\rm th}$ percentile credible intervals. The ANPE posterior distributions are broader than those obtained using the MCMC sampler, but the resulting light-curve predictions are consistent with both the likelihood-based analysis and the data. 
This shows that while both methods can recover data that is a noisy realization of the emulator and the ANPE is overly conservative, the likelihood-based analysis suffers from systematic bias in a few parameters.

\subsubsection{Latin Hypercube Simulation Study}
\label{subsubsec:LHS}

To assess the broader performance of both methods across parameter space, we perform an injection study with 50 synthetic data sets. The parameters for each test are selected using Latin hypercube sampling, which has better space-filling properties than random prior sampling \citep{Stein_1987}. Synthetic light curves are generated using the GP predictive mean with observational uncertainties drawn from the data model described in \cref{subsec:forward_model}. Both the ANPE and likelihood-based analysis are applied to each synthetic data set, enabling a direct comparison of posterior recovery under identical data and model assumptions.

Similar to the validation procedure in \cref{subsec:SBI_valid}, we evaluate posterior recovery using the MAP estimate and normalized posterior bias. For each synthetic data set, we compute the normalized bias using the ANPE and likelihood-based posteriors.

\begin{figure}
    \centering
    \includegraphics[width=\linewidth]{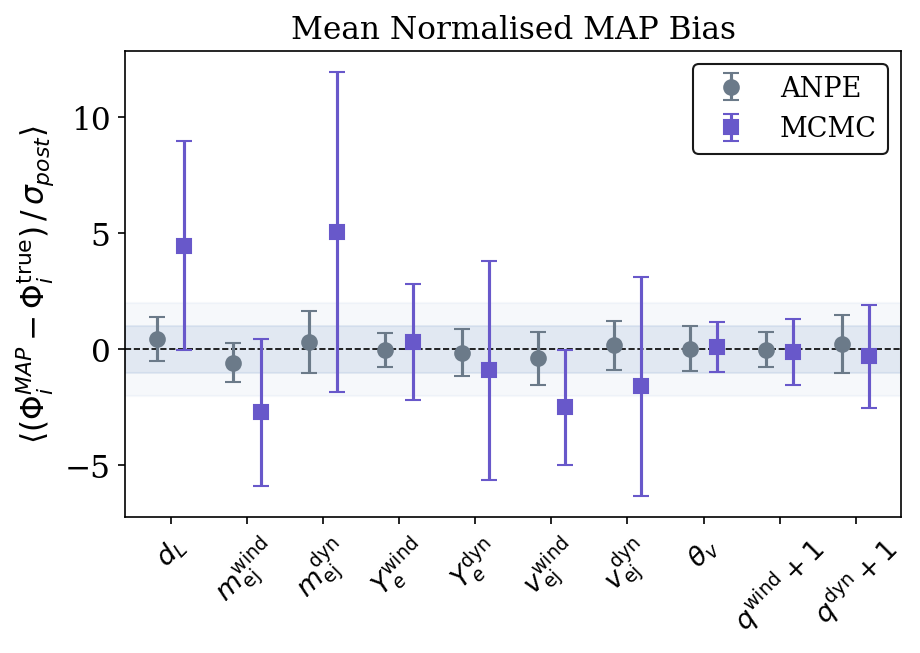}
    \caption{Mean normalized MAP bias across 50 Latin-hypercube injection-recovery tests spanning the adopted prior range. Results are shown for ANPE (grey) and MCMC (purple). Shaded regions indicate absolute normalized biases of one and two posterior standard deviations.}
    \label{fig:injection_study}
\end{figure}

\cref{tab:injection_bias} and \cref{fig:injection_study} summarize the results. The ANPE achieves a MAP bias less than $1\sigma_{\rm post}$ for all parameters. By contrast, the likelihood-based analysis shows biases exceeding $2\sigma_{\rm post}$ for luminosity distance, wind ejecta mass, dynamical ejecta mass, and wind ejecta velocity. The dynamical ejecta velocity bias also exceeds $1\sigma_{\rm post}$.

\begin{table}[ht]
    \centering
    \caption{Mean normalized MAP bias, $\left\langle(\Phi_i^{\rm MAP}-\Phi_i^{\rm true})/\sigma_{{\rm post},i}\right\rangle$, for each model parameter across 50 Latin-hypercube injection-recovery tests spanning the adopted prior range. Results are shown for ANPE and MCMC. Values with absolute magnitude greater than unity indicate that the mean signed MAP offset exceeds one standard deviation of the marginal posterior.}
    \label{tab:injection_bias}
    \begin{tabular}{lcc} \hline \hline
        Parameter \hspace{0.25cm} &  \hspace{0.25cm}  ANPE \hspace{0.25cm}    \hspace{0.25cm}  &   MCMC \hspace{0.25cm} \\ \hline
        $d_L$ [Mpc]                                &      0.462 &      4.469 \\
        $m_{\rm ej}^{\rm wind}$ [$M_\odot$]        &   $-$0.574 &   $-$2.717 \\
        $m_{\rm ej}^{\rm dyn}$ [$M_\odot$]         &      0.327 &      5.057 \\
        $Y_e^{\rm wind}$                           &   $-$0.029 &      0.300 \\
        $Y_e^{\rm dyn}$                            &   $-$0.140 &   $-$0.901 \\
        $v_{\rm ej}^{\rm wind}$ [$c$]              &   $-$0.393 &   $-$2.506 \\
        $v_{\rm ej}^{\rm dyn}$ [$c$]               &      0.175 &   $-$1.589 \\
        $\theta_v$ [$^{\circ}$]                    &      0.035 &      0.093 \\
        $q^{\rm wind}+1$                           &   $-$0.022 &   $-$0.119 \\
        $q^{\rm dyn}+1$                            &      0.222 &   $-$0.305 \\
        \hline \hline
    \end{tabular}
\end{table}

\subsection{Application to AT2017gfo}
\label{subsec:at2017gfo}

\begin{figure*}[ht]
    \centering
    \includegraphics[width=\textwidth]{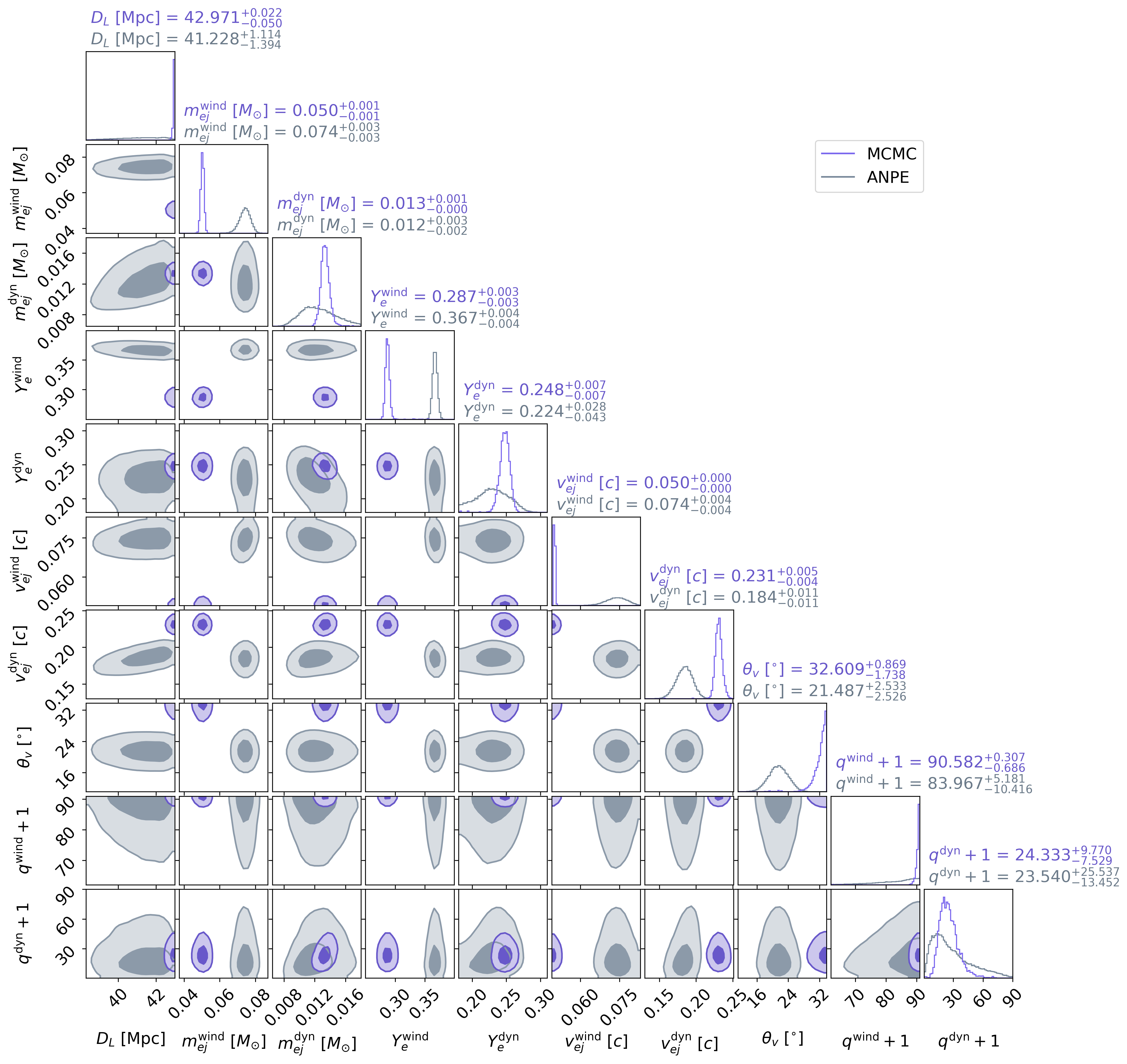}
    \caption{Posterior distributions from AT2017gfo using an MCMC sampler (purple) and the ANPE (grey). Titles report the median and the 16th and 84th percentiles of each posterior. Contours indicate the 1$\sigma$ and 2$\sigma$ credible regions.}
    \label{fig:corner_at2017gfo}
\end{figure*}

In order to compare the performance of ANPE and likelihood-based methods on real data, we apply both inference methods to the real data obtained for AT2017gfo. All 25 available photometric bands are included in the analysis. The resulting posterior distributions are shown in \cref{fig:corner_at2017gfo} and summarized in \cref{tab:pe_summary}. In the event of likelihood misspecification, we expect that the problems with the likelihood-based posteriors evident in the simulation study (\cref{sec:likelihood}) to be amplified.

\renewcommand{\arraystretch}{1.25}
\begin{table}[ht]
    \centering
    \caption{Parameter estimation summary for AT2017gfo.}
    \label{tab:pe_summary}
    \begin{tabular}{ccc} \hline \hline
        parameter & Likelihood-Based & SBI  \\ \hline
        $D_{L}$ [Mpc] & $42.971^{+0.022}_{-0.050}$ & $41.228^{+1.114}_{-1.394}$  \\
        $m_{\rm ej}^{\rm wind}$ $[{\rm M}_{\odot}]$ & $0.0503^{+0.0009}_{-0.0010}$ & $0.0742^{+0.0029}_{-0.0031}$  \\
        $m_{\rm ej}^{\rm dyn}$ $[{\rm M}_{\odot}]$ & $0.0133^{+0.0006}_{-0.0005}$ & $0.0124^{+0.0026}_{-0.0021}$  \\
        $Y_e^{\rm wind}$ & $0.2870^{+0.0033}_{-0.0031}$ & $0.3668^{+0.0040}_{-0.0038}$  \\
        $Y_e^{\rm dyn}$ & $0.2479^{+0.0068}_{-0.0073}$ & $0.2235^{+0.0281}_{-0.0436}$  \\
        $v_{\rm ej}^{\rm wind}$ $[c]$ & $0.0501^{+0.0002}_{-0.0001}$ & $0.0740^{+0.0039}_{-0.0042}$  \\
        $v_{\rm ej}^{\rm dyn}$ $[c]$ & $0.2309^{+0.0046}_{-0.0045}$ & $0.1843^{+0.0106}_{-0.0114}$  \\
        $\theta_{v}$ $[^{\circ}]$ & $32.61^{+0.87}_{-1.74}$ & $21.49^{+2.53}_{-2.53}$  \\
        $q^{\rm wind}+1$ & $90.58^{+0.31}_{-0.69}$ & $83.94^{+5.18}_{-10.42}$  \\
        $q^{\rm dyn}+1$ & $24.3^{+9.8}_{-7.5}$ & $23.5^{+25.5}_{-13.5}$  \\
     \hline \hline 
    \end{tabular}
\end{table}

In the likelihood-based analysis, several parameters are driven toward the boundaries of the prior space, including $d_L$, $\theta_{\rm v}$, $v_{\rm ej}^{\rm wind}$ and $q_{\rm wind}$. This is evidence that the likelihood misspecification discussed in \cref{sec:likelihood} is negatively affecting parameter recovery. The pile up of samples at the prior-boundary seen in the $d_L$ posterior in the simulation recovery (\cref{subsec:sim_study}) now occurs for multiple parameters.

The ANPE framework generally produces broader posterior distributions than the likelihood-based analysis, as was the case with the simulation recovery, and the posterior distributions for $d_L$ and $\theta_{\rm v}$ are again largely prior dominated. The independent constraints on the distance and viewing angle from galaxy surface brightness fluctuations \citep{Cantiello_2018} and VLBI measurements \citep{Mooley_2022} discussed in \cref{sec:at2017gfo} and used to define the priors are known to be more reliable than constraints from the kilonova light curves themselves. It is therefore reasonable that, under these priors, the kilonova data provide little additional information about these parameters. Unlike the ANPE posteriors, the likelihood-based posteriors for $d_L$ and $\theta_{\rm v}$, are pulled away from these independent measurements towards prior boundaries.

The wind ejecta from the accretion disk and the dynamical ejecta generated during merger have distinct timescales and compositions, as summarized in \cref{tab:pe_summary}. The broader implications of these inferences are discussed further in \cref{subsec:discussion-astro}. The wind ejecta constitute the majority ($86\%$ for the ANPE) of the ejecta mass: $m_{\rm ej}^{\rm wind} = 0.074 \,{\rm M}_{\odot}$ versus $m_{\rm ej}^{\rm dyn} = 0.012 \,{\rm M}_{\odot}$. The ANPE infers shape parameters $q^{\rm wind}=83.94^{+5.18}_{-10.42}$ and $q^{\rm dyn}=23.5^{+25.5}_{-13.5}$, suggesting that the ejecta are not toroidal. Under the assumed model, the toroidal ($q=0$) and peanut ($q=1.5$) geometries lie outside the $99^{\rm th}$ percentile credible intervals for both components. The wind ejecta geometry is consistent with being spherical ($q=90$), whereas the dynamical ejecta geometry ($q^{\rm dyn} \simeq 24$) excludes both toroidal and spherical extremes.  These geometries are discussed in more detail in \cref{subsec:discussion-astro}. 

Even though the wind dominates ejecta mass, the faster dynamical ejecta carry the majority of the kinetic energy, with the ANPE estimate of $E_{\rm kin}^{\rm dyn}\sim \! 3 \times 10^{50}\,{\rm erg}$ implying $\sim\!60\%$ of the total. The inferred velocities from the ANPE are $v_{\rm ej}^{\rm wind}=0.074\,c$ and $v_{\rm ej}^{\rm dyn}=0.184\,c$. While the velocities inferred by the likelihood-based analysis differ significantly from those inferred by the ANPE ($v_{\rm ej}^{\rm wind}=0.050\,c$, which is at the prior boundary, and $v_{\rm ej}^{\rm dyn}=0.231\,c$), they do still have the property that $v_{\rm ej}^{\rm wind} < v_{\rm ej}^{\rm dyn}$ implying this result is robust.

The likelihood-based electron fraction posteriors do not show direct signs of likelihood misspecification; neither $Y_e^{\rm wind}$ nor $Y_e^{\rm dyn}$ pile up at a prior boundary. The simulation studies also show no systematic bias in these parameters. The significant differences between the two methods ($Y_e^{\rm wind}: 0.37$ vs.\ $0.29$; $Y_e^{\rm dyn}: 0.22$ vs.\ $0.25$) are therefore likely driven by degeneracies with $d_L$, $m_{\rm ej}$, and $v_{\rm ej}$, which exhibit clear evidence of likelihood misspecification.

\begin{figure}[ht]
    \centering
    \includegraphics[width=0.9\linewidth]{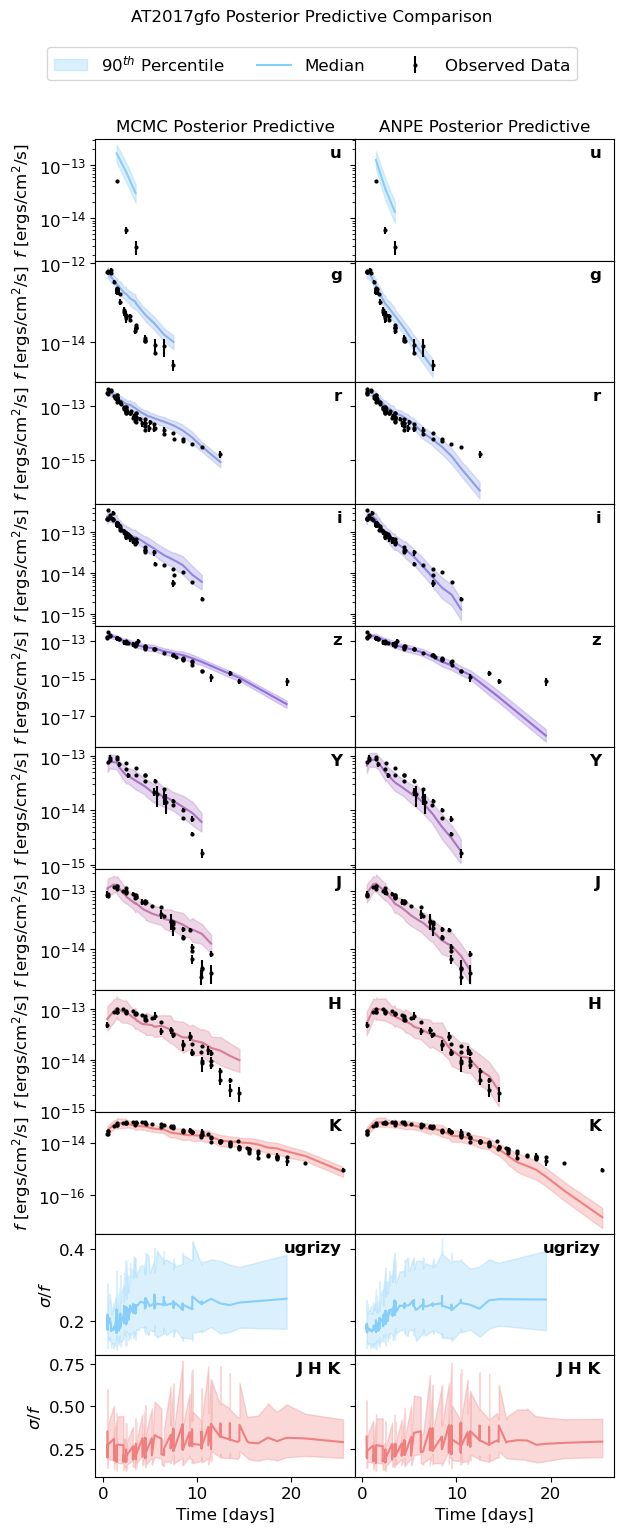}
    \caption{Posterior predictive distributions for AT2017gfo, comparing inference using an MCMC sampler (left) and the ANPE (right). Predicted fluxes are generated from posterior samples, with the median light curve shown as a solid line and the 90th percentile credible interval shown as a shaded band. AT2017gfo is shown in black. The lower panels show the distribution of the flux-to-uncertainty ratio for the reddest bands (\textit{JHK}) and for the remaining optical bands (\textit{ugrizy}).}
    \label{fig:postpred_at2017gfo}
\end{figure}

\Cref{fig:postpred_at2017gfo} shows the posterior predictive light curves for both analyses. For both inference methods, the bluest band ($u$) shows the largest discrepancy with the data. This is likely due to limitations in the \textsc{possis} simulations. Recent work (e.g., \citealt{Gutierrez_2025}) has shown that early UV and blue-band emission is at least in part due to shock cooling and cocoon emission, which are not included in \textsc{possis} \citep{Bulla_2023}. Additionally, the opacities used in \textsc{possis} \citep{Tanaka_2020} may be underestimated when the ejecta are extremely hot \citep{Banerjee_2022}, leading to models that are too bright at early times and in blue bands (see \cref{subsec:possis}). In this case, the \textsc{possis} models underlying the ANPE's training data are missing physics needed to accurately fit the observed data. Since this missing physics is not otherwise accounted for in the training set, the ANPE inherits these limitations and cannot compensate for model misspecification.

In the bluer bands (\textit{u,g,r,i}) the ANPE predictions match the observed data more closely than the likelihood-based ones, which tend to overestimate the brightness (consistent with  \cref{fig:sampling_dist}). 
The ANPE predictions also provide better late-time matches in the \textit{y}, \textit{J}, and \textit{H} bands, while the likelihood-based analysis better matches reddest band (\textit{K}) at $t \gtrsim 15$~d. This is likely because the uncertainty in this band is relatively small and the sampling distribution is more Gaussian. Even though the ANPE and likelihood-based posteriors differ substantially and this difference is reflected in the shape of the predicted light curves, both methods produce reasonable fits to the observed data.

\begin{figure}[ht]
    \centering
    \includegraphics[width=0.99\linewidth]{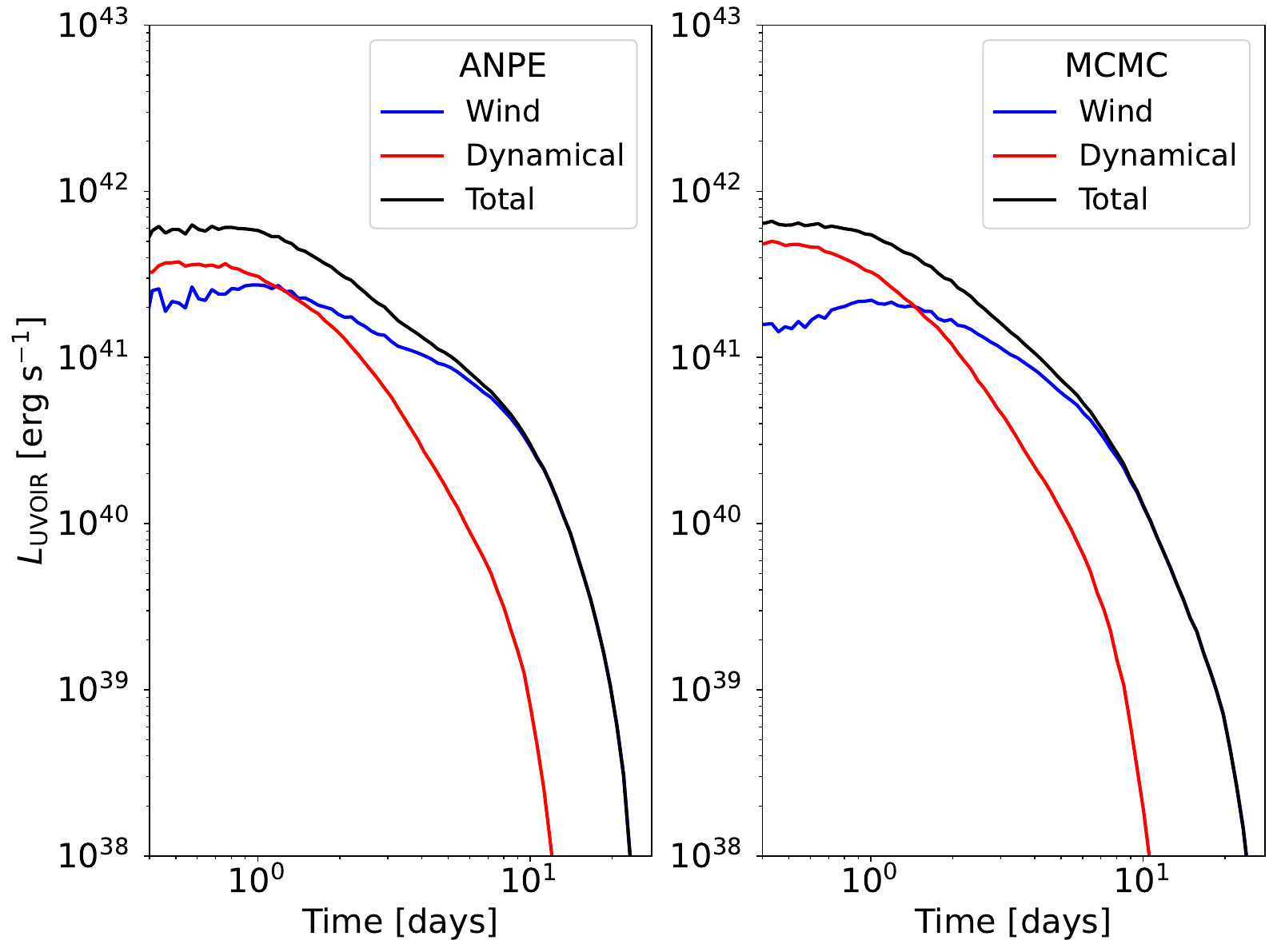}
    \caption{Predicted bolometric light curves for AT2017gfo based on the ANPE (left) and MCMC (right) posteriors. The wind (blue) and dynamical (red) ejecta components are shown separately, with the total luminosity shown in black. Component parameters are taken as the nearest neighbors in the training set to enable computation of luminosity from \textsc{possis} SEDs. The luminosity is integrated over the UVOIR wavelength range. }
    \label{fig:lbol_two_comp}
\end{figure}

Looking at \cref{fig:lbol_two_comp}, we can see that the dynamical ejecta dominate the total luminosity at early times, with the wind component becoming dominant from $\sim\!2$~days. In the likelihood-based analysis, the dynamical component dominates for longer. However, in the ANPE inference, the wind ejecta reach 99\% of the total luminosity approximately 1.1 days later, reflecting differences in the inferred mass and velocity parameters. The relative contribution of the two components is consistent with the wavelength-dependent transition observed in individual photometric bands, where the dynamical ejecta dominate at early times and the wind component takes over between 1--5 days depending on wavelength.

\section{Discussion}
\label{sec:discussion}

We have presented an SBI framework for rapid parameter estimation of kilonova data using a GP emulator trained on 3D Monte Carlo radiative transfer simulations. We compared this approach with a traditional likelihood-based analysis using the same GP-based kilonova emulator and observational data.

\subsection{Methodological lessons}
\label{subsec:discussion-methods}

In simulation studies, the ANPE consistently recovers the injected parameters and produces posterior predictive light curves that match the simulated observations. In the single-injection simulation study, the likelihood-based analysis also recovers most parameters, but the inferred distance and dynamical ejecta mass are skewed toward the upper boundary of the prior range. In the Latin hypercube study, we find that, in addition to distance and dynamical mass, the wind ejecta mass and wind ejecta velocity also exhibit systematic bias.  

This bias arises from the likelihood misspecification discussed in \cref{sec:likelihood}, where the Gaussian diagonal approximation fails to capture the true uncertainty. This effect is most significant where the model uncertainty is largest and the non-Gaussian features of the sampling distribution are more pronounced. For GP emulators, which extrapolate poorly, the likelihood misspecification may be more significant near edges of the training set, which coincide with the prior boundaries in this work. \cref{fig:sampling_dist} shows that smaller model uncertainties (\textit{g} band) lead to more Gaussian sampling distributions than larger ones (\textit{u} and \textit{H} bands).
The relationship between model uncertainty and likelihood misspecification may also explain why the Latin-hypercube study identifies bias in four parameters, whereas the single-injection study identifies only two. The BEO procedure reduced the model uncertainty specifically in the parameter-space region relevant to AT2017gfo.
From Figure~\ref{fig:sampling_dist}, we can also see that the likelihood effectively assumes that the model predictions are fainter than the true distribution, which is reflected in the systematically over-bright light curves predicted by the likelihood-based analysis (\cref{fig:postpred_at2017gfo}). Existing analyses span a spectrum of likelihood approximations, from a single, constant systematic uncertainty to uncorrelated, band- and time-dependent Gaussian emulator variances as in our likelihood-based analysis. The SBI framework presented here goes further by capturing the full non-Gaussian, correlated predictive distribution. 

The effect of likelihood misspecification is more clear when analyzing real data where model misspecification plays a role. Model misspecification arises from missing physics in the \textsc{possis} models such as shock cooling; these limitations are inherited by both the ANPE and the MCMC. This is evident in the posterior predictive light curves for the $u$ band (\cref{fig:postpred_at2017gfo}), where both methods over-predict the flux. 

However, model misspecification affects the two methods differently. In the likelihood-based analysis, when the model predictions differ from the data, the residual term of the likelihood ($\bar{f}_i(\bm{\phi}) - d_i$) increases. This residual is balanced with the uncertainty of the fit ($\sigma_{d,i}^2 + \sigma^2_{m,i}$). As a result, the MCMC sampler favors parameter configurations where the emulator uncertainties compensate for mismatches between the model and the data. In regimes where emulator error is greater than observational uncertainty---which is often the case for kilonova studies that rely on emulators due to the limited number of high-fidelity models---this effect is particularly strong. This effect is clear in the likelihood-based analysis of AT2017gfo, where the parameters that only weakly influence the mean prediction ($d_L$ and $\theta_{\rm v}$) are pushed toward the boundaries of the prior due to the model uncertainty. 

Additionally, our two-component model assumes that the photons in the individual components do not interact and the component light curves can be combined linearly. The accuracy of this assumption is parameter-dependent and may be correlated with GP uncertainty. Notably, both of these are larger near prior boundaries. It is plausible that a correlation between model misspecification and uncertainty could amplify the MCMC's tendency to absorb model error through the variance term, contributing to the pile up of samples at prior boundaries observed for AT2017gfo. Since the ANPE learns the predictive distribution from forward simulations and does not depend explicitly on residuals or uncertainties, it may be more robust to this specific parameter-dependent model misspecification.

The limitations of the approximations are less severe in the simulation study, where the synthetic data are generated from the same model used for inference. In that case, the residuals can be reduced nearly to zero.

The ANPE framework, on the other hand, learns the full emulator uncertainty through the forward modeling process used to generate its training data. This allows it to capture the correlations and non-Gaussian features in the predictive distributions. This results in posteriors that are more robust to large and complex emulator uncertainties that are not captured by a diagonal Gaussian likelihood. This is particularly advantageous given the uncertainties inherent in surrogate models, which are commonly employed in kilonova studies due to the complexity of high-fidelity modeling. This holds true for other areas of astronomy such as cosmology (e.g. Lyman-alpha forest emulators, \citealp{Bird_2019}) that also rely on surrogate models for complex simulations. While the high computational cost of explicitly constructing the full covariance matrix is inherent to GPs, GPs are well suited to this problem due to their ability to provide well-calibrated uncertainty estimates on limited data sets such as those from {\sc possis}. Alternative emulator methods such as neural networks, even methods like Bayesian neural networks that can provide uncertainty estimates, do not produce full covariance matrices.

Another advantage of the ANPE framework is computational efficiency. Once trained, the neural posterior estimator can generate $\sim2\times10^4$ posterior samples in approximately $4$ seconds for a single observation. In contrast, the likelihood-based analysis requires roughly 9 hours (for 1000 walkers and 1400 iterations). All analyses were run using a single NVIDIA HGX A100 GPU card. This speed advantage makes ANPE particularly attractive for large simulation studies and for rapid parameter estimation in future kilonova discoveries.

\subsection{Astrophysical interpretation of AT2017gfo}
\label{subsec:discussion-astro}

While the primary focus of this work is methodological, the parameter estimates for AT2017gfo provide insights into the shape and composition of the merger ejecta. Given the limitations of the likelihood-based analysis discussed above, the numbers quoted here, unless explicitly stated otherwise, come from the ANPE.

A quantitative comparison of our ejecta parameter estimates to published AT2017gfo analyses is complicated by substantial heterogeneity across the literature. These studies employ different radiative transfer codes, composition parameterizations, prior ranges, and assumed ejecta geometries. The analysis by \citet{Anand_2023} is closest to ours in methodological terms, employing a version of the \textsc{possis} 2023 code \citep{Bulla_2023} with an explicit electron fraction parameterization. However, there are key differences. \citet{Anand_2023} employs a self-consistent two component model with a lower ejecta velocity boundary ($v_{\rm ej}^{\rm wind} \geq 0.03c$) and a different ejecta geometry where the ejecta are distributed over all angles.

We do not directly compare to the series of results from the Los Alamos group (e.g., \citealp{Ristic_2022, Peng_2024, King_2025}) or other results that use different models (e.g., \citealp{Kasen_2017, Hinderer_2019, Lukoiute_2022}) as the compounding differences in modeling, composition parameterization, prior ranges, and geometries mean that differences in inferred parameters should be understood as reflections of modeling choices.
 
In this work, the wind ejecta are inferred to be approximately spherical ($q^{\rm wind} \simeq 84$) in geometry, while the dynamical ejecta ($q^{\rm dyn} \simeq 24$) excludes both spherical and toroidal extremes (cf\ the two right-most panels of \cref{fig:cassini}).  Recent work has raised questions about ejecta geometry: \citet{Sneppen_2024} found evidence for spherical ejecta in spectral features at early times. This differs from the predictions of simulations of binary neutron star mergers \citep{Bauswein_2013, Sekiguchi_2015, Bovard_2017,Radice_2018b,Nedora_2022, Foucart_2023,Combi_2023}: dynamical ejecta from shock heating are more isotropic, while dynamical ejecta from the tidal tails and wind ejecta are more concentrated near the equatorial plane. 
However, \citet{Collins_2024} demonstrates that the ejecta photosphere can appear highly spherical even for asymmetric ejecta. Our finding is in tension with \citet{King_2025}, which finds that, under their model, AT2017gfo is best fit by toroidal dynamical and peanut wind ejecta geometries.

We do not assign a physical origin to the two components as recent work has shown that model parameters inferred from light curves may not reflect the true ejecta configuration \citep{Kitamura_2025}. However, we find a more massive, relatively lanthanide-poor component ($Y_e^{\rm wind} \approx 0.37$) and less massive lanthanide-rich component ($Y_e^{\rm dyn} \approx 0.22$). These results are qualitatively consistent with previous studies \citep{Villar_2017, Dietrich_2020, Pang_2023, Anand_2023, Almualla_2022, Peng_2024, King_2025}.  A direct comparison to \citet{Anand_2023} is limited because our model does not account for interactions between components; these interactions can change the inferred composition and mass.

Despite the wind dominating the ejecta mass ($M_{\rm tot} = 0.0866$), the dynamical ejecta carry approximately $60\%$ of the total kinetic energy due to their higher velocity ($v_{\rm ej}^{\rm dyn} / v_{\rm ej}^{\rm wind} \simeq 2.5$). The total kinetic energy of $\approx 6 \times 10^{50}\,{\rm ergs}$ is consistent with estimated energy budgets for kilonova of $\approx 10^{49} - 10^{51}\,{\rm erg}$ \citep{Bauswein_2013, Hotokezaka_2013, Radice_2016, Metzger_2019}. 

\section{Conclusion}

The results in this paper demonstrate that SBI can provide computationally efficient posteriors for kilonovae that are more robust to emulator uncertainty and likelihood misspecification than traditional likelihood-based methods. When applied to AT2017gfo with \textsc{possis} models, the ANPE yields posteriors that are qualitatively consistent with merger simulations and previous analyses, though detailed quantitative comparison is limited by methodological diversity in the literature. The posterior predictive light curves show that the ANPE reasonably reproduces observed AT2017gfo light curves under the assumed model.

As kilonova discovery rates accelerate with new facilities such as JWST, Roman, LSST, and next-generation gravitational-wave detectors, the diversity of observed events will span broad regions of parameter space with varying merger geometries, electron fractions, and mass ratios. Surrogate models trained on limited simulations will inevitably encounter parameter space regions where model uncertainty is large, and traditional likelihood-based approaches will struggle to keep pace with both the data rate and the increased likelihood misspecification. An SBI framework, once trained, can be applied to new observations without retraining, enabling rapid characterization of kilonova properties across diverse observational campaigns. This computational efficiency, combined with robustness to emulator uncertainty and likelihood misspecification, positions SBI as a promising method for maximizing the scientific value of future kilonova observations.

\section*{Author Contributions}

We outline the different contributions below using the CRediT (Contribution Roles Taxonomy) system.
\textbf{SMB:} Conceptualization, Data Curation, Formal Analysis, Investigation, Methodology, Project Administration, Software, Visualization, Writing (original draft), Writing (review and editing). \textbf{MB}: Conceptualization, Investigation, Methodology, Software, Visualization, Writing (original draft).
\textbf{HVP:} Conceptualization, Formal Analysis, Funding Acquisition, Investigation, Methodology, Project Administration, Resources, Supervision, Validation, Visualization, Writing (review and editing).
\textbf{NS:} Conceptualization, Data Curation, Formal Analysis, Investigation, Methodology, Software, Validation, Writing (review and editing).
\textbf{DM:} Formal Analysis, Funding Acquisition, Investigation, Methodology, Validation, Visualization, Writing (review and editing).
\textbf{ST:} Methodology, Software, Writing (review and editing).
\textbf{GJ:} Software.
\textbf{SR:} Conceptualization, Funding Acquisition, Resources, Writing (review and editing).
\textbf{SN:} Conceptualization, Writing (review and editing).

\section*{Acknowledgements}

We thank Justin Alsing, James Alvey, and Oleg Korobkin for  valuable conversations.

SMB is supported by the research project grant ``Fundamental physics from populations of compact object mergers'' funded by VR under Dnr 2021-04195, the research project grant ``Gravity Meets Light'' funded by the Knut and Alice Wallenberg Foundation under Dnr KAW 2019.0112, and by the Netherlands Organization for Scientific Research (NWO) under grant number VI.Veni.242.361. 

MB acknowledges the Department of Physics and Earth Science of the University of Ferrara for the financial support through the FIRD 2025 grant. 

HVP and ST have been supported by funding from the European Research Council (ERC) under the European Union's Horizon 2020 research and innovation programmes (grant agreement no.\ 101018897 CosmicExplorer), and by the research project grant ``Understanding the Dynamic Universe'' funded by the Knut and Alice Wallenberg Foundation under Dnr KAW 2018.0067. HVP was additionally supported by the G\"{o}ran Gustafsson Foundation for Research in Natural Sciences and Medicine.

NS is supported by the Kavli Foundation. 

SR has been supported by the European Research Council (ERC) Advanced Grant INSPIRATION under the European Union's Horizon 2020 research and innovation programme (Grant agreement No.\ 101053985), by Deutsche Forschungsgemeinschaft (DFG, German Research Foundation) under Germany's Excellence Strategy -- EXC 2121 ``Quantum Universe" -- 390833306 and by the Swedish Research Council (VR) under grant number 2020-05044.

SN acknowledges support from the Netherlands Organization for Scientific Research (NWO).

This research utilized the Sunrise HPC facility supported by the Technical Division at the Department of Physics, Stockholm University.

\textit{Software:} 
\texttt{affine}\footnote{\url{https://github.com/justinalsing/affine}} ;
\texttt{corner} \citep{Corner_2016} ;
\texttt{flowfusion}\footnote{\url{https://github.com/Cosmo-Pop/flowfusion}} \citep{Alsing_2024, Thorp_2024, Thorp_2025, Leistedt_2026};
\texttt{GPyTorch} \citep{GPyTorch_2018}\footnote{\url{http://github.com/cornellius-gp/gpytorch}} ;
\texttt{matplotlib} \citep{Matplotlib_2007};
\texttt{NumPy} \citep{NumPy_2020};
\texttt{pandas} \citep{Pandas_2010, McKinney_2010};
\textsc{POSSIS} \citep{Bulla_2019,Bulla_2023}  ;
\texttt{PyTorch} \citep{PyTorch_2019,PyTorch_2024};
\texttt{quantile\_utilities} \citep{QU_2024, Thorp_2024b};
\texttt{RayTune} \citep{RayTune_2018};
\textsc{redback}\footnote{\url{https://github.com/nikhil-sarin/redback}} \citep{Sarin2024};
\texttt{SciPy} \citep{SciPy_2020};
\textsc{sncosmo}\footnote{\url{https://github.com/sncosmo/sncosmo}} \citep{sncosmo_2025};
\texttt{tqdm} \citep{tqdm_2026};
\texttt{torchdiffeq} \citep{torchdiffeq_2018, Chen_2018};
\textsc{wandb} \citep{wandb_2020};

ChatGPT-5.1\footnote{\url{https://chatgpt.com}} was used in the draft to give feedback regarding conciseness and clarity of the text written by the authors. Claude Opus 4.6 and Claude Haiku 4.5\footnote{\url{https://claude.ai}} were used to evaluate the final text for clarity, flow, and redundancy. 

\section*{Data Availability} 

The AT2017gfo data is publicly hosted by the Open Astronomy Catalog~\citep{oac_api} and processed through the open source software package {\sc Redback} \citep{Sarin2024}.

\clearpage

\bibliography{export}{}

@article{Abbott_2017a,
   author = {B. P. Abbott and R. Abbott and T. D. Abbott and F. Acernese and K. Ackley and C. Adams and T. Adams and P. Addesso and R. X. Adhikari and V. B. Adya and others},
   doi = {10.1103/PhysRevLett.119.161101},
   issn = {10797114},
   issue = {16},
   journal = {\prl},
   month = oct,
   pmid = {29099225},
   publisher = {American Physical Society},
   title = {GW170817: Observation of Gravitational Waves from a Binary Neutron Star Inspiral},
   volume = {119},
   year = {2017},
}

@article{Abbott_2017b,
   author = {B. P. Abbott and R. Abbott and T. D. Abbott and F. Acernese and K. Ackley and C. Adams and T. Adams and P. Addesso and R. X. Adhikari and V. B. Adya and others},
   doi = {10.3847/2041-8213/aa920c},
   issn = {2041-8205},
   issue = {2},
   journal = {\apjl},
   month = oct,
   pages = {L13},
   title = {Gravitational Waves and Gamma-Rays from a Binary Neutron Star Merger: GW170817 and GRB 170817A},
   volume = {848},
   url = {https://iopscience.iop.org/article/10.3847/2041-8213/aa920c},
   year = {2017},
}

@article{Abbott_2017c,
   author = {B. P. Abbott and R. Abbott and T. D. Abbott and F. Acernese and K. Ackley and C. Adams and T. Adams and P. Addesso and R. X. Adhikari and V. B. Adya and others},
   doi = {10.3847/2041-8213/aa91c9},
   issn = {2041-8205},
   issue = {2},
   journal = {\apjl},
   month = nov,
   pages = {L12},
   title = {Multi-messenger Observations of a Binary Neutron Star Merger},
   volume = {848},
   url = {https://iopscience.iop.org/article/10.3847/2041-8213/aa91c9},
   year = {2017},
}

@article{Abbott_2017d,
   author = {B. P. Abbott and R. Abbott and T. D. Abbott and F. Acernese and K. Ackley and C. Adams and T. Adams and P. Addesso and R. X. Adhikari and V. B. Adya and others},
   doi = {10.1038/nature24471},
   issn = {14764687},
   issue = {7678},
   journal = {Nature},
   month = nov,
   pages = {85-98},
   pmid = {29094696},
   publisher = {Nature Research},
   title = {A gravitational-wave standard siren measurement of the Hubble constant},
   volume = {551},
   year = {2017},
}

@article{Bauswein_2017b,
   author = {Andreas Bauswein and Oliver Just and Hans-Thomas Janka and Nikolaos Stergioulas},
   doi = {10.3847/2041-8213/aa9994},
   issn = {2041-8205},
   issue = {2},
   journal = {\apjl},
   month = {12},
   pages = {L34},
   publisher = {American Astronomical Society},
   title = {Neutron-star Radius Constraints from GW170817 and Future Detections},
   volume = {850},
   year = {2017},
}

@article{Kasen_2017,
   author = {Daniel Kasen and Brian Metzger and Jennifer Barnes and Eliot Quataert and Enrico Ramirez-Ruiz},
   doi = {10.1038/nature24453},
   issn = {14764687},
   issue = {7678},
   journal = {Nature},
   month = {11},
   pages = {80-84},
   pmid = {29094687},
   publisher = {Nature Publishing Group},
   title = {Origin of the heavy elements in binary neutron-star mergers from a gravitational-wave event},
   volume = {551},
   year = {2017},
}

@article{Hotokezaka_2018,
   author = {{Hotokezaka}, K. and {Nakar}, E. and {Gottlieb}, O. and {Nissanke}, S. and {Masuda}, K. and {Hallinan}, G. and {Mooley}, K.~P. and {Deller}, A.~T.},
        title = "{A Hubble constant measurement from superluminal motion of the jet in GW170817}",
      journal = {NatAs},
         year = 2019,
        month = jul,
       volume = {3},
        pages = {940-944},
          doi = {10.1038/s41550-019-0820-1},
archivePrefix = {arXiv},
       eprint = {1806.10596},
 primaryClass = {astro-ph.CO},
       adsurl = {https://ui.adsabs.harvard.edu/abs/2019NatAs...3..940H}
}

@article{Coughlin_2018,
   author = {Michael W Coughlin and Tim Dietrich and Zoheyr Doctor and Daniel Kasen and Scott Coughlin and Anders Jerkstrand and Giorgos Leloudas and Owen McBrien and Brian D Metzger and Richard O’Shaughnessy and Stephen J Smartt},
   doi = {10.1093/mnras/sty2174},
   issn = {0035-8711},
   issue = {3},
   journal = {\mnras},
   month = {11},
   pages = {3871-3878},
   publisher = {Oxford University Press (OUP)},
   title = {Constraints on the neutron star equation of state from AT2017gfo using radiative transfer simulations},
   volume = {480},
   year = {2018},
}

@article{Annala_2018,
   author = {Eemeli Annala and Tyler Gorda and Aleksi Kurkela and Aleksi Vuorinen},
   doi = {10.1103/PhysRevLett.120.172703},
   issn = {10797114},
   issue = {17},
   journal = {\prl},
   month = {4},
   pmid = {29756823},
   pages = {172703},
   publisher = {American Physical Society},
   title = {Gravitational-Wave Constraints on the Neutron-Star-Matter Equation of State},
   volume = {120},
   year = {2018},
}

@article{Most_2018,
   author = {Elias R. Most and Lukas R. Weih and Luciano Rezzolla and Jürgen Schaffner-Bielich},
   doi = {10.1103/PhysRevLett.120.261103},
   issn = {10797114},
   issue = {26},
   journal = {\prl},
   month = {6},
   pmid = {30004744},
   publisher = {American Physical Society},
   title = {New Constraints on Radii and Tidal Deformabilities of Neutron Stars from GW170817},
   volume = {120},
   pages = {261103},
   year = {2018},
}

@article{Radice_2018a,
   author = {David Radice and Albino Perego and Francesco Zappa and Sebastiano Bernuzzi},
   doi = {10.3847/2041-8213/aaa402},
   issn = {2041-8205},
   issue = {2},
   journal = {\apjl},
   month = {1},
   pages = {L29},
   publisher = {American Astronomical Society},
   title = {GW170817: Joint Constraint on the Neutron Star Equation of State from Multimessenger Observations},
   volume = {852},
   year = {2018},
}

@article{Ruiz_2018,
   author = {Milton Ruiz and Stuart L. Shapiro and Antonios Tsokaros},
   doi = {10.1103/PhysRevD.97.021501},
   issn = {24700029},
   issue = {2},
   journal = {\prd},
   month = {1},
   publisher = {American Physical Society},
   title = {GW170817, general relativistic magnetohydrodynamic simulations, and the neutron star maximum mass},
   volume = {97},
   year = {2018},
   pages = {021501}
}

@article{Coughlin_2019,
   author = {Michael W. Coughlin and Tim Dietrich and Ben Margalit and Brian D. Metzger},
   doi = {10.1093/mnrasl/slz133},
   issn = {17453933},
   issue = {1},
   journal = {\mnras},
   month = {10},
   pages = {L91-L96},
   publisher = {Oxford University Press},
   title = {Multimessenger Bayesian parameter inference of a binary neutron star merger},
   volume = {489},
   year = {2019},
}

@article{Hinderer_2019,
   author = {Tanja Hinderer and Samaya Nissanke and Francois Foucart and Kenta Hotokezaka and Trevor Vincent and Mansi Kasliwal and Patricia Schmidt and Andrew R. Williamson and David A. Nichols and Matthew D. Duez and Lawrence E. Kidder and Harald P. Pfeiffer and Mark A. Scheel},
   doi = {10.1103/PhysRevD.100.063021},
   issn = {24700029},
   issue = {6},
   journal = {\prd},
   month = {9},
   publisher = {American Physical Society},
   title = {Distinguishing the nature of comparable-mass neutron star binary systems with multimessenger observations: GW170817 case study},
   volume = {100},
   year = {2019},
   pages = {063021}
}

@article{Dietrich_2020,
   author = {{Dietrich}, Tim and {Coughlin}, Michael W. and {Pang}, Peter T.~H. and {Bulla}, Mattia and {Heinzel}, Jack and {Issa}, Lina and {Tews}, Ingo and {Antier}, Sarah},
        title = "{Multimessenger constraints on the neutron-star equation of state and the Hubble constant}",
      journal = {Science},
         year = 2020,
        month = dec,
       volume = {370},
       number = {6523},
        pages = {1450-1453},
          doi = {10.1126/science.abb4317},
archivePrefix = {arXiv},
       eprint = {2002.11355},
 primaryClass = {astro-ph.HE},
       adsurl = {https://ui.adsabs.harvard.edu/abs/2020Sci...370.1450D}
}

@article{Raaijmakers_2020,
   author = {G. Raaijmakers and S. K. Greif and T. E. Riley and T. Hinderer and K. Hebeler and A. Schwenk and A. L. Watts and S. Nissanke and S. Guillot and J. M. Lattimer and R. M. Ludlam},
   doi = {10.3847/2041-8213/ab822f},
   issn = {2041-8205},
   issue = {1},
   journal = {\apjl},
   month = {4},
   pages = {L21},
   publisher = {American Astronomical Society},
   title = {Constraining the Dense Matter Equation of State with Joint Analysis of NICER and LIGO/Virgo Measurements},
   volume = {893},
   year = {2020},
}

@article{Pang_2021,
   author = {Peter T. H. Pang and Ingo Tews and Michael W. Coughlin and Mattia Bulla and Chris Van Den Broeck and Tim Dietrich},
   doi = {10.3847/1538-4357/ac19ab},
   issn = {0004-637X},
   issue = {1},
   journal = {\apj},
   month = {11},
   pages = {14},
   publisher = {American Astronomical Society},
   title = {Nuclear Physics Multimessenger Astrophysics Constraints on the Neutron Star Equation of State: Adding NICER’s PSR J0740+6620 Measurement},
   volume = {922},
   year = {2021},
}

@article{Pang_2023,
   author = {Peter T. H. Pang and Tim Dietrich and Michael W. Coughlin and Mattia Bulla and Ingo Tews and Mouza Almualla and Tyler Barna and Ramodgwendé Weizmann Kiendrebeogo and Nina Kunert and Gargi Mansingh and Brandon Reed and Niharika Sravan and Andrew Toivonen and Sarah Antier and Robert O. VandenBerg and Jack Heinzel and Vsevolod Nedora and Pouyan Salehi and Ritwik Sharma and Rahul Somasundaram and Chris Van Den Broeck},
   doi = {10.1038/s41467-023-43932-6},
   issn = {20411723},
   issue = {1},
   journal = {NatCo},
   month = {12},
   pmid = {38123551},
   pages = {8352},
   publisher = {Nature Research},
   title = {An updated nuclear-physics and multi-messenger astrophysics framework for binary neutron star mergers},
   volume = {14},
   year = {2023},
}

@ARTICLE{Rosswog_2018,
       author = {{Rosswog}, S. and {Sollerman}, J. and {Feindt}, U. and {Goobar}, A. and {Korobkin}, O. and {Wollaeger}, R. and {Fremling}, C. and {Kasliwal}, M.~M.},
        title = "{The first direct double neutron star merger detection: Implications for cosmic nucleosynthesis}",
      journal = {\aap},
         year = 2018,
        month = jul,
       volume = {615},
          eid = {A132},
        pages = {A132},
          doi = {10.1051/0004-6361/201732117},
archivePrefix = {arXiv},
       eprint = {1710.05445},
 primaryClass = {astro-ph.HE},
       adsurl = {https://ui.adsabs.harvard.edu/abs/2018A&A...615A.132R}
}

@article{Peng_2024,
       author = {{Peng}, Yinglei and {Risti{\'c}}, Marko and {Kedia}, Atul and {O'Shaughnessy}, Richard and {Fontes}, Christopher J. and {Fryer}, Chris L. and {Korobkin}, Oleg and {Mumpower}, Matthew R. and {Villar}, V. Ashley and {Wollaeger}, Ryan T.},
        title = "{Kilonova light-curve interpolation with neural networks}",
      journal = {PhRvR},
         year = 2024,
        month = jul,
       volume = {6},
       number = {3},
          eid = {033078},
        pages = {033078},
          doi = {10.1103/PhysRevResearch.6.033078},
archivePrefix = {arXiv},
       eprint = {2402.05871},
 primaryClass = {astro-ph.HE},
       adsurl = {https://ui.adsabs.harvard.edu/abs/2024PhRvR...6c3078P}
}

@article{Ragosta_2024,
   author = {{Ragosta}, Fabio and {Ahumada}, Tom{\'a}s and {Piranomonte}, Silvia and {Andreoni}, Igor and {Melandri}, Andrea and {Colombo}, Alberto and {Coughlin}, Michael W.},
        title = "{Kilonova Parameter Estimation with LSST at Vera C. Rubin Observatory}",
      journal = {\apj},
         year = 2024,
        month = may,
       volume = {966},
       number = {2},
          eid = {214},
        pages = {214},
          doi = {10.3847/1538-4357/ad35c1},
archivePrefix = {arXiv},
       eprint = {2403.14016},
 primaryClass = {astro-ph.HE},
       adsurl = {https://ui.adsabs.harvard.edu/abs/2024ApJ...966..214R}
}

@article{Sneppen_2024,
   author = {{Sneppen}, Albert and {Watson}, Darach and {Damgaard}, Rasmus and {Heintz}, Kasper E. and {Vieira}, Nicholas and {V{\"a}is{\"a}nen}, Petri and {Mahoro}, Antoine},
        title = "{Emergence hour-by-hour of r-process features in the kilonova AT2017gfo}",
      journal = {\aap},
         year = 2024,
        month = oct,
       volume = {690},
          eid = {A398},
        pages = {A398},
          doi = {10.1051/0004-6361/202450317},
archivePrefix = {arXiv},
       eprint = {2404.08730},
 primaryClass = {astro-ph.HE},
       adsurl = {https://ui.adsabs.harvard.edu/abs/2024A&A...690A.398S}
}

@article{Savchenko_2017,
   title={INTEGRAL Detection of the First Prompt Gamma-Ray Signal Coincident with the Gravitational-wave Event GW170817},
   volume={848},
   ISSN={2041-8213},
   url={http://dx.doi.org/10.3847/2041-8213/aa8f94},
   DOI={10.3847/2041-8213/aa8f94},
   number={2},
   journal={\apjl},
   publisher={American Astronomical Society},
   author={Savchenko, V. and Ferrigno, C. and Kuulkers, E. and Bazzano, A. and Bozzo, E. and Brandt, S. and Chenevez, J. and Courvoisier, T. J.-L. and Diehl, R. and Domingo, A. and Hanlon, L. and Jourdain, E. and von Kienlin, A. and Laurent, P. and Lebrun, F. and Lutovinov, A. and Martin-Carrillo, A. and Mereghetti, S. and Natalucci, L. and Rodi, J. and Roques, J.-P. and Sunyaev, R. and Ubertini, P.},
   year={2017},
   month=oct, pages={L15} }

@article{Goldstein_2017,
   author = {A. Goldstein and P. Veres and E. Burns and M. S. Briggs and R. Hamburg and D. Kocevski and C. A. Wilson-Hodge and R. D. Preece and S. Poolakkil and O. J. Roberts and C. M. Hui and V. Connaughton and J. Racusin and A. von Kienlin and T. Dal Canton and N. Christensen and T. Littenberg and K. Siellez and L. Blackburn and J. Broida and E. Bissaldi and W. H. Cleveland and M. H. Gibby and M. M. Giles and R. M. Kippen and S. McBreen and J. McEnery and C. A. Meegan and W. S. Paciesas and M. Stanbro},
   doi = {10.3847/2041-8213/aa8f41},
   issn = {2041-8205},
   issue = {2},
   journal = {\apjl},
   month = {10},
   pages = {L14},
   publisher = {American Astronomical Society},
   title = {An Ordinary Short Gamma-Ray Burst with Extraordinary Implications: Fermi-GBM Detection of GRB 170817A},
   volume = {848},
   year = {2017},
}

@article{Coulter_2017,
   author = {D. A. Coulter and R. J. Foley and C. D. Kilpatrick and M. R. Drout and A. L. Piro and B. J. Shappee and M. R. Siebert and J. D. Simon and N. Ulloa and D. Kasen and B. F. Madore and A. Murguia-Berthier and Y. C. Pan and J. X. Prochaska and E. Ramirez-Ruiz and A. Rest and C. Rojas-Bravo},
   doi = {10.1126/science.aap9811},
   issn = {10959203},
   issue = {6370},
   journal = {Science},
   month = {12},
   pages = {1556-1558},
   pmid = {29038368},
   publisher = {American Association for the Advancement of Science},
   title = {Swope Supernova Survey 2017a (SSS17a), the optical counterpart to a gravitational wave source},
   volume = {358},
   year = {2017},
}

@article{Villar_2017,
   author = {{Villar}, V.~A. and {Guillochon}, J. and {Berger}, E. and {Metzger}, B.~D. and {Cowperthwaite}, P.~S. and {Nicholl}, M. and {Alexander}, K.~D. and {Blanchard}, P.~K. and {Chornock}, R. and {Eftekhari}, T. and {Fong}, W. and {Margutti}, R. and {Williams}, P.~K.~G.},
        title = "{The Combined Ultraviolet, Optical, and Near-infrared Light Curves of the Kilonova Associated with the Binary Neutron Star Merger GW170817: Unified Data Set, Analytic Models, and Physical Implications}",
      journal = {\apjl},
         year = 2017,
        month = dec,
       volume = {851},
       number = {1},
          eid = {L21},
        pages = {L21},
          doi = {10.3847/2041-8213/aa9c84},
archivePrefix = {arXiv},
       eprint = {1710.11576},
 primaryClass = {astro-ph.HE},
       adsurl = {https://ui.adsabs.harvard.edu/abs/2017ApJ...851L..21V}
}

@article{Guidorzi_2017,
   author = {C. Guidorzi and R. Margutti and D. Brout and D. Scolnic and W. Fong and K. D. Alexander and P. S. Cowperthwaite and J. Annis and E. Berger and P. K. Blanchard and R. Chornock and D. L. Coppejans and T. Eftekhari and J. A. Frieman and D. Huterer and M. Nicholl and M. Soares-Santos and G. Terreran and V. A. Villar and P. K. G. Williams},
   doi = {10.3847/2041-8213/aaa009},
   month = {10},
   title = {Improved constraints on H0 from a combined analysis of gravitational-wave and electromagnetic emission from GW170817},
   year = {2017},
   journal = {\apjl},
   volume = {851},
   pages = {L36}
}

@article{Wang_2020,
       author = {{Wang}, Hao and {Giannios}, Dimitrios},
        title = "{Multimessenger Parameter Estimation of GW170817: From Jet Structure to the Hubble Constant}",
      journal = {\apj},
         year = 2021,
        month = feb,
       volume = {908},
       number = {2},
          eid = {200},
        pages = {200},
          doi = {10.3847/1538-4357/abd39c},
archivePrefix = {arXiv},
       eprint = {2009.04427},
 primaryClass = {astro-ph.HE},
       adsurl = {https://ui.adsabs.harvard.edu/abs/2021ApJ...908..200W}
}

@article{Coughlin_2020a,
   author = {Michael W. Coughlin and Tim Dietrich and Jack Heinzel and Nandita Khetan and Sarah Antier and Mattia Bulla and Nelson Christensen and David A. Coulter and Ryan J. Foley},
   doi = {10.1103/PhysRevResearch.2.022006},
   issn = {26431564},
   issue = {2},
   journal = {PhRvR},
   month = {4},
   publisher = {American Physical Society},
   title = {Standardizing kilonovae and their use as standard candles to measure the Hubble constant},
   volume = {2},
   year = {2020},
   pages = {022006}
}

@article{Capano_2019,
   author = {Collin D. Capano and Ingo Tews and Stephanie M. Brown and Ben Margalit and Soumi De and Sumit Kumar and Duncan A. Brown and Badri Krishnan and Sanjay Reddy},
   doi = {10.1038/s41550-020-1014-6},
   month = {8},
   title = {Stringent constraints on neutron-star radii from multimessenger observations and nuclear theory},
   url = {http://arxiv.org/abs/1908.10352 http://dx.doi.org/10.1038/s41550-020-1014-6},
   year = {2019},
   journal = {NatAs},
   volume = {4},
   pages = {625}
}

@article{Kasliwal_2022,
   author = {Mansi M. Kasliwal and Daniel Kasen and Ryan M. Lau and Daniel A. Perley and Stephan Rosswog and Eran O. Ofek and Kenta Hotokezaka and Ranga Ram Chary and Jesper Sollerman and Ariel Goobar and David L. Kaplan},
   doi = {10.1093/mnrasl/slz007},
   issn = {17453933},
   issue = {1},
   journal = {\mnras},
   month = {2},
   pages = {L7-L12},
   publisher = {Oxford University Press},
   title = {Spitzer mid-infrared detections of neutron star merger GW170817 suggests synthesis of the heaviest elements},
   volume = {510},
   year = {2022},
}

@article{Cowperthwaite_2017,
   author = {P. S. Cowperthwaite and E. Berger and V. A. Villar and B. D. Metzger and M. Nicholl and R. Chornock and P. K. Blanchard and W. Fong and R. Margutti and M. Soares-Santos and K. D. Alexander and S. Allam and J. Annis and D. Brout and D. A. Brown and R. E. Butler and H.-Y. Chen and H. T. Diehl and Z. Doctor and M. R. Drout and T. Eftekhari and B. Farr and D. A. Finley and R. J. Foley and J. A. Frieman and C. L. Fryer and J. García-Bellido and M. S. S. Gill and J. Guillochon and K. Herner and D. E. Holz and D. Kasen and R. Kessler and J. Marriner and T. Matheson and E. H. Neilsen and E. Quataert and A. Palmese and A. Rest and M. Sako and D. M. Scolnic and N. Smith and D. L. Tucker and P. K. G. Williams and E. Balbinot and J. L. Carlin and E. R. Cook and F. Durret and T. S. Li and P. A. A. Lopes and A. C. C. Lourenço and J. L. Marshall and G. E. Medina and J. Muir and R. R. Muñoz and M. Sauseda and D. J. Schlegel and L. F. Secco and A. K. Vivas and W. Wester and A. Zenteno and Y. Zhang and T. M. C. Abbott and M. Banerji and K. Bechtol and A. Benoit-Lévy and E. Bertin and E. Buckley-Geer and D. L. Burke and D. Capozzi and A. Carnero Rosell and M. Carrasco Kind and F. J. Castander and M. Crocce and C. E. Cunha and C. B. D’Andrea and L. N. da Costa and C. Davis and D. L. DePoy and S. Desai and J. P. Dietrich and A. Drlica-Wagner and T. F. Eifler and A. E. Evrard and E. Fernandez and B. Flaugher and P. Fosalba and E. Gaztanaga and D. W. Gerdes and T. Giannantonio and D. A. Goldstein and D. Gruen and R. A. Gruendl and G. Gutierrez and K. Honscheid and B. Jain and D. J. James and T. Jeltema and M. W. G. Johnson and M. D. Johnson and S. Kent and E. Krause and R. Kron and K. Kuehn and N. Nuropatkin and O. Lahav and M. Lima and H. Lin and M. A. G. Maia and M. March and P. Martini and R. G. McMahon and F. Menanteau and C. J. Miller and R. Miquel and J. J. Mohr and E. Neilsen and R. C. Nichol and R. L. C. Ogando and A. A. Plazas and N. Roe and A. K. Romer and A. Roodman and E. S. Rykoff and E. Sanchez and V. Scarpine and R. Schindler and M. Schubnell and I. Sevilla-Noarbe and M. Smith and R. C. Smith and F. Sobreira and E. Suchyta and M. E. C. Swanson and G. Tarle and D. Thomas and R. C. Thomas and M. A. Troxel and V. Vikram and A. R. Walker and R. H. Wechsler and J. Weller and B. Yanny and J. Zuntz},
   doi = {10.3847/2041-8213/aa8fc7},
   issn = {20418213},
   issue = {2},
   journal = {\apjl},
   month = {10},
   pages = {L17},
   publisher = {American Astronomical Society},
   title = {The Electromagnetic Counterpart of the Binary Neutron Star Merger LIGO/Virgo GW170817. II. UV, Optical, and Near-infrared Light Curves and Comparison to Kilonova Models},
   volume = {848},
   year = {2017},
}

@article{Evans_2017,
   author = {{Evans}, P.~A. and {Cenko}, S.~B. and {Kennea}, J.~A. and {Emery}, S.~W.~K. and {Kuin}, N.~P.~M. and {Korobkin}, O. and {Wollaeger}, R.~T. and {Fryer}, C.~L. and {Madsen}, K.~K. and {Harrison}, F.~A. and {Xu}, Y. and {Nakar}, E. and {Hotokezaka}, K. and {Lien}, A. and {Campana}, S. and {Oates}, S.~R. and {Troja}, E. and {Breeveld}, A.~A. and {Marshall}, F.~E. and {Barthelmy}, S.~D. and {Beardmore}, A.~P. and {Burrows}, D.~N. and {Cusumano}, G. and {D'A{\`\i}}, A. and {D'Avanzo}, P. and {D'Elia}, V. and {de Pasquale}, M. and {Even}, W.~P. and {Fontes}, C.~J. and {Forster}, K. and {Garcia}, J. and {Giommi}, P. and {Grefenstette}, B. and {Gronwall}, C. and {Hartmann}, D.~H. and {Heida}, M. and {Hungerford}, A.~L. and {Kasliwal}, M.~M. and {Krimm}, H.~A. and {Levan}, A.~J. and {Malesani}, D. and {Melandri}, A. and {Miyasaka}, H. and {Nousek}, J.~A. and {O'Brien}, P.~T. and {Osborne}, J.~P. and {Pagani}, C. and {Page}, K.~L. and {Palmer}, D.~M. and {Perri}, M. and {Pike}, S. and {Racusin}, J.~L. and {Rosswog}, S. and {Siegel}, M.~H. and {Sakamoto}, T. and {Sbarufatti}, B. and {Tagliaferri}, G. and {Tanvir}, N.~R. and {Tohuvavohu}, A.},
        title = "{Swift and NuSTAR observations of GW170817: Detection of a blue kilonova}",
      journal = {Science},
         year = 2017,
        month = dec,
       volume = {358},
       number = {6370},
        pages = {1565-1570},
          doi = {10.1126/science.aap9580},
archivePrefix = {arXiv},
       eprint = {1710.05437},
 primaryClass = {astro-ph.HE},
       adsurl = {https://ui.adsabs.harvard.edu/abs/2017Sci...358.1565E}
}

@article{Bulla_2023,
   author = {Mattia Bulla},
   doi = {10.1093/mnras/stad232},
   issn = {13652966},
   issue = {2},
   journal = {\mnras},
   month = {4},
   pages = {2558-2570},
   publisher = {Oxford University Press},
   title = {The critical role of nuclear heating rates, thermalization efficiencies, and opacities for kilonova modelling and parameter inference},
   volume = {520},
   year = {2023},
}

@article{Bulla_2019,
   author = {Mattia Bulla},
   doi = {10.1093/mnras/stz2495},
   issn = {13652966},
   issue = {4},
   journal = {\mnras},
   month = {11},
   pages = {5037-5045},
   publisher = {Oxford University Press},
   title = {PossiS: Predicting spectra, light curves, and polarization for multidimensional models of supernovae and kilonovae},
   volume = {489},
   year = {2019},
}

@article{Lupton_1999,
   title={A Modified Magnitude System that Produces Well-Behaved Magnitudes, Colors, and Errors Even for Low Signal-to-Noise Ratio Measurements},
   volume={118},
   ISSN={0004-6256},
   url={http://dx.doi.org/10.1086/301004},
   DOI={10.1086/301004},
   number={3},
   journal={\aj},
   publisher={American Astronomical Society},
   author={Lupton, Robert H. and Gunn, James E. and Szalay, Alexander S.},
   year={1999},
   month=sep, 
   pages={1406-1410} }

@article{Korobkin_2021,
   title={Axisymmetric Radiative Transfer Models of Kilonovae},
   volume={910},
   ISSN={1538-4357},
   url={http://dx.doi.org/10.3847/1538-4357/abe1b5},
   DOI={10.3847/1538-4357/abe1b5},
   number={2},
   journal={\apj},
   publisher={American Astronomical Society},
   author={Korobkin, Oleg and Wollaeger, Ryan T. and Fryer, Christopher L. and Hungerford, Aimee L. and Rosswog, Stephan and Fontes, Christopher J. and Mumpower, Matthew R. and Chase, Eve A. and Even, Wesley P. and Miller, Jonah and Misch, G. Wendell and Lippuner, Jonas},
   year={2021},
   month=apr, pages={116} }

@inproceedings{Kingma_2017,
      author       = {Diederik P. Kingma and
                  Jimmy Ba},
  editor       = {Yoshua Bengio and
                  Yann LeCun},
  title        = {Adam: {A} Method for Stochastic Optimization},
  booktitle    = {3rd International Conference on Learning Representations},
  address      = {San Diego, CA, USA},
  year         = {2015},
archivePrefix = {arXiv},
    eprint = {1412.6980},
  bibsource    = {dblp computer science bibliography, https://dblp.org},
adsurl = {https://ui.adsabs.harvard.edu/abs/2014arXiv1412.6980K}
}

@inproceedings{Gardner_2018,
  author = {Gardner, Jacob and Pleiss, Geoff and Weinberger, Kilian and Bindel, David and Wilson, Andrew G},
 booktitle = {Advances in Neural Information Processing Systems},
 editor = {S. Bengio and H. Wallach and H. Larochelle and K. Grauman and N. Cesa-Bianchi and R. Garnett},
 pages = {7587-7597},
 publisher = {Curran Associates, Inc.},
 title = {GPyTorch: Blackbox Matrix-Matrix Gaussian Process Inference with GPU Acceleration},
 url = {https://proceedings.neurips.cc/paper_files/paper/2018/file/27e8e17134dd7083b050476733207ea1-Paper.pdf},
 volume = {31},
 year = {2018}
}

@article{Andreoni_2021,
   title={Optimizing Cadences with Realistic Light-curve Filtering for Serendipitous Kilonova Discovery with Vera Rubin Observatory},
   volume={258},
   ISSN={1538-4365},
   url={http://dx.doi.org/10.3847/1538-4365/ac3bae},
   DOI={10.3847/1538-4365/ac3bae},
   number={1},
   journal={\apjs},
   publisher={American Astronomical Society},
   author={Andreoni, Igor and Coughlin, Michael W. and Almualla, Mouza and Bellm, Eric C. and Bianco, Federica B. and Bulla, Mattia and Cucchiara, Antonino and Dietrich, Tim and Goobar, Ariel and Kool, Erik C. and Li, Xiaolong and Ragosta, Fabio and Sagués-Carracedo, Ana and Singer, Leo P.},
   year={2021},
   month=dec, pages={5} }

@INPROCEEDINGS{Reitze_2019,
      author = {{Reitze}, David and {Adhikari}, Rana X. and {Ballmer}, Stefan and {Barish}, Barry and {Barsotti}, Lisa and {Billingsley}, GariLynn and {Brown}, Duncan A. and {Chen}, Yanbei and {Coyne}, Dennis and {Eisenstein}, Robert and {Evans}, Matthew and {Fritschel}, Peter and {Hall}, Evan D. and {Lazzarini}, Albert and {Lovelace}, Geoffrey and {Read}, Jocelyn and {Sathyaprakash}, B.~S. and {Shoemaker}, David and {Smith}, Joshua and {Torrie}, Calum and {Vitale}, Salvatore and {Weiss}, Rainer and {Wipf}, Christopher and {Zucker}, Michael},
        title = "{Cosmic Explorer: The U.S. Contribution to Gravitational-Wave Astronomy beyond LIGO}",
     booktitle = {Astro2020 APC White Papers},
    series = {BAAS},
         year = 2019,
       volume = {51},
        month = sep,
          eid = {35},
        pages = {35},
          doi = {10.48550/arXiv.1907.04833},
archivePrefix = {arXiv},
       eprint = {1907.04833},
 primaryClass = {astro-ph.IM},
       adsurl = {https://ui.adsabs.harvard.edu/abs/2019BAAS...51g..35R}
}

@ARTICLE{Colombo_2025,
       author = {{Colombo}, Alberto and {Salafia}, Om Sharan and {Ghirlanda}, Giancarlo and {Iacovelli}, Francesco and {Mancarella}, Michele and {Broekgaarden}, Floor S. and {Nava}, Lara and {Giacomazzo}, Bruno and {Colpi}, Monica},
        title = "{Multi-messenger observations in the Einstein Telescope era: Binary neutron star and black hole─neutron star mergers}",
      journal = {\aap},
         year = 2025,
        month = dec,
       volume = {704},
          eid = {A260},
        pages = {A260},
          doi = {10.1051/0004-6361/202554326},
archivePrefix = {arXiv},
       eprint = {2503.00116},
 primaryClass = {astro-ph.HE},
       adsurl = {https://ui.adsabs.harvard.edu/abs/2025A&A...704A.260C}
}

@article{Setzer_2023,
   title={Modelling populations of kilonovae},
   volume={520},
   ISSN={1365-2966},
   url={http://dx.doi.org/10.1093/mnras/stad257},
   DOI={10.1093/mnras/stad257},
   number={2},
   journal={\mnras},
   publisher={Oxford University Press (OUP)},
   author={Setzer, Christian N and Peiris, Hiranya V and Korobkin, Oleg and Rosswog, Stephan},
   year={2023},
   month=jan, pages={2829--2842} }

@article{Lukoiute_2022,
   title={<tt>KilonovaNet</tt>: Surrogate models of kilonova spectra with conditional variational autoencoders},
   volume={516},
   ISSN={1365-2966},
   url={http://dx.doi.org/10.1093/mnras/stac2342},
   DOI={10.1093/mnras/stac2342},
   number={1},
   journal={\mnras},
   publisher={Oxford University Press (OUP)},
   author={Lukošiute, K and Raaijmakers, G and Doctor, Z and Soares-Santos, M and Nord, B},
   year={2022},
   month=aug, pages={1137-1148} }

@article{Ristic_2022,
  title = {Interpolating detailed simulations of kilonovae: Adaptive learning and parameter inference applications},
  author = {Ristic, M. and Champion, E. and O'Shaughnessy, R. and Wollaeger, R. and Korobkin, O. and Chase, E. A. and Fryer, C. L. and Hungerford, A. L. and Fontes, C. J.},
  journal = {PhRvR},
  volume = {4},
  issue = {1},
  pages = {013046},
  numpages = {17},
  year = {2022},
  month = {Jan},
  publisher = {American Physical Society},
  doi = {10.1103/PhysRevResearch.4.013046},
  url = {https://link.aps.org/doi/10.1103/PhysRevResearch.4.013046}
}

@article{Anand_2023,
      author = {{Anand}, Shreya and {Pang}, Peter T.~H. and {Bulla}, Mattia and {Coughlin}, Michael W. and {Dietrich}, Tim and {Healy}, Brian and {Hussenot-Desenonges}, Thomas and {Jegou du Laz}, Theophile and {Kasliwal}, Mansi M. and {Kunert}, Nina and {Markin}, Ivan and {Mooley}, Kunal and {Nedora}, Vsevolod and {Neuweiler}, Anna},
        title = "{Chemical Distribution of the Dynamical Ejecta in the Neutron Star Merger GW170817}",
      journal = {arXiv e-prints},
         year = 2023,
        month = jul,
          eid = {arXiv:2307.11080},
        pages = {arXiv:2307.11080},
          doi = {10.48550/arXiv.2307.11080},
archivePrefix = {arXiv},
       eprint = {2307.11080},
 primaryClass = {astro-ph.HE},
       adsurl = {https://ui.adsabs.harvard.edu/abs/2023arXiv230711080A}
}

@article{King_2025,
      author = {{King}, Brendan L. and {De}, Soumi and {Korobkin}, Oleg and {Coughlin}, Michael W. and {Pang}, Peter T.~H. and {Strother}, Terrance T.},
        title = "{Inferring Neutron Star Merger Ejecta Morphology with Kilonovae}",
      journal = {\pasp},
         year = 2025,
        month = oct,
       volume = {137},
       number = {10},
          eid = {104507},
        pages = {104507},
          doi = {10.1088/1538-3873/ae10df},
archivePrefix = {arXiv},
       eprint = {2505.16876},
 primaryClass = {astro-ph.HE},
       adsurl = {https://ui.adsabs.harvard.edu/abs/2025PASP..137j4507K}
}

@article{Alsing_2019,
   title={Fast likelihood-free cosmology with neural density estimators and active learning},
   ISSN={1365-2966},
   url={http://dx.doi.org/10.1093/mnras/stz1960},
   DOI={10.1093/mnras/stz1960},
   journal={\mnras},
   publisher={Oxford University Press (OUP)},
   author={Alsing, Justin and Charnock, Tom and Feeney, Stephen and Wandelt, Benjamin},
   year={2019},
   volume = {488},
   pages={4440},
   month=jul }

@ARTICLE{Kasen_2013,
       author = {{Kasen}, Daniel and {Badnell}, N.~R. and {Barnes}, Jennifer},
        title = "{Opacities and Spectra of the r-process Ejecta from Neutron Star Mergers}",
      journal = {\apj},
         year = 2013,
        month = sep,
       volume = {774},
       number = {1},
          eid = {25},
        pages = {25},
          doi = {10.1088/0004-637X/774/1/25},
archivePrefix = {arXiv},
       eprint = {1303.5788},
 primaryClass = {astro-ph.HE},
       adsurl = {https://ui.adsabs.harvard.edu/abs/2013ApJ...774...25K}
}

@ARTICLE{Cranmer_2020,
       author = {{Cranmer}, Kyle and {Brehmer}, Johann and {Louppe}, Gilles},
        title = "{The frontier of simulation-based inference}",
      journal = {Proc.\ Natl.\ Acad.\ Sci.},
         year = 2020,
        month = dec,
       volume = {117},
       number = {48},
        pages = {30055-30062},
          doi = {10.1073/pnas.1912789117},
archivePrefix = {arXiv},
       eprint = {1911.01429},
 primaryClass = {stat.ML},
       adsurl = {https://ui.adsabs.harvard.edu/abs/2020PNAS..11730055C}
}

@article{Papamakarios_2021,
  author  = {George Papamakarios and Eric Nalisnick and Danilo Jimenez Rezende and Shakir Mohamed and Balaji Lakshminarayanan},
  title   = {Normalizing Flows for Probabilistic Modeling and Inference},
  journal = {J.\ Machine Learning Res.},
  year    = {2021},
  volume  = {22},
  number  = {57},
  pages   = {1--64},
  url     = {http://jmlr.org/papers/v22/19-1028.html}
}

@ARTICLE{Kobyzev_2021,
  author={Kobyzev, Ivan and Prince, Simon J.D. and Brubaker, Marcus A.},
  journal={IEEE Trans.\ Pattern Analysis Machine Intelligence}, 
  title={Normalizing Flows: An Introduction and Review of Current Methods}, 
  year={2021},
  volume={43},
  number={11},
  pages={3964-3979},
  doi={10.1109/TPAMI.2020.2992934}
  }

@article{Yang_2023, 
author = {Yang, Ling and Zhang, Zhilong and Song, Yang and Hong, Shenda and Xu, Runsheng and Zhao, Yue and Zhang, Wentao and Cui, Bin and Yang, Ming-Hsuan}, 
title = {Diffusion Models: A Comprehensive Survey of Methods and Applications}, 
year = {2023}, 
issue_date = {April 2024}, 
publisher = {Association for Computing Machinery}, 
address = {New York, NY, USA}, 
volume = {56}, 
number = {4}, 
doi = {10.1145/3626235}, 
journal = {ACM Comput.\ Surv.}, 
month = nov, 
pages = {105}, 
numpages = {39}
}

@ARTICLE{Arruda_2025,
       author = {{Arruda}, Jonas and {Bracher}, Niels and {K{\"o}the}, Ullrich and {Hasenauer}, Jan and {Radev}, Stefan T.},
        title = "{Diffusion Models in Simulation-Based Inference: A Tutorial Review}",
      journal = {arXiv e-prints},
         year = 2025,
        month = dec,
          eid = {arXiv:2512.20685},
        pages = {arXiv:2512.20685},
          doi = {10.48550/arXiv.2512.20685},
archivePrefix = {arXiv},
       eprint = {2512.20685},
 primaryClass = {stat.ML},
       adsurl = {https://ui.adsabs.harvard.edu/abs/2025arXiv251220685A}
}

@ARTICLE{Tanaka_2013,
       author = {{Tanaka}, Masaomi and {Hotokezaka}, Kenta},
        title = "{Radiative Transfer Simulations of Neutron Star Merger Ejecta}",
      journal = {\apj},
         year = 2013,
        month = oct,
       volume = {775},
       number = {2},
          eid = {113},
        pages = {113},
          doi = {10.1088/0004-637X/775/2/113},
archivePrefix = {arXiv},
       eprint = {1306.3742},
 primaryClass = {astro-ph.HE},
       adsurl = {https://ui.adsabs.harvard.edu/abs/2013ApJ...775..113T}
}

@ARTICLE{Wollaeger_2014,
       author = {{Wollaeger}, Ryan T. and {van Rossum}, Daniel R.},
        title = "{Radiation Transport for Explosive Outflows: Opacity Regrouping}",
      journal = {\apjs},
         year = 2014,
        month = oct,
       volume = {214},
       number = {2},
          eid = {28},
        pages = {28},
          doi = {10.1088/0067-0049/214/2/28},
archivePrefix = {arXiv},
       eprint = {1407.3833},
 primaryClass = {astro-ph.HE},
       adsurl = {https://ui.adsabs.harvard.edu/abs/2014ApJS..214...28W}
}

@software{sncosmo_2025,
    author = {Barbary, Kyle and
      Bailey, Stephen and
      Barentsen, Geert and
      Barclay, Tom and
      Biswas, Rahul and
      Boone, Kyle and
      Craig, Matt and
      Feindt, Ulrich and
      Friesen, Brian and
      Goldstein, Danny and
      Jha, Saurabh W. and
      Jones, David O. and
      Mondon, Florian and
      Papadogiannakis, Seméli and
      Perrefort, Daniel and
      Pierel, Justin and
      Rodney, Steve and
      Rose, Benjamin and
      Saunders, Clare and
      Sipőcz, Brigitta and
      Sofiatti, Caroline and
      Thomas, Rollin C. and
      van Santen, Jakob and
      Vincenzi, Maria and
      Wang, David and
      Wood-Vasey, Michael},
  title        = {SNCosmo},
  month        = mar,
  year         = 2025,
  publisher    = {Zenodo},
  version      = {v2.12.1},
  doi          = {10.5281/zenodo.15019859},
  url          = {https://doi.org/10.5281/zenodo.15019859},
  swhid        = {swh:1:dir:335092d3d6716aebad6d052fc6219ae600062484
                   ;origin=https://doi.org/10.5281/zenodo.592747;visit=swh:1:snp:22689df264588628a8d487675f3a1a7715c77b26;anchor=swh:1:rel:dccc5b60d0073ddc18c513bc1969c0
                   9cba697fdb;path=sncosmo-sncosmo-6c34401
                  },
}

@article{Auer_2002a,
    author = {Auer, Peter and Cesa-Bianchi, Nicolò and Fischer, Paul},
    year = {2002},
    month = {05},
    pages = {235-256},
    title = {Finite-time Analysis of the Multiarmed Bandit Problem},
    volume = {47},
    journal = {Machine Learning},
    doi = {10.1023/A:1013689704352}
}

@article{Auer_2002b,
author = {Auer, Peter},
year = {2002},
month = {01},
pages = {397-422},
title = {Using Confidence Bounds for Exploitation-Exploration Trade-offs.},
volume = {3},
journal = {J.\ Machine Learning Res.},
url = {https://www.jmlr.org/papers/v3/auer02a.html}
}

@inproceedings{Dani_2008,
    author = {Dani, Varsha and Hayes, Thomas P. and Kakade, Sham M.},
    year = {2008},
    month = {01},
    pages = {355-366},
    title = {Stochastic Linear Optimization under Bandit Feedback},
    booktitle = {21st Annual Conference on Learning Theory},
    url = {https://www.learningtheory.org/colt2008/papers/80-Dani.pdf},
    editor = {Servedio, R and Zhang, T},
    address = {Helsinki, Finland}
    }

@article{Rogers_2019,
   title={Bayesian emulator optimisation for cosmology: application to the Lyman-alpha forest},
   volume={2019(02)},
   ISSN={1475-7516},
   url={http://dx.doi.org/10.1088/1475-7516/2019/02/031},
   DOI={10.1088/1475-7516/2019/02/031},
   number={02},
   journal={\jcap},
   publisher={IOP Publishing},
   author={Rogers, Keir K. and Peiris, Hiranya V. and Pontzen, Andrew and Bird, Simeon and Verde, Licia and Font-Ribera, Andreu},
   year={2019},
   month=feb, pages={031} }

@article{Mooley_2022,
   title={Optical superluminal motion measurement in the neutron-star merger GW170817},
   volume={610},
   ISSN={1476-4687},
   url={http://dx.doi.org/10.1038/s41586-022-05145-7},
   DOI={10.1038/s41586-022-05145-7},
   number={7931},
   journal={Nature},
   publisher={Springer Science and Business Media LLC},
   author={Mooley, Kunal P. and Anderson, Jay and Lu, Wenbin},
   year={2022},
   month=oct, pages={273-276} }

@inproceedings{Lipman_2023,
    author = {{Lipman}, Yaron and {Chen}, Ricky T.~Q. and {Ben-Hamu}, Heli and {Nickel}, Maximilian and {Le}, Matt},
    title = "{Flow Matching for Generative Modeling}",
    editor = {Liu, Y},
    address = {Kigali, Rwanda},
    booktitle = {11th International Conference on Learning Representations},
     year = {2023},
      eid = {arXiv:2210.02747},
    archivePrefix = {arXiv},
    eprint = {2210.02747},
    primaryClass = {cs.LG},
      url = {https://openreview.net/forum?id=PqvMRDCJT9t},
    adsurl = {https://ui.adsabs.harvard.edu/abs/2022arXiv221002747L}
}

@inproceedings{Loshchilov_2019,
    author     = {Ilya Loshchilov and Frank Hutter},
    editor = {Tara Sainath},
    title        = {Decoupled Weight Decay Regularization},
    booktitle    = {7th International Conference on Learning Representations},
    address = {New Orleans, LA, USA},
    year         = {2019},
    url          = {https://openreview.net/forum?id=Bkg6RiCqY7},
    eprint = {1711.05101},
    archivePrefix={arXiv},
    primaryClass={cs.LG},
}

@article{Alsing_2024,
   title={pop-cosmos: A Comprehensive Picture of the Galaxy Population from COSMOS Data},
   volume={274},
   ISSN={1538-4365},
   url={http://dx.doi.org/10.3847/1538-4365/ad5c69},
   DOI={10.3847/1538-4365/ad5c69},
   number={1},
   journal={\apjs},
   publisher={American Astronomical Society},
   author={Alsing, Justin and Thorp, Stephen and Deger, Sinan and Peiris, Hiranya V. and Leistedt, Boris and Mortlock, Daniel and Leja, Joel},
   year={2024},
   month=sep, pages={12} }

@article{Thorp_2025,
   title={<tt>pop-cosmos</tt>: Insights from Generative Modeling of a Deep, Infrared-selected Galaxy Population},
   volume={993},
   ISSN={1538-4357},
   url={http://dx.doi.org/10.3847/1538-4357/ae0936},
   DOI={10.3847/1538-4357/ae0936},
   number={2},
   journal={\apj},
   publisher={American Astronomical Society},
   author={Thorp, Stephen and Peiris, Hiranya V. and Jagwani, Gurjeet and Deger, Sinan and Alsing, Justin and Leistedt, Boris and Mortlock, Daniel J. and Halder, Anik and Leja, Joel},
   year={2025},
   month=nov, pages={240} }

@article{Thorp_2024,
   title={pop-cosmos: Scaleable Inference of Galaxy Properties and Redshifts with a Data-driven Population Model},
   volume={975},
   ISSN={1538-4357},
   url={http://dx.doi.org/10.3847/1538-4357/ad7736},
   DOI={10.3847/1538-4357/ad7736},
   number={1},
   journal={\apj},
   publisher={American Astronomical Society},
   author={Thorp, Stephen and Alsing, Justin and Peiris, Hiranya V. and Deger, Sinan and Mortlock, Daniel J. and Leistedt, Boris and Leja, Joel and Loureiro, Arthur},
   year={2024},
   month=oct, pages={145} }

@article{Jhawar_2025,
   title={Data-driven approach for modeling the temporal and spectral evolution of kilonova systematic uncertainties},
   volume={111},
   pages={043046},
   ISSN={2470-0029},
   url={http://dx.doi.org/10.1103/PhysRevD.111.043046},
   DOI={10.1103/physrevd.111.043046},
   number={4},
   journal={\prd},
   publisher={American Physical Society (APS)},
   author={Jhawar, Sahil and Wouters, Thibeau and Pang, Peter T. H. and Bulla, Mattia and Coughlin, Michael W. and Dietrich, Tim},
   year={2025},
   month=feb }

@article{Desai_2025,
    author = {Desai, M M and Chatterjee, D and Jhawar, S and Harris, P and Katsavounidis, E and Coughlin, M W},
    title = {Rapid parameter estimation for kilonovae using likelihood-free inference},
    journal = {\mnras},
    volume = {541},
    number = {3},
    pages = {2619-2630},
    year = {2025},
    month = {06},
    issn = {0035-8711},
    doi = {10.1093/mnras/staf1045},
    url = {https://doi.org/10.1093/mnras/staf1045},
    eprint = {https://academic.oup.com/mnras/article-pdf/541/3/2619/63712510/staf1045.pdf},
}

@inproceedings{Lueckmann_2017,
  author       = {Jan{-}Matthis Lueckmann and
                  Pedro J. Gon{\c{c}}alves and
                  Giacomo Bassetto and
                  Kaan {\"{O}}cal and
                  Marcel Nonnenmacher and
                  Jakob H. Macke},
  editor       = {Isabelle Guyon and
                  Ulrike von Luxburg and
                  Samy Bengio and
                  Hanna M. Wallach and
                  Rob Fergus and
                  S. V. N. Vishwanathan and
                  Roman Garnett},
  title        = {Flexible statistical inference for mechanistic models of neural dynamics},
  booktitle    = {Advances in Neural Information Processing Systems},
  publisher    = {Curran Associates, Inc.},
  volume       = {30},
  pages        = {1289--1299},
  year         = {2017},
  url          = {https://proceedings.neurips.cc/paper/2017/hash/addfa9b7e234254d26e9c7f2af1005cb-Abstract.html},
  bibsource    = {dblp computer science bibliography, https://dblp.org},
}

@phdthesis{Papamakarios_2019a,
      title={Neural Density Estimation and Likelihood-free Inference}, 
      school = {University of Edinburgh},
      author={George Papamakarios},
      year={2019},
      eprint={1910.13233},
      archivePrefix={arXiv},
      primaryClass={stat.ML},
      url={https://arxiv.org/abs/1910.13233}, 
}

@inproceedings{Papamakarios_2016,
      author = {Papamakarios, George and Murray, Iain},
 booktitle = {Advances in Neural Information Processing Systems},
 editor = {D. Lee and M. Sugiyama and U. Luxburg and I. Guyon and R. Garnett},
 pages = {1036-1044},
 publisher = {Curran Associates, Inc.},
 title = {Fast \epsilon -free Inference of Simulation Models with Bayesian Conditional Density Estimation},
 url = {https://proceedings.neurips.cc/paper_files/paper/2016/file/6aca97005c68f1206823815f66102863-Paper.pdf},
 volume = {29},
 year = {2016}
}

@article{Cantiello_2018,
   title={A Precise Distance to the Host Galaxy of the Binary Neutron Star Merger GW170817 Using Surface Brightness Fluctuations},
   volume={854},
   ISSN={2041-8213},
   url={http://dx.doi.org/10.3847/2041-8213/aaad64},
   DOI={10.3847/2041-8213/aaad64},
   number={2},
   journal={\apjl},
   publisher={American Astronomical Society},
   author={Cantiello, Michele and Jensen, J. B. and Blakeslee, J. P. and Berger, E. and Levan, A. J. and Tanvir, N. R. and Raimondo, G. and Brocato, E. and Alexander, K. D. and Blanchard, P. K. and Branchesi, M. and Cano, Z. and Chornock, R. and Covino, S. and Cowperthwaite, P. S. and D’Avanzo, P. and Eftekhari, T. and Fong, W. and Fruchter, A. S. and Grado, A. and Hjorth, J. and Holz, D. E. and Lyman, J. D. and Mandel, I. and Margutti, R. and Nicholl, M. and Villar, V. A. and Williams, P. K. G.},
   year={2018},
   month=feb, pages={L31} }

@article{Gutierrez_2025,
  title = {Cocoon shock breakout emission from binary neutron star mergers},
  author = {Guti\'errez, Eduardo M. and Bhattacharya, Mukul and Radice, David and Murase, Kohta and Bernuzzi, Sebastiano},
  journal = {\prd},
  volume = {111},
  issue = {6},
  pages = {063031},
  numpages = {26},
  year = {2025},
  month = {Mar},
  publisher = {American Physical Society},
  doi = {10.1103/PhysRevD.111.063031},
  url = {https://link.aps.org/doi/10.1103/PhysRevD.111.063031}
}

@article{Tanvir_2017,
   title={The Emergence of a Lanthanide-rich Kilonova Following the Merger of Two Neutron Stars},
   volume={848},
   ISSN={2041-8213},
   url={http://dx.doi.org/10.3847/2041-8213/aa90b6},
   DOI={10.3847/2041-8213/aa90b6},
   number={2},
   journal={\apjl},
   publisher={American Astronomical Society},
   author={Tanvir, N. R. and Levan, A. J. and González-Fernández, C. and Korobkin, O. and Mandel, I. and Rosswog, S. and Hjorth, J. and D’Avanzo, P. and Fruchter, A. S. and Fryer, C. L. and Kangas, T. and Milvang-Jensen, B. and Rosetti, S. and Steeghs, D. and Wollaeger, R. T. and Cano, Z. and Copperwheat, C. M. and Covino, S. and D’Elia, V. and de Ugarte Postigo, A. and Evans, P. A. and Even, W. P. and Fairhurst, S. and Jaimes, R. Figuera and Fontes, C. J. and Fujii, Y. I. and Fynbo, J. P. U. and Gompertz, B. P. and Greiner, J. and Hodosan, G. and Irwin, M. J. and Jakobsson, P. and Jørgensen, U. G. and Kann, D. A. and Lyman, J. D. and Malesani, D. and McMahon, R. G. and Melandri, A. and O’Brien, P. T. and Osborne, J. P. and Palazzi, E. and Perley, D. A. and Pian, E. and Piranomonte, S. and Rabus, M. and Rol, E. and Rowlinson, A. and Schulze, S. and Sutton, P. and Thöne, C. C. and Ulaczyk, K. and Watson, D. and Wiersema, K. and Wijers, R. A. M. J.},
   year={2017},
   month=oct, pages={L27} }

@article{Troja_2019,
    author = {{Troja}, E. and {Castro-Tirado}, A.~J. and {Becerra Gonz{\'a}lez}, J. and {Hu}, Y. and {Ryan}, G.~S. and {Cenko}, S.~B. and {Ricci}, R. and {Novara}, G. and {S{\'a}nchez-R{\'a}mirez}, R. and {Acosta-Pulido}, J.~A. and {Ackley}, K.~D. and {Caballero Garc{\'\i}a}, M.~D. and {Eikenberry}, S.~S. and {Guziy}, S. and {Jeong}, S. and {Lien}, A.~Y. and {M{\'a}rquez}, I. and {Pandey}, S.~B. and {Park}, I.~H. and {Sakamoto}, T. and {Tello}, J.~C. and {Sokolov}, I.~V. and {Sokolov}, V.~V. and {Tiengo}, A. and {Valeev}, A.~F. and {Zhang}, B.~B. and {Veilleux}, S.},
        title = "{The afterglow and kilonova of the short GRB 160821B}",
      journal = {\mnras},
         year = 2019,
        month = oct,
       volume = {489},
       number = {2},
        pages = {2104-2116},
          doi = {10.1093/mnras/stz2255},
archivePrefix = {arXiv},
       eprint = {1905.01290},
 primaryClass = {astro-ph.HE},
       adsurl = {https://ui.adsabs.harvard.edu/abs/2019MNRAS.489.2104T}
}

@article{Tanvir_2013,
   title={A ‘kilonova’ associated with the short-duration γ-ray burst GRB 130603B},
   volume={500},
   ISSN={1476-4687},
   url={http://dx.doi.org/10.1038/nature12505},
   DOI={10.1038/nature12505},
   number={7464},
   journal={Nature},
   publisher={Springer Science and Business Media LLC},
   author={Tanvir, N. R. and Levan, A. J. and Fruchter, A. S. and Hjorth, J. and Hounsell, R. A. and Wiersema, K. and Tunnicliffe, R. L.},
   year={2013},
   month=aug, pages={547-549} }

@article{Rastinejad_2022,
   title={A kilonova following a long-duration gamma-ray burst at 350 Mpc},
   volume={612},
   ISSN={1476-4687},
   url={http://dx.doi.org/10.1038/s41586-022-05390-w},
   DOI={10.1038/s41586-022-05390-w},
   number={7939},
   journal={Nature},
   publisher={Springer Science and Business Media LLC},
   author={Rastinejad, Jillian C. and Gompertz, Benjamin P. and Levan, Andrew J. and Fong, Wen-fai and Nicholl, Matt and Lamb, Gavin P. and Malesani, Daniele B. and Nugent, Anya E. and Oates, Samantha R. and Tanvir, Nial R. and de Ugarte Postigo, Antonio and Kilpatrick, Charles D. and Moore, Christopher J. and Metzger, Brian D. and Ravasio, Maria Edvige and Rossi, Andrea and Schroeder, Genevieve and Jencson, Jacob and Sand, David J. and Smith, Nathan and Fernández, José Feliciano Agüí and Berger, Edo and Blanchard, Peter K. and Chornock, Ryan and Cobb, Bethany E. and De Pasquale, Massimiliano and Fynbo, Johan P. U. and Izzo, Luca and Kann, D. Alexander and Laskar, Tanmoy and Marini, Ester and Paterson, Kerry and Escorial, Alicia Rouco and Sears, Huei M. and Thöne, Christina C.},
   year={2022},
   month=dec, pages={223-227} }

@article{Troja_2018b,
   author = {{Troja}, E. and {Ryan}, G. and {Piro}, L. and {van Eerten}, H. and {Cenko}, S.~B. and {Yoon}, Y. and {Lee}, S.-K. and {Im}, M. and {Sakamoto}, T. and {Gatkine}, P. and {Kutyrev}, A. and {Veilleux}, S.},
        title = "{A luminous blue kilonova and an off-axis jet from a compact binary merger at z = 0.1341}",
      journal = {NatCo},
         year = 2018,
        month = oct,
       volume = {9},
          eid = {4089},
        pages = {4089},
          doi = {10.1038/s41467-018-06558-7},
archivePrefix = {arXiv},
       eprint = {1806.10624},
 primaryClass = {astro-ph.HE},
       adsurl = {https://ui.adsabs.harvard.edu/abs/2018NatCo...9.4089T}
}

@ARTICLE{Sarin2024,
       author = {{Sarin}, Nikhil and {H{\"u}bner}, Moritz and {Omand}, Conor M.~B. and {Setzer}, Christian N. and et al.},
        title = "{REDBACK: a Bayesian inference software package for electromagnetic transients}",
      journal = {\mnras},
         year = 2024,
        month = jun,
       volume = {531},
       number = {1},
        pages = {1203-1227},
          doi = {10.1093/mnras/stae1238},
archivePrefix = {arXiv},
       eprint = {2308.12806},
 primaryClass = {astro-ph.HE},
       adsurl = {https://ui.adsabs.harvard.edu/abs/2024MNRAS.531.1203S}
}

@ARTICLE{Ivezic_2019,
       author = {{Ivezi{\'c}}, {\v{Z}}eljko and {Kahn}, Steven M. and {Tyson}, J. Anthony and {Abel}, Bob and {Acosta}, Emily and {Allsman}, Robyn and {Alonso}, David and {AlSayyad}, Yusra and {Anderson}, Scott F. and {Andrew}, John and {Angel}, James Roger P. and {Angeli}, George Z. and {Ansari}, Reza and {Antilogus}, Pierre and {Araujo}, Constanza and {Armstrong}, Robert and {Arndt}, Kirk T. and {Astier}, Pierre and {Aubourg}, {\'E}ric and {Auza}, Nicole and {Axelrod}, Tim S. and {Bard}, Deborah J. and {Barr}, Jeff D. and {Barrau}, Aurelian and {Bartlett}, James G. and {Bauer}, Amanda E. and {Bauman}, Brian J. and {Baumont}, Sylvain and {Bechtol}, Ellen and {Bechtol}, Keith and {Becker}, Andrew C. and {Becla}, Jacek and {Beldica}, Cristina and {Bellavia}, Steve and {Bianco}, Federica B. and {Biswas}, Rahul and {Blanc}, Guillaume and {Blazek}, Jonathan and {Blandford}, Roger D. and {Bloom}, Josh S. and {Bogart}, Joanne and {Bond}, Tim W. and {Booth}, Michael T. and {Borgland}, Anders W. and {Borne}, Kirk and {Bosch}, James F. and {Boutigny}, Dominique and {Brackett}, Craig A. and {Bradshaw}, Andrew and {Brandt}, William Nielsen and {Brown}, Michael E. and {Bullock}, James S. and {Burchat}, Patricia and {Burke}, David L. and {Cagnoli}, Gianpietro and {Calabrese}, Daniel and {Callahan}, Shawn and {Callen}, Alice L. and {Carlin}, Jeffrey L. and {Carlson}, Erin L. and {Chandrasekharan}, Srinivasan and {Charles-Emerson}, Glenaver and {Chesley}, Steve and {Cheu}, Elliott C. and {Chiang}, Hsin-Fang and {Chiang}, James and {Chirino}, Carol and {Chow}, Derek and {Ciardi}, David R. and {Claver}, Charles F. and {Cohen-Tanugi}, Johann and {Cockrum}, Joseph J. and {Coles}, Rebecca and {Connolly}, Andrew J. and {Cook}, Kem H. and {Cooray}, Asantha and {Covey}, Kevin R. and {Cribbs}, Chris and {Cui}, Wei and {Cutri}, Roc and {Daly}, Philip N. and {Daniel}, Scott F. and {Daruich}, Felipe and {Daubard}, Guillaume and {Daues}, Greg and {Dawson}, William and {Delgado}, Francisco and {Dellapenna}, Alfred and {de Peyster}, Robert and {de Val-Borro}, Miguel and {Digel}, Seth W. and {Doherty}, Peter and {Dubois}, Richard and {Dubois-Felsmann}, Gregory P. and {Durech}, Josef and {Economou}, Frossie and {Eifler}, Tim and {Eracleous}, Michael and {Emmons}, Benjamin L. and {Fausti Neto}, Angelo and {Ferguson}, Henry and {Figueroa}, Enrique and {Fisher-Levine}, Merlin and {Focke}, Warren and {Foss}, Michael D. and {Frank}, James and {Freemon}, Michael D. and {Gangler}, Emmanuel and {Gawiser}, Eric and {Geary}, John C. and {Gee}, Perry and {Geha}, Marla and {Gessner}, Charles J.~B. and {Gibson}, Robert R. and {Gilmore}, D. Kirk and {Glanzman}, Thomas and {Glick}, William and {Goldina}, Tatiana and {Goldstein}, Daniel A. and {Goodenow}, Iain and {Graham}, Melissa L. and {Gressler}, William J. and {Gris}, Philippe and {Guy}, Leanne P. and {Guyonnet}, Augustin and {Haller}, Gunther and {Harris}, Ron and {Hascall}, Patrick A. and {Haupt}, Justine and {Hernandez}, Fabio and {Herrmann}, Sven and {Hileman}, Edward and {Hoblitt}, Joshua and {Hodgson}, John A. and {Hogan}, Craig and {Howard}, James D. and {Huang}, Dajun and {Huffer}, Michael E. and {Ingraham}, Patrick and {Innes}, Walter R. and {Jacoby}, Suzanne H. and {Jain}, Bhuvnesh and {Jammes}, Fabrice and {Jee}, M. James and {Jenness}, Tim and {Jernigan}, Garrett and {Jevremovi{\'c}}, Darko and {Johns}, Kenneth and {Johnson}, Anthony S. and {Johnson}, Margaret W.~G. and {Jones}, R. Lynne and {Juramy-Gilles}, Claire and {Juri{\'c}}, Mario and {Kalirai}, Jason S. and {Kallivayalil}, Nitya J. and {Kalmbach}, Bryce and {Kantor}, Jeffrey P. and {Karst}, Pierre and {Kasliwal}, Mansi M. and {Kelly}, Heather and {Kessler}, Richard and {Kinnison}, Veronica and {Kirkby}, David and {Knox}, Lloyd and {Kotov}, Ivan V. and {Krabbendam}, Victor L. and {Krughoff}, K. Simon and {Kub{\'a}nek}, Petr and {Kuczewski}, John and {Kulkarni}, Shri and {Ku}, John and {Kurita}, Nadine R. and {Lage}, Craig S. and {Lambert}, Ron and {Lange}, Travis and {Langton}, J. Brian and {Le Guillou}, Laurent and {Levine}, Deborah and {Liang}, Ming and {Lim}, Kian-Tat and {Lintott}, Chris J. and {Long}, Kevin E. and {Lopez}, Margaux and {Lotz}, Paul J. and {Lupton}, Robert H. and {Lust}, Nate B. and {MacArthur}, Lauren A. and {Mahabal}, Ashish and {Mandelbaum}, Rachel and {Markiewicz}, Thomas W. and {Marsh}, Darren S. and {Marshall}, Philip J. and {Marshall}, Stuart and {May}, Morgan and {McKercher}, Robert and {McQueen}, Michelle and {Meyers}, Joshua and {Migliore}, Myriam and {Miller}, Michelle and {Mills}, David J.},
        title = "{LSST: From Science Drivers to Reference Design and Anticipated Data Products}",
      journal = {\apj},
         year = 2019,
        month = mar,
       volume = {873},
       number = {2},
          eid = {111},
        pages = {111},
          doi = {10.3847/1538-4357/ab042c},
archivePrefix = {arXiv},
       eprint = {0805.2366},
 primaryClass = {astro-ph},
       adsurl = {https://ui.adsabs.harvard.edu/abs/2019ApJ...873..111I}
}

@article{Bellm_2018,
   title={The Zwicky Transient Facility: System Overview, Performance, and First Results},
   volume={131},
   ISSN={1538-3873},
   url={http://dx.doi.org/10.1088/1538-3873/aaecbe},
   DOI={10.1088/1538-3873/aaecbe},
   number={995},
   journal={\pasp},
   publisher={IOP Publishing},
   author={Bellm, Eric C. and Kulkarni, Shrinivas R. and Graham, Matthew J. and Dekany, Richard and Smith, Roger M. and Riddle, Reed and Masci, Frank J. and Helou, George and Prince, Thomas A. and Adams, Scott M. and Barbarino, C. and Barlow, Tom and Bauer, James and Beck, Ron and Belicki, Justin and Biswas, Rahul and Blagorodnova, Nadejda and Bodewits, Dennis and Bolin, Bryce and Brinnel, Valery and Brooke, Tim and Bue, Brian and Bulla, Mattia and Burruss, Rick and Cenko, S. Bradley and Chang, Chan-Kao and Connolly, Andrew and Coughlin, Michael and Cromer, John and Cunningham, Virginia and De, Kishalay and Delacroix, Alex and Desai, Vandana and Duev, Dmitry A. and Eadie, Gwendolyn and Farnham, Tony L. and Feeney, Michael and Feindt, Ulrich and Flynn, David and Franckowiak, Anna and Frederick, S. and Fremling, C. and Gal-Yam, Avishay and Gezari, Suvi and Giomi, Matteo and Goldstein, Daniel A. and Golkhou, V. Zach and Goobar, Ariel and Groom, Steven and Hacopians, Eugean and Hale, David and Henning, John and Ho, Anna Y. Q. and Hover, David and Howell, Justin and Hung, Tiara and Huppenkothen, Daniela and Imel, David and Ip, Wing-Huen and Ivezić, Željko and Jackson, Edward and Jones, Lynne and Juric, Mario and Kasliwal, Mansi M. and Kaspi, S. and Kaye, Stephen and Kelley, Michael S. P. and Kowalski, Marek and Kramer, Emily and Kupfer, Thomas and Landry, Walter and Laher, Russ R. and Lee, Chien-De and Lin, Hsing Wen and Lin, Zhong-Yi and Lunnan, Ragnhild and Giomi, Matteo and Mahabal, Ashish and Mao, Peter and Miller, Adam A. and Monkewitz, Serge and Murphy, Patrick and Ngeow, Chow-Choong and Nordin, Jakob and Nugent, Peter and Ofek, Eran and Patterson, Maria T. and Penprase, Bryan and Porter, Michael and Rauch, Ludwig and Rebbapragada, Umaa and Reiley, Dan and Rigault, Mickael and Rodriguez, Hector and Roestel, Jan van and Rusholme, Ben and Santen, Jakob van and Schulze, S. and Shupe, David L. and Singer, Leo P. and Soumagnac, Maayane T. and Stein, Robert and Surace, Jason and Sollerman, Jesper and Szkody, Paula and Taddia, F. and Terek, Scott and Van Sistine, Angela and van Velzen, Sjoert and Vestrand, W. Thomas and Walters, Richard and Ward, Charlotte and Ye, Quan-Zhi and Yu, Po-Chieh and Yan, Lin and Zolkower, Jeffry},
   year={2018},
   month=dec, pages={018002} }

@article{Stratta_2025,
   title={The Puzzling Long GRB 191019A: Evidence for Kilonova Light},
   volume={979},
   ISSN={1538-4357},
   url={http://dx.doi.org/10.3847/1538-4357/ad9b7b},
   DOI={10.3847/1538-4357/ad9b7b},
   number={2},
   journal={\apj},
   publisher={American Astronomical Society},
   author={Stratta, G. and Nicuesa Guelbenzu, A. M. and Klose, S. and Rossi, A. and Singh, P. and Palazzi, E. and Guidorzi, C. and Camisasca, A. and Bernuzzi, S. and Rau, A. and Bulla, M. and Ragosta, F. and Maiorano, E. and Paris, D.},
   year={2025},
   month=jan, pages={159} }

@article{Levan_2023,
   title={Heavy-element production in a compact object merger observed by JWST},
   volume={626},
   ISSN={1476-4687},
   url={http://dx.doi.org/10.1038/s41586-023-06759-1},
   DOI={10.1038/s41586-023-06759-1},
   number={8000},
   journal={Nature},
   publisher={Springer Science and Business Media LLC},
   author={Levan, Andrew J. and Gompertz, Benjamin P. and Salafia, Om Sharan and Bulla, Mattia and Burns, Eric and Hotokezaka, Kenta and Izzo, Luca and Lamb, Gavin P. and Malesani, Daniele B. and Oates, Samantha R. and Ravasio, Maria Edvige and Rouco Escorial, Alicia and Schneider, Benjamin and Sarin, Nikhil and Schulze, Steve and Tanvir, Nial R. and Ackley, Kendall and Anderson, Gemma and Brammer, Gabriel B. and Christensen, Lise and Dhillon, Vikram S. and Evans, Phil A. and Fausnaugh, Michael and Fong, Wen-fai and Fruchter, Andrew S. and Fryer, Chris and Fynbo, Johan P. U. and Gaspari, Nicola and Heintz, Kasper E. and Hjorth, Jens and Kennea, Jamie A. and Kennedy, Mark R. and Laskar, Tanmoy and Leloudas, Giorgos and Mandel, Ilya and Martin-Carrillo, Antonio and Metzger, Brian D. and Nicholl, Matt and Nugent, Anya and Palmerio, Jesse T. and Pugliese, Giovanna and Rastinejad, Jillian and Rhodes, Lauren and Rossi, Andrea and Saccardi, Andrea and Smartt, Stephen J. and Stevance, Heloise F. and Tohuvavohu, Aaron and van der Horst, Alexander and Vergani, Susanna D. and Watson, Darach and Barclay, Thomas and Bhirombhakdi, Kornpob and Breedt, Elmé and Breeveld, Alice A. and Brown, Alexander J. and Campana, Sergio and Chrimes, Ashley A. and D’Avanzo, Paolo and D’Elia, Valerio and De Pasquale, Massimiliano and Dyer, Martin J. and Galloway, Duncan K. and Garbutt, James A. and Green, Matthew J. and Hartmann, Dieter H. and Jakobsson, Páll and Kerry, Paul and Kouveliotou, Chryssa and Langeroodi, Danial and Le Floc’h, Emeric and Leung, James K. and Littlefair, Stuart P. and Munday, James and O’Brien, Paul and Parsons, Steven G. and Pelisoli, Ingrid and Sahman, David I. and Salvaterra, Ruben and Sbarufatti, Boris and Steeghs, Danny and Tagliaferri, Gianpiero and Thöne, Christina C. and de Ugarte Postigo, Antonio and Kann, David Alexander},
   year={2023},
   month=oct, pages={737-741} }

@article{Ristic_2023,
   author = {{Risti{\'c}}, Marko and {O'Shaughnessy}, Richard and {Villar}, V. Ashley and {Wollaeger}, Ryan T. and {Korobkin}, Oleg and {Fryer}, Chris L. and {Fontes}, Christopher J. and {Kedia}, Atul},
        title = "{Interpolated kilonova spectra models: Examining the effects of a phenomenological, blue component in the fitting of AT2017gfo spectra}",
      journal = {PhRvR},
         year = 2023,
        month = nov,
       volume = {5},
       number = {4},
          eid = {043106},
        pages = {043106},
          doi = {10.1103/PhysRevResearch.5.043106},
archivePrefix = {arXiv},
       eprint = {2304.06699},
 primaryClass = {astro-ph.HE},
       adsurl = {https://ui.adsabs.harvard.edu/abs/2023PhRvR...5d3106R}
}

@article{Tanaka_2020,
   title={Systematic opacity calculations for kilonovae},
   volume={496},
   ISSN={1365-2966},
   url={http://dx.doi.org/10.1093/mnras/staa1576},
   DOI={10.1093/mnras/staa1576},
   number={2},
   journal={\mnras},
   publisher={Oxford University Press (OUP)},
   author={Tanaka, Masaomi and Kato, Daiji and Gaigalas, Gediminas and Kawaguchi, Kyohei},
   year={2020},
   month=jun, pages={1369-1392} }

@article{Banerjee_2022,
   title={Opacity of the Highly Ionized Lanthanides and the Effect on the Early Kilonova},
   volume={934},
   ISSN={1538-4357},
   url={http://dx.doi.org/10.3847/1538-4357/ac7565},
   DOI={10.3847/1538-4357/ac7565},
   number={2},
   journal={\apj},
   publisher={American Astronomical Society},
   author={Banerjee, Smaranika and Tanaka, Masaomi and Kato, Daiji and Gaigalas, Gediminas and Kawaguchi, Kyohei and Domoto, Nanae},
   year={2022},
   month=jul, pages={117} }

@ARTICLE{Bulla_2015,
       author = {{Bulla}, M. and {Sim}, S.~A. and {Kromer}, M.},
        title = "{Polarization spectral synthesis for Type Ia supernova explosion models}",
      journal = {\mnras},
         year = 2015,
        month = jun,
       volume = {450},
       number = {1},
        pages = {967-981},
          doi = {10.1093/mnras/stv657},
archivePrefix = {arXiv},
       eprint = {1503.07002},
 primaryClass = {astro-ph.HE},
       adsurl = {https://ui.adsabs.harvard.edu/abs/2015MNRAS.450..967B}
}

@ARTICLE{Rosswog_2024,
       author = {{Rosswog}, Stephan and {Korobkin}, Oleg},
        title = "{Heavy Elements and Electromagnetic Transients from Neutron Star Mergers}",
      journal = {AnP},
         year = 2024,
        month = feb,
       volume = {536},
       number = {2},
          eid = {2200306},
        pages = {2200306},
          doi = {10.1002/andp.202200306},
archivePrefix = {arXiv},
       eprint = {2208.14026},
 primaryClass = {astro-ph.HE},
       adsurl = {https://ui.adsabs.harvard.edu/abs/2024AnP...53600306R}
}

@ARTICLE{Wollaeger_2018,
       author = {{Wollaeger}, Ryan T. and {Korobkin}, Oleg and {Fontes}, Christopher J. and {Rosswog}, Stephan K. and {Even}, Wesley P. and {Fryer}, Christopher L. and {Sollerman}, Jesper and {Hungerford}, Aimee L. and {van Rossum}, Daniel R. and {Wollaber}, Allan B.},
        title = "{Impact of ejecta morphology and composition on the electromagnetic signatures of neutron star mergers}",
      journal = {\mnras},
         year = 2018,
        month = aug,
       volume = {478},
       number = {3},
        pages = {3298-3334},
          doi = {10.1093/mnras/sty1018},
archivePrefix = {arXiv},
       eprint = {1705.07084},
 primaryClass = {astro-ph.HE},
       adsurl = {https://ui.adsabs.harvard.edu/abs/2018MNRAS.478.3298W}
}

@ARTICLE{Barnes_2016,
       author = {{Barnes}, Jennifer and {Kasen}, Daniel and {Wu}, Meng-Ru and {Mart{\'\i}nez-Pinedo}, Gabriel},
        title = "{Radioactivity and Thermalization in the Ejecta of Compact Object Mergers and Their Impact on Kilonova Light Curves}",
      journal = {\apj},
         year = 2016,
        month = oct,
       volume = {829},
       number = {2},
          eid = {110},
        pages = {110},
          doi = {10.3847/0004-637X/829/2/110},
archivePrefix = {arXiv},
       eprint = {1605.07218},
 primaryClass = {astro-ph.HE},
       adsurl = {https://ui.adsabs.harvard.edu/abs/2016ApJ...829..110B}
}

@ARTICLE{Sarin_2024,
       author = {{Sarin}, Nikhil and {Rosswog}, Stephan},
        title = "{Cautionary Tales on Heating-rate Prescriptions in Kilonovae}",
      journal = {\apjl},
         year = 2024,
        month = sep,
       volume = {973},
       number = {1},
          eid = {L24},
        pages = {L24},
          doi = {10.3847/2041-8213/ad739d},
archivePrefix = {arXiv},
       eprint = {2404.07271},
 primaryClass = {astro-ph.HE},
       adsurl = {https://ui.adsabs.harvard.edu/abs/2024ApJ...973L..24S}
}

@ARTICLE{Brethauer_2024,
       author = {{Brethauer}, D. and {Kasen}, D. and {Margutti}, R. and {Chornock}, R.},
        title = "{Impact of Systematic Modeling Uncertainties on Kilonova Property Estimation}",
      journal = {\apj},
         year = 2024,
        month = nov,
       volume = {975},
       number = {2},
          eid = {213},
        pages = {213},
          doi = {10.3847/1538-4357/ad7d83},
archivePrefix = {arXiv},
       eprint = {2408.02731},
 primaryClass = {astro-ph.HE},
       adsurl = {https://ui.adsabs.harvard.edu/abs/2024ApJ...975..213B}
}

@ARTICLE{Banerjee_2020,
       author = {{Banerjee}, Smaranika and {Tanaka}, Masaomi and {Kawaguchi}, Kyohei and {Kato}, Daiji and {Gaigalas}, Gediminas},
        title = "{Simulations of Early Kilonova Emission from Neutron Star Mergers}",
      journal = {\apj},
         year = 2020,
        month = sep,
       volume = {901},
       number = {1},
          eid = {29},
        pages = {29},
          doi = {10.3847/1538-4357/abae61},
archivePrefix = {arXiv},
       eprint = {2008.05495},
 primaryClass = {astro-ph.HE},
       adsurl = {https://ui.adsabs.harvard.edu/abs/2020ApJ...901...29B}
}

@ARTICLE{Pognan_2022,
       author = {{Pognan}, Quentin and {Jerkstrand}, Anders and {Grumer}, Jon},
        title = "{NLTE effects on kilonova expansion opacities}",
      journal = {\mnras},
         year = 2022,
        month = jul,
       volume = {513},
       number = {4},
        pages = {5174-5197},
          doi = {10.1093/mnras/stac1253},
archivePrefix = {arXiv},
       eprint = {2202.09245},
 primaryClass = {astro-ph.HE},
       adsurl = {https://ui.adsabs.harvard.edu/abs/2022MNRAS.513.5174P}
}

@ARTICLE{Kawaguchi_2018,
       author = {{Kawaguchi}, Kyohei and {Shibata}, Masaru and {Tanaka}, Masaomi},
        title = "{Radiative Transfer Simulation for the Optical and Near-infrared Electromagnetic Counterparts to GW170817}",
      journal = {\apjl},
         year = 2018,
        month = oct,
       volume = {865},
       number = {2},
          eid = {L21},
        pages = {L21},
          doi = {10.3847/2041-8213/aade02},
archivePrefix = {arXiv},
       eprint = {1806.04088},
 primaryClass = {astro-ph.HE},
       adsurl = {https://ui.adsabs.harvard.edu/abs/2018ApJ...865L..21K}
}

@ARTICLE{Kato_2024,
       author = {{Kato}, Daiji and {Tanaka}, Masaomi and {Gaigalas}, Gediminas and {Kitovien{\.{e}}}, Laima and {Rynkun}, Pavel},
        title = "{Systematic opacity calculations for kilonovae - II. Improved atomic data for singly ionized lanthanides}",
      journal = {\mnras},
         year = 2024,
        month = dec,
       volume = {535},
       number = {3},
        pages = {2670-2686},
          doi = {10.1093/mnras/stae2504},
archivePrefix = {arXiv},
       eprint = {2501.13286},
 primaryClass = {astro-ph.HE},
       adsurl = {https://ui.adsabs.harvard.edu/abs/2024MNRAS.535.2670K}
}

@inproceedings{Chen_2018,
 author = {Chen, Ricky T. Q. and Rubanova, Yulia and Bettencourt, Jesse and Duvenaud, David K},
 booktitle = {Advances in Neural Information Processing Systems},
 editor = {S. Bengio and H. Wallach and H. Larochelle and K. Grauman and N. Cesa-Bianchi and R. Garnett},
 volume = {31},
 pages = {6572--6583},
 title = {Neural Ordinary Differential Equations},
 year = {2018},
publisher = {Curran Associates, Inc.},
       url = {https://papers.nips.cc/paper_files/paper/2018/file/69386f6bb1dfed68692a24c8686939b9-Paper.pdf}
}

@INPROCEEDINGS{Grathwohl_2019,
       author = {{Grathwohl}, Will and {Chen}, Ricky T.~Q. and {Bettencourt}, Jesse and {Sutskever}, Ilya and {Duvenaud}, David},
        title = "{FFJORD: Free-form Continuous Dynamics for Scalable Reversible Generative Models}",
       editor = {Tara Sainath},
    booktitle = {7th International Conference on Learning Representations},
      address = {New Orleans, LA, USA},
         year = 2019,
        month = oct,
          eid = {arXiv:1810.01367},
archivePrefix = {arXiv},
       eprint = {1810.01367},
 primaryClass = {cs.LG},
       adsurl = {https://ui.adsabs.harvard.edu/abs/2018arXiv181001367G}
}

@ARTICLE{Bird_2019,
       author = {{Bird}, Simeon and {Rogers}, Keir K. and {Peiris}, Hiranya V. and {Verde}, Licia and {Font-Ribera}, Andreu and {Pontzen}, Andrew},
        title = "{An emulator for the Lyman-{\ensuremath{\alpha}} forest}",
      journal = {\jcap},
         year = 2019,
        month = feb,
       volume = {2019(02)},
          eid = {050},
        pages = {050},
          doi = {10.1088/1475-7516/2019/02/050},
archivePrefix = {arXiv},
       eprint = {1812.04654},
 primaryClass = {astro-ph.CO},
       adsurl = {https://ui.adsabs.harvard.edu/abs/2019JCAP...02..050B}
}

@article{Gardner_2023,
   title={The James Webb Space Telescope Mission},
   volume={135},
   ISSN={1538-3873},
   url={http://dx.doi.org/10.1088/1538-3873/acd1b5},
   DOI={10.1088/1538-3873/acd1b5},
   number={1048},
   journal={\pasp},
   publisher={IOP Publishing},
   author={Gardner, Jonathan P. and others},
   year={2023},
   month=jun, pages={068001} }

@INPROCEEDINGS{Schlieder_2024,
       author = {{Schlieder}, Joshua E. and {Barclay}, Thomas and {Barnes}, Amethyst and {Bray}, Evan and {Choi}, Ami and {Cromey}, Benjamin and {Delker}, Thomas and {Finch}, Timothy and {Frater}, Eric H. and {Hill}, Robert J. and {Kruk}, Jeffrey and {Lasco}, Jeffrey and {Louie}, Dana R. and {Malhotra}, Sangeeta and {McEnery}, Julie E. and {Mosby}, Gregory and {Paine}, Jennie and {Perkins}, Jeremy S. and {Rauscher}, Bernard J. and {Rhoads}, James E. and {Rizzo}, Maxime and {Sabatke}, Derek and {Schweickart}, Rusty and {Shukis}, Diana and {Switzer}, Eric R. and {Wollack}, Edward J. and {Zellem}, Robert T. and {Zimmerman}, Neil T.},
        title = "{Survey science with the Nancy Grace Roman Space Telescope Wide Field Instrument}",
    booktitle = {Space Telescopes and Instrumentation 2024: Optical, Infrared, and Millimeter Wave},
         year = 2024,
       editor = {{Coyle}, Laura E. and {Matsuura}, Shuji and {Perrin}, Marshall D.},
       series = {SPIE Conference Series},
       volume = {13092},
        month = aug,
          eid = {130920S},
        pages = {130920S},
          doi = {10.1117/12.3020622},
       adsurl = {https://ui.adsabs.harvard.edu/abs/2024SPIE13092E..0SS}
}

@article{Metzger_2019,
   title={Kilonovae},
   volume={23},
   pages={1},
   ISSN={1433-8351},
   url={http://dx.doi.org/10.1007/s41114-019-0024-0},
   DOI={10.1007/s41114-019-0024-0},
   number={1},
   journal={LRR},
   publisher={Springer Science and Business Media LLC},
   author={Metzger, Brian D.},
   year={2019},
   month=Dec }

@article{Darc_2024,
   author = {P. Darc and Clecio R. Bom and Bernardo M. O. Fraga and Charlie D. Kilpatrick},
   doi = {10.3847/1538-4357/ad53c7},
   issue = {1},
   journal = {\apj},
   month = {8},
   pages = {82},
   publisher = {American Astronomical Society},
   title = {Kilonova Spectral Inverse Modelling with Simulation-Based Inference: An Amortized Neural Posterior Estimation Analysis},
   volume = {971},
   year = {2024},
}

@article{Speagle_2020,
   title={dynesty: a dynamic nested sampling package for estimating Bayesian posteriors and evidences},
   volume={493},
   ISSN={1365-2966},
   url={http://dx.doi.org/10.1093/mnras/staa278},
   DOI={10.1093/mnras/staa278},
   number={3},
   journal={\mnras},
   publisher={Oxford University Press (OUP)},
   author={Speagle, Joshua S},
   year={2020},
   month=Feb, pages={3132-3158} }

@article{Cole_2022,
   title={Fast and credible likelihood-free cosmology with truncated marginal neural ratio estimation},
   volume={2022(09)},
   ISSN={1475-7516},
   url={http://dx.doi.org/10.1088/1475-7516/2022/09/004},
   DOI={10.1088/1475-7516/2022/09/004},
   number={09},
   journal={\jcap},
   publisher={IOP Publishing},
   author={Cole, Alex and Miller, Benjamin K. and Witte, Samuel J. and Cai, Maxwell X. and Grootes, Meiert W. and Nattino, Francesco and Weniger, Christoph},
   year={2022},
   month=sep, pages={004} }

@dataset{Villar_2018,
       author = {{Villar}, V.~A. and {Guillochon}, J. and {Berger}, E. and {Metzger}, B.~D. and {Cowperthwaite}, P.~S. and {Nicholl}, M. and {Alexander}, K.~D. and {Blanchard}, P.~K. and {Chornock}, R. and {Eftekhari}, T. and {Fong}, W. and {Margutti}, R. and {Williams}, P.~K.~G.},
        title = "{UV-NIR obs.\ compilation of GW170817 counterpart: J/ApJ/851/L21 (Villar+17)}",
         year = 2018,
        month = jul,
          publisher = {VizieR},
          doi = {10.26093/cds/vizier.18519021},
       adsurl = {https://ui.adsabs.harvard.edu/abs/2018yCat..18519021V}
}

@article{Mekid_2008,
    title = {Propagation of uncertainty: Expressions of second and third order uncertainty with third and fourth moments},
    journal = {Measurement},
    volume = {41},
    number = {6},
    pages = {600-609},
    year = {2008},
    issn = {0263-2241},
    doi = {https://doi.org/10.1016/j.measurement.2007.07.004},
    url = {https://www.sciencedirect.com/science/article/pii/S0263224107000681},
    author = {S. Mekid and D. Vaja}}

@INPROCEEDINGS{Almualla_2022,
       author = {{Almualla}, Mouza and {Ning}, Yuhong and {Bulla}, Mattia and {Dietrich}, Tim and {Coughlin}, Michael and {Guessoum}, Nidhal},
        title = "{Using Neural Networks to Perform Rapid High-Dimensional Kilonova Parameter Inference}",
    booktitle = {44th COSPAR Scientific Assembly. Held 16-24 July},
         year = 2022,
       volume = {44},
        month = jul,
        pages = {1973},
       adsurl = {https://ui.adsabs.harvard.edu/abs/2022cosp...44.1973A}
}

@ARTICLE{oac_api,
       author = {{Guillochon}, James and {Parrent}, Jerod and {Kelley}, Luke Zoltan and {Margutti}, Raffaella},
        title = "{An Open Catalog for Supernova Data}",
      journal = {\apj},
         year = 2017,
        month = jan,
       volume = {835},
       number = {1},
          eid = {64},
        pages = {64},
          doi = {10.3847/1538-4357/835/1/64},
archivePrefix = {arXiv},
       eprint = {1605.01054},
 primaryClass = {astro-ph.SR},
       adsurl = {https://ui.adsabs.harvard.edu/abs/2017ApJ...835...64G}
}

@article{Andreoni_2017,
   title={Follow Up of GW170817 and Its Electromagnetic Counterpart by Australian-Led Observing Programmes},
   volume={34},
   pages = {e069},
   ISSN={1448-6083},
   url={http://dx.doi.org/10.1017/pasa.2017.65},
   DOI={10.1017/pasa.2017.65},
   journal={\pasa},
   publisher={Cambridge University Press (CUP)},
   author={Andreoni, I. and Ackley, K. and Cooke, J. and Acharyya, A. and Allison, J. R. and Anderson, G. E. and Ashley, M. C. B. and Baade, D. and Bailes, M. and Bannister, K. and Beardsley, A. and Bessell, M. S. and Bian, F. and Bland, P. A. and Boer, M. and Booler, T. and Brandeker, A. and Brown, I. S. and Buckley, D. A. H. and Chang, S.-W. and Coward, D. M. and Crawford, S. and Crisp, H. and Crosse, B. and Cucchiara, A. and Cupák, M. and de Gois, J. S. and Deller, A. and Devillepoix, H. A. R. and Dobie, D. and Elmer, E. and Emrich, D. and Farah, W. and Farrell, T. J. and Franzen, T. and Gaensler, B. M. and Galloway, D. K. and Gendre, B. and Giblin, T. and Goobar, A. and Green, J. and Hancock, P. J. and Hartig, B. A. D. and Howell, E. J. and Horsley, L. and Hotan, A. and Howie, R. M. and Hu, L. and Hu, Y. and James, C. W. and Johnston, S. and Johnston-Hollitt, M. and Kaplan, D. L. and Kasliwal, M. and Keane, E. F. and Kenney, D. and Klotz, A. and Lau, R. and Laugier, R. and Lenc, E. and Li, X. and Liang, E. and Lidman, C. and Luvaul, L. C. and Lynch, C. and Ma, B. and Macpherson, D. and Mao, J. and McClelland, D. E. and McCully, C. and Möller, A. and Morales, M. F. and Morris, D. and Murphy, T. and Noysena, K. and Onken, C. A. and Orange, N. B. and Osłowski, S. and Pallot, D. and Paxman, J. and Potter, S. B. and Pritchard, T. and Raja, W. and Ridden-Harper, R. and Romero-Colmenero, E. and Sadler, E. M. and Sansom, E. K. and Scalzo, R. A. and Schmidt, B. P. and Scott, S. M. and Seghouani, N. and Shang, Z. and Shannon, R. M. and Shao, L. and Shara, M. M. and Sharp, R. and Sokolowski, M. and Sollerman, J. and Staff, J. and Steele, K. and Sun, T. and Suntzeff, N. B. and Tao, C. and Tingay, S. and Towner, M. C. and Thierry, P. and Trott, C. and Tucker, B. E. and Väisänen, P. and Krishnan, V. Venkatraman and Walker, M. and Wang, L. and Wang, X. and Wayth, R. and Whiting, M. and Williams, A. and Williams, T. and Wolf, C. and Wu, C. and Wu, X. and Yang, J. and Yuan, X. and Zhang, H. and Zhou, J. and Zovaro, H.},
   year={2017} }

@article{Arcavi_2017,
   title={Optical emission from a kilonova following a gravitational-wave-detected neutron-star merger},
   volume={551},
   ISSN={1476-4687},
   url={http://dx.doi.org/10.1038/nature24291},
   DOI={10.1038/nature24291},
   number={7678},
   journal={Nature},
   publisher={Springer Science and Business Media LLC},
   author={Arcavi, Iair and Hosseinzadeh, Griffin and Howell, D. Andrew and McCully, Curtis and Poznanski, Dovi and Kasen, Daniel and Barnes, Jennifer and Zaltzman, Michael and Vasylyev, Sergiy and Maoz, Dan and Valenti, Stefano},
   year={2017},
   month=Oct, pages={64-66} }

@ARTICLE{Diaz_2017,
       author = {{D{\'\i}az}, M.~C. and {Macri}, L.~M. and {Garcia Lambas}, D. and {Mendes de Oliveira}, C. and {Nilo Castell{\'o}n}, J.~L. and {Ribeiro}, T. and {S{\'a}nchez}, B. and {Schoenell}, W. and {Abramo}, L.~R. and {Akras}, S. and {Alcaniz}, J.~S. and {Artola}, R. and {Beroiz}, M. and {Bonoli}, S. and {Cabral}, J. and {Camuccio}, R. and {Castillo}, M. and {Chavushyan}, V. and {Coelho}, P. and {Colazo}, C. and {Costa-Duarte}, M.~V. and {Cuevas Larenas}, H. and {DePoy}, D.~L. and {Dom{\'\i}nguez Romero}, M. and {Dultzin}, D. and {Fern{\'a}ndez}, D. and {Garc{\'\i}a}, J. and {Girardini}, C. and {Gon{\c{c}}alves}, D.~R. and {Gon{\c{c}}alves}, T.~S. and {Gurovich}, S. and {Jim{\'e}nez-Teja}, Y. and {Kanaan}, A. and {Lares}, M. and {Lopes de Oliveira}, R. and {L{\'o}pez-Cruz}, O. and {Marshall}, J.~L. and {Melia}, R. and {Molino}, A. and {Padilla}, N. and {Pe{\~n}uela}, T. and {Placco}, V.~M. and {Qui{\~n}ones}, C. and {Ram{\'\i}rez Rivera}, A. and {Renzi}, V. and {Riguccini}, L. and {R{\'\i}os-L{\'o}pez}, E. and {Rodriguez}, H. and {Sampedro}, L. and {Schneiter}, M. and {Sodr{\'e}}, L. and {Starck}, M. and {Torres-Flores}, S. and {Tornatore}, M. and {Zadro{\.z}ny}, A.},
        title = "{Observations of the First Electromagnetic Counterpart to a Gravitational-wave Source by the TOROS Collaboration}",
      journal = {\apjl},
         year = 2017,
        month = oct,
       volume = {848},
       number = {2},
          eid = {L29},
        pages = {L29},
          doi = {10.3847/2041-8213/aa9060},
archivePrefix = {arXiv},
       eprint = {1710.05844},
 primaryClass = {astro-ph.HE},
       adsurl = {https://ui.adsabs.harvard.edu/abs/2017ApJ...848L..29D}
}

@article{Drout_2017,
    author = {M. R. Drout  and A. L. Piro  and B. J. Shappee  and C. D. Kilpatrick  and J. D. Simon  and C. Contreras  and D. A. Coulter  and R. J. Foley  and M. R. Siebert  and N. Morrell  and K. Boutsia  and F. Di Mille  and T. W.-S. Holoien  and D. Kasen  and J. A. Kollmeier  and B. F. Madore  and A. J. Monson  and A. Murguia-Berthier  and Y.-C. Pan  and J. X. Prochaska  and E. Ramirez-Ruiz  and A. Rest  and C. Adams  and K. Alatalo  and E. Bañados  and J. Baughman  and T. C. Beers  and R. A. Bernstein  and T. Bitsakis  and A. Campillay  and T. T. Hansen  and C. R. Higgs  and A. P. Ji  and G. Maravelias  and J. L. Marshall  and C. Moni Bidin  and J. L. Prieto  and K. C. Rasmussen  and C. Rojas-Bravo  and A. L. Strom  and N. Ulloa  and J. Vargas-González  and Z. Wan  and D. D. Whitten },
    title = {Light curves of the neutron star merger GW170817/SSS17a: Implications for r-process nucleosynthesis},
    journal = {Science},
    volume = {358},
    number = {6370},
    pages = {1570-1574},
    year = {2017},
    doi = {10.1126/science.aaq0049},
    URL = {https://www.science.org/doi/abs/10.1126/science.aaq0049},
    eprint = {https://www.science.org/doi/pdf/10.1126/science.aaq0049}}

@article{Hu_2017,
   title={Optical observations of LIGO source GW 170817 by the Antarctic Survey Telescopes at Dome A, Antarctica},
   volume={62},
   ISSN={2095-9273},
   url={http://dx.doi.org/10.1016/j.scib.2017.10.006},
   DOI={10.1016/j.scib.2017.10.006},
   number={21},
   journal={Science Bulletin},
   publisher={Elsevier BV},
   author={Hu, Lei and Wu, Xuefeng and Andreoni, Igor and B. Ashley, Michael C. and Cooke, Jeff and Cui, Xiangqun and Du, Fujia and Dai, Zigao and Gu, Bozhong and Hu, Yi and Lu, Haiping and Li, Xiaoyan and Li, Zhengyang and Liang, Ensi and Liu, Liangduan and Ma, Bin and Shang, Zhaohui and Sun, Tianrui and Suntzeff, N.B. and Tao, Charling and Uddin, Syed A. and Wang, Lifan and Wang, Xiaofeng and Wen, Haikun and Xiao, Di and Xu, Jin and Yang, Ji and Yang, Shihai and Yuan, Xiangyan and Zhou, Hongyan and Zhang, Hui and Zhou, Jilin and Zhu, Zonghong},
   year={2017},
   month=Nov, pages={1433-1438} }

@article{Kasliwal_2017,
    author = {M. M. Kasliwal  and E. Nakar  and L. P. Singer  and D. L. Kaplan  and D. O. Cook  and A. Van Sistine  and R. M. Lau  and C. Fremling  and O. Gottlieb  and J. E. Jencson  and S. M. Adams  and U. Feindt  and K. Hotokezaka  and S. Ghosh  and D. A. Perley  and P.-C. Yu  and T. Piran  and J. R. Allison  and G. C. Anupama  and A. Balasubramanian  and K. W. Bannister  and J. Bally  and J. Barnes  and S. Barway  and E. Bellm  and V. Bhalerao  and D. Bhattacharya  and N. Blagorodnova  and J. S. Bloom  and P. R. Brady  and C. Cannella  and D. Chatterjee  and S. B. Cenko  and B. E. Cobb  and C. Copperwheat  and A. Corsi  and K. De  and D. Dobie  and S. W. K. Emery  and P. A. Evans  and O. D. Fox  and D. A. Frail  and C. Frohmaier  and A. Goobar  and G. Hallinan  and F. Harrison  and G. Helou  and T. Hinderer  and A. Y. Q. Ho  and A. Horesh  and W.-H. Ip  and R. Itoh  and D. Kasen  and H. Kim  and N. P. M. Kuin  and T. Kupfer  and C. Lynch  and K. Madsen  and P. A. Mazzali  and A. A. Miller  and K. Mooley  and T. Murphy  and C.-C. Ngeow  and D. Nichols  and S. Nissanke  and P. Nugent  and E. O. Ofek  and H. Qi  and R. M. Quimby  and S. Rosswog  and F. Rusu  and E. M. Sadler  and P. Schmidt  and J. Sollerman  and I. Steele  and A. R. Williamson  and Y. Xu  and L. Yan  and Y. Yatsu  and C. Zhang  and W. Zhao },
    title = {Illuminating gravitational waves: A concordant picture of photons from a neutron star merger},
    journal = {Science},
    volume = {358},
    number = {6370},
    pages = {1559-1565},
    year = {2017},
    doi = {10.1126/science.aap9455},
    URL = {https://www.science.org/doi/abs/10.1126/science.aap9455},
    eprint = {https://www.science.org/doi/pdf/10.1126/science.aap9455}}

@ARTICLE{Valenti_2017,
       author = {{Valenti}, Stefano and {Sand}, David J. and {Yang}, Sheng and {Cappellaro}, Enrico and {Tartaglia}, Leonardo and {Corsi}, Alessandra and {Jha}, Saurabh W. and {Reichart}, Daniel E. and {Haislip}, Joshua and {Kouprianov}, Vladimir},
        title = "{The Discovery of the Electromagnetic Counterpart of GW170817: Kilonova AT 2017gfo/DLT17ck}",
      journal = {\apjl},
         year = 2017,
        month = oct,
       volume = {848},
       number = {2},
          eid = {L24},
        pages = {L24},
          doi = {10.3847/2041-8213/aa8edf},
archivePrefix = {arXiv},
       eprint = {1710.05854},
 primaryClass = {astro-ph.HE},
       adsurl = {https://ui.adsabs.harvard.edu/abs/2017ApJ...848L..24V}
}

@ARTICLE{Lipunov_2017,
       author = {{Lipunov}, V.~M. and {Gorbovskoy}, E. and {Kornilov}, V.~G. and {. Tyurina}, N. and {Balanutsa}, P. and {Kuznetsov}, A. and {Vlasenko}, D. and {Kuvshinov}, D. and {Gorbunov}, I. and {Buckley}, D.~A.~H. and {Krylov}, A.~V. and {Podesta}, R. and {Lopez}, C. and {Podesta}, F. and {Levato}, H. and {Saffe}, C. and {Mallamachi}, C. and {Potter}, S. and {Budnev}, N.~M. and {Gress}, O. and {Ishmuhametova}, Yu. and {Vladimirov}, V. and {Zimnukhov}, D. and {Yurkov}, V. and {Sergienko}, Yu. and {Gabovich}, A. and {Rebolo}, R. and {Serra-Ricart}, M. and {Israelyan}, G. and {Chazov}, V. and {Wang}, Xiaofeng and {Tlatov}, A. and {Panchenko}, M.~I.},
        title = "{MASTER Optical Detection of the First LIGO/Virgo Neutron Star Binary Merger GW170817}",
      journal = {\apjl},
         year = 2017,
        month = nov,
       volume = {850},
       number = {1},
          eid = {L1},
        pages = {L1},
          doi = {10.3847/2041-8213/aa92c0},
archivePrefix = {arXiv},
       eprint = {1710.05461},
 primaryClass = {astro-ph.HE},
       adsurl = {https://ui.adsabs.harvard.edu/abs/2017ApJ...850L...1L}
}

@article{Pian_2017,
   title={Spectroscopic identification of r-process nucleosynthesis in a double neutron-star merger},
   volume={551},
   ISSN={1476-4687},
   url={http://dx.doi.org/10.1038/nature24298},
   DOI={10.1038/nature24298},
   number={7678},
   journal={Nature},
   publisher={Springer Science and Business Media LLC},
   author={Pian, E. and D’Avanzo, P. and Benetti, S. and Branchesi, M. and Brocato, E. and Campana, S. and Cappellaro, E. and Covino, S. and D’Elia, V. and Fynbo, J. P. U. and Getman, F. and Ghirlanda, G. and Ghisellini, G. and Grado, A. and Greco, G. and Hjorth, J. and Kouveliotou, C. and Levan, A. and Limatola, L. and Malesani, D. and Mazzali, P. A. and Melandri, A. and Møller, P. and Nicastro, L. and Palazzi, E. and Piranomonte, S. and Rossi, A. and Salafia, O. S. and Selsing, J. and Stratta, G. and Tanaka, M. and Tanvir, N. R. and Tomasella, L. and Watson, D. and Yang, S. and Amati, L. and Antonelli, L. A. and Ascenzi, S. and Bernardini, M. G. and Boër, M. and Bufano, F. and Bulgarelli, A. and Capaccioli, M. and Casella, P. and Castro-Tirado, A. J. and Chassande-Mottin, E. and Ciolfi, R. and Copperwheat, C. M. and Dadina, M. and De Cesare, G. and Di Paola, A. and Fan, Y. Z. and Gendre, B. and Giuffrida, G. and Giunta, A. and Hunt, L. K. and Israel, G. L. and Jin, Z.-P. and Kasliwal, M. M. and Klose, S. and Lisi, M. and Longo, F. and Maiorano, E. and Mapelli, M. and Masetti, N. and Nava, L. and Patricelli, B. and Perley, D. and Pescalli, A. and Piran, T. and Possenti, A. and Pulone, L. and Razzano, M. and Salvaterra, R. and Schipani, P. and Spera, M. and Stamerra, A. and Stella, L. and Tagliaferri, G. and Testa, V. and Troja, E. and Turatto, M. and Vergani, S. D. and Vergani, D.},
   year={2017},
   month=Oct, pages={67-70} }

@article{Pozanenko_2018,
   title={GRB 170817A Associated with GW170817: Multi-frequency Observations and Modeling of Prompt Gamma-Ray Emission},
   volume={852},
   ISSN={2041-8213},
   url={http://dx.doi.org/10.3847/2041-8213/aaa2f6},
   DOI={10.3847/2041-8213/aaa2f6},
   number={2},
   journal={\apjl},
   publisher={American Astronomical Society},
   author={Pozanenko, A. S. and Barkov, M. V. and Minaev, P. Yu. and Volnova, A. A. and Mazaeva, E. D. and Moskvitin, A. S. and Krugov, M. A. and Samodurov, V. A. and Loznikov, V. M. and Lyutikov, M.},
   year={2018},
   month=Jan, pages={L30} }

@article{Shappee_2017,
   title={Early spectra of the gravitational wave source GW170817: Evolution of a neutron star merger},
   volume={358},
   ISSN={1095-9203},
   url={http://dx.doi.org/10.1126/science.aaq0186},
   DOI={10.1126/science.aaq0186},
   number={6370},
   journal={Science},
   publisher={American Association for the Advancement of Science (AAAS)},
   author={Shappee, B. J. and Simon, J. D. and Drout, M. R. and Piro, A. L. and Morrell, N. and Prieto, J. L. and Kasen, D. and Holoien, T. W.-S. and Kollmeier, J. A. and Kelson, D. D. and Coulter, D. A. and Foley, R. J. and Kilpatrick, C. D. and Siebert, M. R. and Madore, B. F. and Murguia-Berthier, A. and Pan, Y.-C. and Prochaska, J. X. and Ramirez-Ruiz, E. and Rest, A. and Adams, C. and Alatalo, K. and Bañados, E. and Baughman, J. and Bernstein, R. A. and Bitsakis, T. and Boutsia, K. and Bravo, J. R. and Di Mille, F. and Higgs, C. R. and Ji, A. P. and Maravelias, G. and Marshall, J. L. and Placco, V. M. and Prieto, G. and Wan, Z.},
   year={2017},
   month=Dec, pages={1574-1578} }

@article{Smartt_2017,
   title={A kilonova as the electromagnetic counterpart to a gravitational-wave source},
   volume={551},
   ISSN={1476-4687},
   url={http://dx.doi.org/10.1038/nature24303},
   DOI={10.1038/nature24303},
   number={7678},
   journal={Nature},
   publisher={Springer Science and Business Media LLC},
   author={Smartt, S. J. and Chen, T.-W. and Jerkstrand, A. and Coughlin, M. and Kankare, E. and Sim, S. A. and Fraser, M. and Inserra, C. and Maguire, K. and Chambers, K. C. and Huber, M. E. and Krühler, T. and Leloudas, G. and Magee, M. and Shingles, L. J. and Smith, K. W. and Young, D. R. and Tonry, J. and Kotak, R. and Gal-Yam, A. and Lyman, J. D. and Homan, D. S. and Agliozzo, C. and Anderson, J. P. and Angus, C. R. and Ashall, C. and Barbarino, C. and Bauer, F. E. and Berton, M. and Botticella, M. T. and Bulla, M. and Bulger, J. and Cannizzaro, G. and Cano, Z. and Cartier, R. and Cikota, A. and Clark, P. and De Cia, A. and Della Valle, M. and Denneau, L. and Dennefeld, M. and Dessart, L. and Dimitriadis, G. and Elias-Rosa, N. and Firth, R. E. and Flewelling, H. and Flörs, A. and Franckowiak, A. and Frohmaier, C. and Galbany, L. and González-Gaitán, S. and Greiner, J. and Gromadzki, M. and Guelbenzu, A. Nicuesa and Gutiérrez, C. P. and Hamanowicz, A. and Hanlon, L. and Harmanen, J. and Heintz, K. E. and Heinze, A. and Hernandez, M.-S. and Hodgkin, S. T. and Hook, I. M. and Izzo, L. and James, P. A. and Jonker, P. G. and Kerzendorf, W. E. and Klose, S. and Kostrzewa-Rutkowska, Z. and Kowalski, M. and Kromer, M. and Kuncarayakti, H. and Lawrence, A. and Lowe, T. B. and Magnier, E. A. and Manulis, I. and Martin-Carrillo, A. and Mattila, S. and McBrien, O. and Müller, A. and Nordin, J. and O’Neill, D. and Onori, F. and Palmerio, J. T. and Pastorello, A. and Patat, F. and Pignata, G. and Podsiadlowski, Ph. and Pumo, M. L. and Prentice, S. J. and Rau, A. and Razza, A. and Rest, A. and Reynolds, T. and Roy, R. and Ruiter, A. J. and Rybicki, K. A. and Salmon, L. and Schady, P. and Schultz, A. S. B. and Schweyer, T. and Seitenzahl, I. R. and Smith, M. and Sollerman, J. and Stalder, B. and Stubbs, C. W. and Sullivan, M. and Szegedi, H. and Taddia, F. and Taubenberger, S. and Terreran, G. and van Soelen, B. and Vos, J. and Wainscoat, R. J. and Walton, N. A. and Waters, C. and Weiland, H. and Willman, M. and Wiseman, P. and Wright, D. E. and Wyrzykowski, Ł. and Yaron, O.},
   year={2017},
   month=Oct, pages={75–79} }

@article{Troja_2017,
   title={The X-ray counterpart to the gravitational-wave event GW170817},
   volume={551},
   ISSN={1476-4687},
   url={http://dx.doi.org/10.1038/nature24290},
   DOI={10.1038/nature24290},
   number={7678},
   journal={Nature},
   publisher={Springer Science and Business Media LLC},
   author={Troja, E. and Piro, L. and van Eerten, H. and Wollaeger, R. T. and Im, M. and Fox, O. D. and Butler, N. R. and Cenko, S. B. and Sakamoto, T. and Fryer, C. L. and Ricci, R. and Lien, A. and Ryan, R. E. and Korobkin, O. and Lee, S.-K. and Burgess, J. M. and Lee, W. H. and Watson, A. M. and Choi, C. and Covino, S. and D’Avanzo, P. and Fontes, C. J. and González, J. Becerra and Khandrika, H. G. and Kim, J. and Kim, S.-L. and Lee, C.-U. and Lee, H. M. and Kutyrev, A. and Lim, G. and Sánchez-Ramírez, R. and Veilleux, S. and Wieringa, M. H. and Yoon, Y.},
   year={2017},
   month=Oct, pages={71-74} }

@ARTICLE{Utsumi_2017,
       author = {{Utsumi}, Yousuke and {Tanaka}, Masaomi and {Tominaga}, Nozomu and {Yoshida}, Michitoshi and {Barway}, Sudhanshu and {Nagayama}, Takahiro and {Zenko}, Tetsuya and {Aoki}, Kentaro and {Fujiyoshi}, Takuya and {Furusawa}, Hisanori and {Kawabata}, Koji S. and {Koshida}, Shintaro and {Lee}, Chien-Hsiu and {Morokuma}, Tomoki and {Motohara}, Kentaro and {Nakata}, Fumiaki and {Ohsawa}, Ryou and {Ohta}, Kouji and {Okita}, Hirofumi and {Tajitsu}, Akito and {Tanaka}, Ichi and {Terai}, Tsuyoshi and {Yasuda}, Naoki and {Abe}, Fumio and {Asakura}, Yuichiro and {Bond}, Ian A. and {Miyazaki}, Shota and {Sumi}, Takahiro and {Tristram}, Paul J. and {Honda}, Satoshi and {Itoh}, Ryosuke and {Itoh}, Yoichi and {Kawabata}, Miho and {Morihana}, Kumiko and {Nagashima}, Hiroki and {Nakaoka}, Tatsuya and {Ohshima}, Tomohito and {Takahashi}, Jun and {Takayama}, Masaki and {Aoki}, Wako and {Baar}, Stefan and {Doi}, Mamoru and {Finet}, Francois and {Kanda}, Nobuyuki and {Kawai}, Nobuyuki and {Kim}, Ji Hoon and {Kuroda}, Daisuke and {Liu}, Wei and {Matsubayashi}, Kazuya and {Murata}, Katsuhiro L. and {Nagai}, Hiroshi and {Saito}, Tomoki and {Saito}, Yoshihiko and {Sako}, Shigeyuki and {Sekiguchi}, Yuichiro and {Tamura}, Yoichi and {Tanaka}, Masayuki and {Uemura}, Makoto and {Yamaguchi}, Masaki S.},
        title = "{J-GEM observations of an electromagnetic counterpart to the neutron star merger GW170817}",
      journal = {\pasj},
         year = 2017,
        month = dec,
       volume = {69},
       number = {6},
          eid = {101},
        pages = {101},
          doi = {10.1093/pasj/psx118},
archivePrefix = {arXiv},
       eprint = {1710.05848},
 primaryClass = {astro-ph.HE},
       adsurl = {https://ui.adsabs.harvard.edu/abs/2017PASJ...69..101U}
}

@article{LAT_2017,
       author = {{Ajello}, M. and {Allafort}, A. and {Axelsson}, M. and {Baldini}, L. and {Barbiellini}, G. and {Baring}, M.~G. and {Bastieri}, D. and {Bellazzini}, R. and {Berenji}, B. and {Bissaldi}, E. and {Blandford}, R.~D. and {Bloom}, E.~D. and {Bonino}, R. and {Bottacini}, E. and {Brandt}, T.~J. and {Bregeon}, J. and {Bruel}, P. and {Buehler}, R. and {Burnett}, T.~H. and {Buson}, S. and {Cameron}, R.~A. and {Caputo}, R. and {Caraveo}, P.~A. and {Casandjian}, J.~M. and {Cavazzuti}, E. and {Chekhtman}, A. and {Cheung}, C.~C. and {Chiang}, J. and {Chiaro}, G. and {Ciprini}, S. and {Cohen-Tanugi}, J. and {Cominsky}, L.~R. and {Costantin}, D. and {Cuoco}, A. and {Cutini}, S. and {D'Ammando}, F. and {de Palma}, F. and {Di Lalla}, N. and {Di Mauro}, M. and {Di Venere}, L. and {Dubois}, R. and {Dumora}, D. and {Favuzzi}, C. and {Ferrara}, E.~C. and {Franckowiak}, A. and {Fukazawa}, Y. and {Funk}, S. and {Fusco}, P. and {Gargano}, F. and {Gasparrini}, D. and {Giglietto}, N. and {Gill}, R. and {Giordano}, F. and {Giroletti}, M. and {Glanzman}, T. and {Granot}, J. and {Green}, D. and {Grenier}, I.~A. and {Grondin}, M.-H. and {Guillemot}, L. and {Guiriec}, S. and {Harding}, A.~K. and {Hays}, E. and {Horan}, D. and {Imazato}, F. and {J{\'o}hannesson}, G. and {Kamae}, T. and {Kensei}, S. and {Kocevski}, D. and {Kuss}, M. and {La Mura}, G. and {Larsson}, S. and {Latronico}, L. and {Li}, J. and {Longo}, F. and {Loparco}, F. and {Lovellette}, M.~N. and {Lubrano}, P. and {Magill}, J.~D. and {Maldera}, S. and {Manfreda}, A. and {Mazziotta}, M.~N. and {Michelson}, P.~F. and {Mizuno}, T. and {Moiseev}, A.~A. and {Monzani}, M.~E. and {Moretti}, E. and {Morselli}, A. and {Moskalenko}, I.~V. and {Negro}, M. and {Nuss}, E. and {Ojha}, R. and {Omodei}, N. and {Orlando}, E. and {Ormes}, J.~F. and {Palatiello}, M. and {Paliya}, V.~S. and {Paneque}, D. and {Persic}, M. and {Pesce-Rollins}, M. and {Petrosian}, V. and {Piron}, F. and {Porter}, T.~A. and {Principe}, G. and {Racusin}, J.~L. and {Rain{\`o}}, S. and {Rando}, R. and {Razzano}, M. and {Razzaque}, S. and {Reimer}, A. and {Reimer}, O. and {Ritz}, S. and {Rochester}, L.~S. and {Ryde}, F. and {Saz Parkinson}, P.~M. and {Sgr{\`o}}, C. and {Siskind}, E.~J. and {Spada}, F. and {Spandre}, G. and {Spinelli}, P. and {Suson}, D.~J. and {Tajima}, H. and {Takahashi}, M. and {Tak}, D. and {Thayer}, J.~G. and {Thayer}, J.~B. and {Torres}, D.~F. and {Torresi}, E. and {Tosti}, G. and {Troja}, E. and {Valverde}, J. and {Venters}, T.~M. and {Vianello}, G. and {Wood}, K. and {Yang}, C. and {Zaharijas}, G.},
        title = "{Fermi-LAT Observations of LIGO/Virgo Event GW170817}",
      journal = {\apj},
         year = 2018,
        month = jul,
       volume = {861},
       number = {2},
          eid = {85},
        pages = {85},
          doi = {10.3847/1538-4357/aac515},
       adsurl = {https://ui.adsabs.harvard.edu/abs/2018ApJ...861...85A}
}

@article{Scolnic_2019,
      title={The Next Generation of Cosmological Measurements with Type Ia Supernovae}, 
      author={Dan Scolnic and Saul Perlmutter and Greg Aldering and Dillon Brout and Tamara Davis and Alex Filippenko and Ryan Foley and Renee Hlozek and Rebekah Hounsell and Saurabh Jha and David Jones and Pat Kelly and Rick Kessler and Alex Kim and David Rubin and Adam Riess and Steven Rodney and Justin Roberts-Pierel and Christopher Stubbs and Yun Wang and Jacobo Asorey and Arturo Avelino and Chetan Bavdhankar and Peter J. Brown and Anthony Challinor and Christophe Balland and Asantha Cooray and Suhail Dhawan and Georgios Dimitriadis and Cora Dvorkin and Julien Guy and Will Handley and Ryan E. Keeley and Jean-Paul Kneib and Benjamin LHuillier and Massimiliano Lattanzi and Kaisey Mandel and James Mertens and Mickael Rigault and Pavel Motloch and Suvodip Mukherjee and Gautham Narayan and Andrei Nomerotski and Lyman Page and Levon Pogosian and Giuseppe Puglisi and Marco Raveri and Nicolas Regnault and Armin Rest and Cesar Rojas-Bravo and Masao Sako and Feng Shi and Srivatsan Sridhar and Aritoki Suzuki and Yu-Dai Tsai and W. M. Wood-Vasey and Yannick Copin and Gong-Bo Zhao and Ningfeng Zhu},
      year = {2019},
      journal = {arXiv e-prints},
      pages = {arXiv:1903.05128},
      doi = {10.48550/arXiv.1903.05128},
archivePrefix = {arXiv},
       eprint = {1903.05128},
 primaryClass = {astro-ph.CO},
       adsurl = {https://ui.adsabs.harvard.edu/abs/2019astro2020T.270S}
}

@article{Ashton_2022,
   title={Nested sampling for physical scientists},
   volume={2},
   pages={39},
   ISSN={2662-8449},
   url={http://dx.doi.org/10.1038/s43586-022-00121-x},
   DOI={10.1038/s43586-022-00121-x},
   number={1},
   journal={NRvMP},
   publisher={Springer Science and Business Media LLC},
   author={Ashton, Greg and Bernstein, Noam and Buchner, Johannes and Chen, Xi and Csányi, Gábor and Fowlie, Andrew and Feroz, Farhan and Griffiths, Matthew and Handley, Will and Habeck, Michael and Higson, Edward and Hobson, Michael and Lasenby, Anthony and Parkinson, David and Pártay, Livia B. and Pitkin, Matthew and Schneider, Doris and Speagle, Joshua S. and South, Leah and Veitch, John and Wacker, Philipp and Wales, David J. and Yallup, David},
   year={2022},
   month=May }

@INPROCEEDINGS{Skilling_2004,
       author = {{Skilling}, John},
        title = "{Nested Sampling}",
    booktitle = {Bayesian Inference and Maximum Entropy Methods in Science and Engineering: 24th International Workshop on Bayesian Inference and Maximum Entropy Methods in Science and Engineering},
         year = 2004,
       editor = {{Fischer}, Rainer and {Preuss}, Roland and {Toussaint}, Udo Von},
       series = {AIP Conference Series},
       volume = {735},
        month = nov,
    publisher = {AIP},
        pages = {395-405},
          doi = {10.1063/1.1835238},
       adsurl = {https://ui.adsabs.harvard.edu/abs/2004AIPC..735..395S}
}

@article{Skilling_2006,
    author = {John Skilling},
    title = {{Nested sampling for general Bayesian computation}},
    volume = {1},
    journal = {Bayesian Analysis},
    number = {4},
    publisher = {International Society for Bayesian Analysis},
    pages = {833 -- 859},
    year = {2006},
    doi = {10.1214/06-BA127},
    URL = {https://doi.org/10.1214/06-BA127}
}

@article{Feroz_2009,
   title={MultiNest: an efficient and robust Bayesian inference tool for cosmology and particle physics},
   volume={398},
   ISSN={1365-2966},
   url={http://dx.doi.org/10.1111/j.1365-2966.2009.14548.x},
   DOI={10.1111/j.1365-2966.2009.14548.x},
   number={4},
   journal={\mnras},
   publisher={Oxford University Press (OUP)},
   author={Feroz, F. and Hobson, M. P. and Bridges, M.},
   year={2009},
   month=Oct, pages={1601-1614} }

@article{Handley_2015,
   title={polychord: next-generation nested sampling},
   volume={453},
   ISSN={1365-2966},
   url={http://dx.doi.org/10.1093/mnras/stv1911},
   DOI={10.1093/mnras/stv1911},
   number={4},
   journal={\mnras},
   publisher={Oxford University Press (OUP)},
   author={Handley, W. J. and Hobson, M. P. and Lasenby, A. N.},
   year={2015},
   month=sep, pages={4384-4398} }

@inproceedings{Geyer_2011,
    author = {Geyer, C.~J.},
    editor = {Brooks, S. and Gelman, A. and Jones, G.~L. and Meng, X.-L.},
    title = {Introduction to Markov Chain Monte Carlo},
    booktitle = {Handbook of Markov Chain Monte Carlo},
    publisher = {Chapman \& Hall/CRC},
    address = {New York},
    year = {2011},
    chapter = {1},
    edition = {1st},
    pages = {3--48},
    doi = {10.1201/b10905-2}
}

@ARTICLE{Metzger_2010,
       author = {{Metzger}, B.~D. and {Mart{\'\i}nez-Pinedo}, G. and {Darbha}, S. and {Quataert}, E. and {Arcones}, A. and {Kasen}, D. and {Thomas}, R. and {Nugent}, P. and {Panov}, I.~V. and {Zinner}, N.~T.},
        title = "{Electromagnetic counterparts of compact object mergers powered by the radioactive decay of r-process nuclei}",
      journal = {\mnras},
         year = 2010,
        month = aug,
       volume = {406},
       number = {4},
        pages = {2650-2662},
          doi = {10.1111/j.1365-2966.2010.16864.x},
archivePrefix = {arXiv},
       eprint = {1001.5029},
 primaryClass = {astro-ph.HE},
       adsurl = {https://ui.adsabs.harvard.edu/abs/2010MNRAS.406.2650M}
}

@article{kulkarni_2005,
       author = {{Kulkarni}, S.~R.},
        title = "{Modeling Supernova-like Explosions Associated with Gamma-ray Bursts with Short Durations}",
      journal = {arXiv e-prints},
         year = 2005,
        month = oct,
        pages = {astro\textendash{}ph/0510256},
          doi = {10.48550/arXiv.astro-ph/0510256},
archivePrefix = {arXiv},
       eprint = {astro-ph/0510256},
 primaryClass = {astro-ph},
       adsurl = {https://ui.adsabs.harvard.edu/abs/2005astro.ph.10256K}
}

@ARTICLE{Li_1998,
       author = {{Li}, Li-Xin and {Paczy{\'n}ski}, Bohdan},
        title = "{Transient Events from Neutron Star Mergers}",
      journal = {\apjl},
         year = 1998,
        month = nov,
       volume = {507},
       number = {1},
        pages = {L59-L62},
          doi = {10.1086/311680},
archivePrefix = {arXiv},
       eprint = {astro-ph/9807272},
 primaryClass = {astro-ph},
       adsurl = {https://ui.adsabs.harvard.edu/abs/1998ApJ...507L..59L}
}

@ARTICLE{Lattimer_1976,
       author = {{Lattimer}, J.~M. and {Schramm}, D.~N.},
        title = "{The tidal disruption of neutron stars by black holes in close binaries.}",
      journal = {\apj},
         year = 1976,
        month = dec,
       volume = {210},
        pages = {549-567},
          doi = {10.1086/154860},
       adsurl = {https://ui.adsabs.harvard.edu/abs/1976ApJ...210..549L}
}

@ARTICLE{Nakar_2020,
       author = {{Nakar}, Ehud},
        title = "{The electromagnetic counterparts of compact binary mergers}",
      journal = {\physrep},
         year = 2020,
        month = nov,
       volume = {886},
        pages = {1-84},
          doi = {10.1016/j.physrep.2020.08.008},
archivePrefix = {arXiv},
       eprint = {1912.05659},
 primaryClass = {astro-ph.HE},
       adsurl = {https://ui.adsabs.harvard.edu/abs/2020PhR...886....1N}
}

@article{Dormand_1980,
title = {A family of embedded Runge-Kutta formulae},
journal = {JCoAM},
volume = {6},
number = {1},
pages = {19-26},
year = {1980},
issn = {0377-0427},
doi = {10.1016/0771-050X(80)90013-3},
author = {J.R. Dormand and P.J. Prince},
}

@ARTICLE{Kerzendorf_2014,
       author = {{Kerzendorf}, Wolfgang E. and {Sim}, Stuart A.},
        title = "{A spectral synthesis code for rapid modelling of supernovae}",
      journal = {\mnras},
         year = 2014,
        month = may,
       volume = {440},
       number = {1},
        pages = {387-404},
          doi = {10.1093/mnras/stu055},
archivePrefix = {arXiv},
       eprint = {1401.5469},
 primaryClass = {astro-ph.SR},
       adsurl = {https://ui.adsabs.harvard.edu/abs/2014MNRAS.440..387K}
}

@article{Rogers_2021,
   title = "{General framework for cosmological dark matter bounds using N -body simulations}",
   volume={103},
   ISSN={2470-0029},
   url={http://dx.doi.org/10.1103/PhysRevD.103.043526},
   DOI={10.1103/physrevd.103.043526},
   number={4},
   journal={\prd},
   publisher={American Physical Society (APS)},
   author={Rogers, Keir K. and Peiris, Hiranya V.},
   year={2021},
   pages = {043526},
   month=feb }

@article{Pedersen_2021,
   title={An emulator for the Lyman-α forest in beyond-ΛCDM cosmologies},
   volume={2021(05)},
   ISSN={1475-7516},
   url={http://dx.doi.org/10.1088/1475-7516/2021/05/033},
   DOI={10.1088/1475-7516/2021/05/033},
   number={05},
   journal={\jcap},
   publisher={IOP Publishing},
   author={Pedersen, Christian and Font-Ribera, Andreu and Rogers, Keir K. and McDonald, Patrick and Peiris, Hiranya V. and Pontzen, Andrew and Slosar, Anže},
   year={2021},
   month=may, 
   pages={033} }

@ARTICLE{Leistedt_2026,
       author = {{Leistedt}, Boris and {Peiris}, Hiranya V. and {Halder}, Anik and {Thorp}, Stephen and {Mortlock}, Daniel J. and {Loureiro}, Arthur and {Alsing}, Justin and {Jagwani}, Gurjeet and {Tudorache}, Madalina N. and {Deger}, Sinan and {Leja}, Joel and {Van den Bussche}, Benedict and {Wright}, Angus H. and {Li}, Shun-Sheng and {Kuijken}, Konrad and {Hildebrandt}, Hendrik},
        title = "{pop-cosmos: Forward modeling KiDS-1000 redshift distributions using realistic galaxy populations}",
      journal = {arXiv e-prints},
         year = 2026,
        month = feb,
          eid = {arXiv:2602.03935},
        pages = {arXiv:2602.03935},
          doi = {10.48550/arXiv.2602.03935},
archivePrefix = {arXiv},
       eprint = {2602.03935},
 primaryClass = {astro-ph.CO},
       adsurl = {https://ui.adsabs.harvard.edu/abs/2026arXiv260203935L}
}

@article{Corner_2016,
      doi = {10.21105/joss.00024},
      url = {https://doi.org/10.21105/joss.00024},
      year  = {2016},
      month = {jun},
      publisher = {The Open Journal},
      volume = {1},
      number = {2},
      pages = {24},
      author = {Daniel Foreman-Mackey},
      title = {corner.py: Scatterplot matrices in Python},
      journal = {The Journal of Open Source Software}
    }

@software{Pandas_2010,
    author = {{The pandas development team}},
    doi = {10.5281/zenodo.3509134},
    license = {BSD-3-Clause},
    title = {{pandas-dev/pandas: Pandas}},
    url = {https://github.com/pandas-dev/pandas},
    year = {2010}
}
\bibliographystyle{aasjournalv7}
\setlength{\tabcolsep}{4pt}  
\appendix
\vspace{-0.5cm}  
\section{Notation}
\label{app:notation}

A summary of the notation used in this paper is presented in \cref{tab:notation}. For clarity, the table is broken up into four sections: kilonova models, likelihood-based Bayesian inference, Gaussian process emulation, and simulation based inference with flow matching.
\vspace{-0.25cm}  
\renewcommand{\arraystretch}{1.}
\begin{table*}[!h]
    \centering
    \caption{Notation and symbols used throughout the paper. Parameters are organized by category.}
    \label{tab:notation}
    \begin{subtable}{0.49 \textwidth}
        \centering
        \caption*{\underline{\textit{kilonova models}}}
        \label{tab:notation:kilonova}
        \begin{tabular}{p{1.5cm}p{4.5cm}}
        \hline \hline
        symbol & description \\
        \hline
        ${\rm wind}$ & wind ejecta \\
        ${\rm dyn}$ & dynamical ejecta \\
        $m_{\rm ej}$ & single component ejecta mass \\
        $M_{\rm tot}$ & total ejecta mass \\
        $Y_{\rm e}$ & electron fraction \\
        $v_{\rm ej}$ & avg.\ ejecta velocity \\
        $q$ & shape parameter \\
        $\theta_{\rm v}$ & viewing angle \\
        $\rho_0$ & density at reference time $t_0$ \\
        $v_0$ & velocity scale \\
        $v_r$ & \textsc{possis} model coordinate \\
        $\theta$ & \textsc{possis} model coordinate \\
        $d_L$ & luminosity distance \\
        $\bm{\phi}$ & one-component param.\ vector \\
        $\bm{\Phi}$ & two-component param.\ vector \\
        $n_{\rm grid}$ & grid cells per side \\
        $t_0$ & simulation reference time \\
        $t$ & time post-merger \\
        $\beta$ & photometric band \\
        $\mu_\beta$ & asinh magnitude \\
        $\Gamma_\beta$ & photon flux \\
        $\Gamma_{\beta,0}$ & AB zero-point photon flux \\
        $f_\beta$ & energy flux \\
        $f_{\beta,0}$ & AB zero-point energy flux \\
        \hline \hline
        \end{tabular}
    \end{subtable}
    \hfill
    \begin{subtable}{0.49\textwidth}
        \centering
        \caption*{\underline{\textit{Gaussian process emulator}}}
        \label{tab:notation:gp}
        \begin{tabular}{p{1.5cm}p{4.5cm}}
        \hline \hline
        symbol & description \\
        \hline
        $N$ & number of training simulations \\
        $J$ & number of time points \\
        $B$ & number of photometric bands \\
        $*$ & test point \\
        $\bm{K}_{\beta}$ & GP covariance matrix \\
        $\bm{\mathcal{K}}$ & stacked covariances \\
        $\bm{\psi}_\beta$ & GP hyperparameters \\
        $\sigma_{0,\beta}^2$ & GP amplitude \\
        $\ell_{\beta,1:5}$ & GP length scales \\
        $\sigma_{w,\beta}^2$ & GP white-noise \\
        $k_\beta(\bm{\phi}, \bm{\phi}')$ & GP kernel \\
        $\delta_{mn}$ & Kronecker delta \\
        $\bar{\mu}_{\beta}$ & GP mean magnitude \\
        $\bar{f}$ & GP mean flux \\
        $\Tilde{\bm{\mu}}_\beta$ & normalized magnitudes \\
        $\mathrm{Var}()$ & variance \\
        $\mathrm{Cov}()$ & covariance \\
        $\bm{Y}_\beta$ & GP training set \\
        $\bm{\mathcal{Y}}$ & stacked training sets \\
        $\mathcal{A}(\bm{\Phi})$ & acquisition function \\
        $\alpha$ & exploration/exploitation ratio \\
        $\mathcal{L}(\bm{\psi}_\beta)$ & GP loss (marginal log likelihood) \\
        \hline \hline
        \end{tabular}
    \end{subtable}
    
    \vspace{0.3cm}
    
    \begin{subtable}{0.49\textwidth}
        \centering
        \caption*{\underline{\textit{likelihood-based Bayesian inference}}}
        \label{tab:notation:likelihood}
        \begin{tabular}{p{1.5cm}p{4.5cm}}
        \hline \hline
        symbol & description \\
        \hline
        $\bm{\sigma_{m}}$ & model uncertainty \\
        $\Sigma$ & uncertainty matrix of observed data \\
        $\mathcal{L}(\bm{d}_{\rm obs}|\bm{\Phi})$ & likelihood \\
        $n_{\rm obs}$ & number of observations \\
        $\mathcal{P}(\bm{\Phi}|\bm{d})$ & posterior \\
        $\bm{d}$ & data vector \\
        $\bm{d}_{\rm obs}$ & observed AT2017gfo data \\
        \hline \hline
        \end{tabular}
    \end{subtable}
    \hfill
    \begin{subtable}{0.49\textwidth}
        \centering
        \caption*{\underline{\textit{simulation-based inference (flow matching)}}}
        \label{tab:notation:sbi}
        \begin{tabular}{p{1.5cm}p{4.5cm}}
        \hline \hline
        symbol & description \\
        \hline
        $\tau$ & pseudo-time \\
        $\bm{z}$ & latent variable \\
        $\bm{z}_0$ & target distribution \\
        $\bm{z}_1$ & base distribution \\
        $\bm{u}$ & target velocity field \\
        $\bm{v}_{\vartheta}(\bm{z}_\tau, \tau, \tilde{\bm{f}})$ & learned velocity field \\
        $\hat{\bm{f}}$ & light curve (model flux + noise) \\
        $\tilde{\bm{f}}$ & normalized light curve \\
        $\mathcal{L}(\vartheta)$ & flow matching loss \\
        $\vartheta$ & neural network parameters \\
        \hline \hline
        \end{tabular}
    \end{subtable}
    
\end{table*}

\end{document}